\documentclass[fleqn,usenatbib]{mnras}

\DeclareRobustCommand{\VAN}[3]{#2}
\let\VANthebibliography\thebibliography
\def\thebibliography{\DeclareRobustCommand{\VAN}[3]{##3}\VANthebibliography}

\DeclareSymbolFont{CMletters}{OML}{cmm}{m}{it}

\usepackage{graphicx}
\usepackage{amsmath}
\usepackage{amssymb}

\usepackage{newtxtext}
\usepackage[varvw]{newtxmath}
\usepackage[T1]{fontenc}
\usepackage{bm}

\DeclareMathSymbol{v}{\mathord}{CMletters}{`v}

\title[Haloes in SFDM with Repulsive Self-Interaction]
{Core-envelope haloes in scalar field dark matter with repulsive self-interaction: 
fluid dynamics beyond the de Broglie wavelength}

\author[Dawoodbhoy, Shapiro, \& Rindler-Daller]{
Taha Dawoodbhoy,$^{1}$\thanks{E-mail: tahad@astro.as.utexas.edu}
Paul R. Shapiro,$^{1}$\thanks{E-mail: shapiro@astro.as.utexas.edu}
and Tanja Rindler-Daller$^{2}$\thanks{E-mail: tanja.rindler-daller@univie.ac.at}
\\
$^{1}$Department of Astronomy and Texas Cosmology Center, University of Texas at Austin, Austin, TX 78712-1083, USA\\
$^{2}$Institut f\"ur Astrophysik, Universit\"atssternwarte Wien, University of Vienna, T\"urkenschanzstr.17, A-1180 Vienna, Austria
}

\date{Accepted XXX. Received YYY; in original form ZZZ}

\pubyear{2021}

\begin{document}
\label{firstpage}
\pagerange{\pageref{firstpage}--\pageref{lastpage}}
\maketitle

\begin{abstract}
Scalar Field Dark Matter (SFDM) comprised of ultralight bosons 
has attracted great interest as an alternative to standard, collisionless Cold Dark Matter (CDM) 
because of its novel structure-formation dynamics, described by the coupled Schr\"odinger-Poisson equations. 
In the free-field (``fuzzy'') limit of SFDM (FDM), structure is inhibited below the 
de~Broglie wavelength, but resembles CDM on larger scales.  
Virialized haloes have ``solitonic'' cores of radius $\sim\lambda_\text{deB}$, 
surrounded by CDM-like envelopes. When a strong enough repulsive self-interaction (SI) 
is also present, structure can be inhibited below a second length scale, $\lambda_\text{SI}$, 
with $\lambda_\text{SI}> \lambda_\text{deB}$ – called the Thomas-Fermi (TF) regime. 
FDM dynamics differs from CDM because of quantum pressure, and SFDM-TF differs further 
by adding SI pressure.  In the small-$\lambda_\text{deB}$ limit, 
however, we can model all three by fluid conservation equations for a compressible, $\gamma=5/3$ ideal 
gas, with ideal gas pressure sourced by internal velocity dispersion and, 
for the TF regime, an added SI pressure, $P_\text{SI}\propto \rho^2$.  We use these fluid equations to simulate halo 
formation from gravitational collapse in 1D, spherical symmetry, demonstrating for the first time that SFDM-TF haloes form with
cores the size  of $R_\text{TF}$, the radius of an SI-pressure-supported $(n=1)$-polytrope, 
surrounded by CDM-like envelopes.  In comparison with rotation curves of dwarf galaxies in the 
local Universe, SFDM-TF haloes pass the [``too-big-to-fail'' + ``cusp-core'']-test 
if $R_\text{TF}\gtrsim 1$ kpc.  
\end{abstract}




\section{Introduction}

\subsection{Overview}
\label{sec:Overview}

The quest for the origin and nature of cosmic dark matter has led to a recent explosion of interest in models based on a scalar field 
comprised of ultralight bosons ($10^{-3} \text{ eV} \gtrsim mc^2 \gtrsim 10^{-23}$ eV), which we shall refer
to as Scalar Field Dark Matter (SFDM).  SFDM is an alternative to 
the standard Cold Dark Matter (CDM) model, in which weakly-interacting massive ($mc^2 \gtrsim 1$ GeV) particles (WIMPs), produced as thermal relics of 
the Big Bang, behave as a non-relativistic, collisionless gas of point masses, interacting only gravitationally, once their abundance froze out when their weak interactions became slower than Hubble expansion.
This CDM model has been generally successful at matching observations
of the Universe and the formation of galaxies and structure
on large scales, but faces a challenge from apparent discrepancies between
its predictions and observations of small-scale structure.  The latter
include, for example, density profiles inside galactic
haloes which diverge toward the center, while observations prefer profiles that flatten (the ``cusp-core'' problem), 
and over-prediction of the number of sub-haloes that survive the hierarchical assembly of small galaxies into
large ones, when compared with the observed satellite dwarf galaxies
of the Milky Way and the Local Group (the ``missing satellites'' and
``too-big-to-fail'' problems) \citep[see, e.g.,][and references therein, for a review]{BBK17}.
Meanwhile, attempts to detect the putative WIMPs, either
directly in the laboratory, or indirectly by astronomical
signals from their decay or annihilation, have thus far failed
and, in so doing, significantly reduced the space of allowed parameters \citep[see, e.g.,][]{Strigari13,Schumann19}.   
For these reasons, attention has increased
to alternatives to standard CDM which might share its success 
on large scales, match observations on galactic 
scales better, and side-step the challenge to WIMPs from their null-detection in
dark matter searches.  

Structure formation in SFDM differs from that
in the non-relativistic, collisionless gas of CDM particles, because
SFDM is a Bose-Einstein condensate (BEC) and quantum superfluid.
As we shall emphasize in what follows, this distinction is most important
on small scales, while on large scales, the behavior of SFDM approaches
that of CDM, as required for it to be a viable alternative
that preserves the successes of the latter.  There are two possible length 
scales which characterize the scale below which the dynamics of SFDM
differs from that of CDM and structure is suppressed, depending on
the properties of the scalar field. One is the de~Broglie wavelength, which depends on the boson mass ($m$) and their characteristic velocity ($v$),
$\lambda_\text{deB} = h/mv$, and which can be astronomically large for ultralight bosons. 
The second is the scale $\lambda_\text{SI}$, which results from the presence 
of a possible repulsive\footnote{
Self-interaction can also be attractive, but in that
case, it would not suppress small-scale structure,
so we do not consider this possibility here.} quartic self-interaction (SI).
In the free-field limit
(in which SI is absent), sometimes referred to as Fuzzy Dark Matter
(FDM), structure formation is suppressed on scales below $\lambda_\text{deB}$. 
On the other hand, in the presence of SI with self-coupling strength $g$, $\lambda_\text{SI} \propto \sqrt{g/m^2}$,
and if the interaction is strong enough that 
$\lambda_\text{SI} \gg \lambda_\text{deB}$, structure is suppressed below $\lambda_\text{SI}$,
instead.  We shall refer to this latter case as the Thomas-Fermi (TF)
regime.  If the SFDM model should avoid the small-scale structure problems of CDM, either $\lambda_\text{deB}$ or $\lambda_\text{SI}$ should be $\sim 1$ kpc. 

Aside from the suppression of small-scale structure, the novel gravitational
dynamics of SFDM,
which obeys the nonlinear Schr\"odinger equation (NLSE) (also known as 
the Gross-Pitaevski equation), distinguishes it in other ways 
from CDM, which obeys the collisionless Boltzmann equation (CBE) 
(also known as the Vlasov equation), instead. As we shall describe in more detail later, the NLSE
can be rewritten, without loss of generality, in terms of
``quantum hydrodynamics'' (QHD) equations for the conservation of mass and momentum
that resemble those of compressible fluid dynamics.  However, in the 
QHD case, the force term associated with 
a standard thermal gas pressure is replaced by
two different quantities, one that reflects the kinetic term and one
the SI-potential term
in the NLSE: a ``quantum pressure'' tensor
$\bm{\Pi}$, and a scalar ``self-interaction pressure''
$P_\text{SI}$.  The former depends upon
spatial derivatives of the density, while the latter is proportional
to the square of that density.  This $P_\text{SI}$ corresponds, therefore, to
a gas with an isotropic pressure that obeys a
polytropic equation of state with index $n = 1$.

Several novel features of the structures that arise from 
gravitational instability in SFDM which distinguish it from CDM
can be understood from this description. 
In the TF regime, for example, if we neglect the quantum pressure term altogether,
the solution for hydrostatic equilibrium in spherical symmetry is
the well known analytical solution of the Lane-Emden equation for an 
($n = 1$)-polytrope, with a finite radius which depends only 
on the value of $g/m^2$ (independent of the total mass inside the sphere), 
and a flattened core.  In the free-field, FDM limit, on the other hand, 
the equation of hydrostatic equilibrium in
spherical symmetry yields a numerical solution (corresponding to the 
ground-state solution of the NLSE), with a finite-mass profile that extends to infinity, 
and a flattened core that contains most of that
mass within a radius $R \sim \lambda_\text{deB} = h/mv$, for $v \sim \sqrt{GM/R}$.
As such, $R \propto M^{-1}$.

This distinguishes SFDM from CDM, 
where the latter tends to form virialized objects supported against gravity
by the random motions of particle orbits, ``thermalized'' by violent relaxation
arising from density inhomogeneities during collapse.  For these CDM ``haloes'', 
the central density profile is found by N-body simulations 
to diverge toward the center,
as steeply as $\sim r^{-1}$ to $r^{-1.5}$, steepening at large radius to
fall off asymptotically as $r^{-3}$. 
It is customary to describe this CDM profile in terms of a
fitting formula referred to as the NFW profile \citep*{NFW97}, 
although the innermost slope in high-resolution N-body simulations departs
somewhat from its shape of $\sim r^{-1}$. By convention, the total mass in this
CDM profile is taken to be that inside the radius within which the
average density is related to the critical density of the
background Universe from which it formed by gravitational collapse, 
by an overdensity factor $\Delta_\text{crit}$ (e.g. $\Delta_\text{crit} \simeq 200$ for an Einstein-de Sitter universe 
or at early times in the observed $\Lambda$CDM universe). Apart from their
different density profiles, the mass-radius relationship of 
virialized haloes that form in N-body simulations of
CDM is also quite different from that of the equilibrium objects 
described above for SFDM, in either regime.

The equilibrium objects described above for the two regimes of SFDM,
however, are just the ``solitonic cores'' expected to form when 
gravitational instability leads a larger mass to collapse and virialize. 
As we noted in \citet{RDS14}, the process of forming a virialized
galactic halo in the SFDM model must become similar to that in CDM
on large scales, larger than both $\lambda_\text{deB}$ and $\lambda_\text{SI}$, and, hence,
larger than the radii of these solitonic cores.  We suggested that the
quantum fluid behavior of SFDM under these conditions would result in 
random internal wave motions equivalent to the random
particle orbits of CDM, capable of supporting haloes of size and
mass just as large as those of CDM haloes, in virial equilibrium, 
well beyond those of isolated solitonic cores. Meanwhile, numerical
simulations have been performed in detail for the case of FDM,
finding that all virialized haloes indeed have solitonic cores of the size
of the de~Broglie wavelength (as evaluated inside the haloes), 
just like the ground-state solution of the NLSE described above, which are surrounded by
envelopes with profiles that resemble the NFW profiles of CDM,
supported there by random motions suggestive of quantum turbulence
(\citealp*{SCB14,SNE16}; \citealp{Mocz17}).  Henceforth, we shall refer to this as the
core-envelope structure of virialized FDM haloes.
In what follows, we will demonstrate that a similar core-envelope
structure arises in the virialized objects that form from
gravitational instability and collapse of SFDM in the TF regime, too,
but with its core size set by the radius of the SI polytrope,
instead of the de~Broglie wavelength.

Numerical simulations in 3D like those described above
are computationally more challenging for the TF regime than for the
FDM regime, because they require a much larger dynamic range.
In both cases, it is necessary to resolve scales well below the
de~Broglie wavelength everywhere in the collapsing object, but
in the TF regime, $\lambda_\text{deB}$ is, by definition, already much smaller 
than the size of the polytropic core, which is, in turn, much smaller
than the total virialized object, which is, in turn, much smaller
than the initial linear perturbation from which it forms by gravitational collapse.
To overcome this challenge, we shall take full advantage of the fact
that, for the problem at hand, we are interested in the behavior in
the limit of small de Broglie wavelength.   However, we must do this in
such a way that we retain the full impact of the QHD dynamics we need to follow, 
including full account of the emergent large-scale effects 
of the quantum pressure term. And we will also take advantage of the fact 
that the resulting virialized object can be well-approximated by
solving the problem in 1D, spherical symmetry, thereby optimizing 
our high dynamic range to levels well beyond the current limitations 
of 3D simulation. 

We shall consider the time-dependent problem of dynamically evolving the scalar field 
to form a virialized object as the outcome of 
gravitational instability, from linear perturbation through highly-nonlinear 
collapse.  Our focus will be on the TF regime, in which $\lambda_\text{deB}\ll \lambda_\text{SI}$.  
A naive expectation of what the dynamics of this regime would look like might be found by 
dropping the quantum pressure term in the QHD momentum equation, and relying only on the self-interaction pressure 
(i.e. assuming $P_\text{SI} \gg \Pi$ everywhere). However, as we will show, the self-interaction pressure will be small 
on scales larger than $\lambda_\text{SI}$, and so will be unable to support a CDM-like envelope against gravity. 
The quantum pressure, therefore, cannot be neglected entirely, because it must be the driver of the aforementioned turbulent wave motions
that FDM simulations have shown produce such an envelope, and should dominate over the self-interaction pressure there. 
As we shall show,
the best way to understand how the quantum pressure acquires such a pivotal role is to reformulate SFDM dynamics in phase space and re-derive
the QHD equations by taking momentum moments of a quantum phase space 
distribution function and its equation of motion (obtained via insertion of the NLSE). 
Doing so reveals that quantum pressure is sourced by 
an internal phase-space velocity dispersion, 
which makes it analogous to the effective velocity-dispersion ``pressure'' 
that supports virialized CDM haloes against gravity.

This alternate route to deriving the QHD equations 
also provides a pathway for us to model the large-scale effects of 
quantum pressure in the envelope
without having to resolve the (arbitrarily) small scale of the de~Broglie wavelength in the TF regime. This is done by following the same procedure
involving momentum moments in phase space, but starting from a 
distribution function which is ``smoothed'' in phase space. In the limit of small smoothing scale,
this approach recovers the result with no smoothing.  However, 
when the smoothing scale is much larger than $\lambda_\text{deB}$, 
the corresponding equation of motion (again obtained via insertion of the NLSE) can then be accurately approximated by the CBE, 
thereby formalizing the resemblance between SFDM and CDM on large scales.

This correspondence between the NLSE and the CBE (on scales larger than $\lambda_\text{deB}$)
was first discussed by \citet{WK93} in the context of CDM. They pointed out that
CDM structure formation can be modeled by solving the NLSE in the free-field
limit (i.e. with no self-interaction) as a computational alternative to
simulating the gravitational N-body dynamics of a non-relativistic, collisionless
gas of CDM particles, as long as the quantum analog was tuned to make its
$\lambda_\text{deB}$ much smaller than any scale of interest.  
More recently, this correspondence has been studied further by \cite{Mocz18},
including its implications in the opposite direction, which we consider here, of
describing structure formation (on scales larger than $\lambda_\text{deB}$) in the quantum superfluid of SFDM 
by solving the CBE coupled to the Poisson equation, for the free-field limit
of FDM.   

We shall build upon this previous work and go beyond it by taking momentum moments of the CBE to derive a new
set of fluid conservation equations for SFDM which are similar to those
for an ideal, compressible gas with ratio of specific heats $\gamma =5/3$, 
as previously done for CDM by \citet{AS05}.
The ideal gas pressure of the fluid description in this case
will take explicit account of the large-scale effect of quantum pressure in SFDM.  
Temperature in the ideal gas law will now correspond to the 
internal velocity dispersion derived from the NLSE as described above, 
from momentum moments of the phase space distribution function, 
smoothed on scales large compared to $\lambda_\text{deB}$.
\citet{AS05} needed to make some assumptions
to close the infinite moment hierarchy of the CBE 
(spherical symmetry, skew-free velocity distribution, 
and isotropic velocity dispersion), which were well-justified for 
modeling the gravitational collapse and virialization which lead to 
halo formation in CDM, but which also apply to the SFDM problem
on which we focus here, so we will adopt these assumptions, as well. However, we will modify their derivation 
by replacing the gravitational potential, $\Phi$, with the total 
potential, $V$, which adds to gravity a potential, $V_\text{SI}$,
for the repulsive self-interaction.  
This will result in the appearance of a second pressure in the Euler 
equation which obeys the $(n=1)$-polytropic law 
in addition to the ideal gas pressure described above.

\subsection{Outline}
\label{sec:Outline}

Our paper is structured as follows.
In \S\ref{sec:model}, we describe the SFDM model and the
basic equations which describe its non-relativistic dynamics.  
We discuss the correspondence described above between the NLSE and 
the exact QHD equations derived from it, in \S\ref{sec:NLSE} and \S\ref{sec:QHD}. The phase-space-smoothed reformulation and its 
reduction to the CBE once smoothed on scales larger than $\lambda_\text{deB}$ 
are described in \S\ref{sec:classical} (along with Appendix~\ref{sec:QPT-to-P}). In \S\ref{sec:fluidapprox},
we introduce \textit{the fluid approximation}, valid in this limit of small $\lambda_\text{deB}$, \textit{a new tool that allows us to model halo formation in SFDM 
by hydrodynamical conservation equations identical to those for a compressible, ideal gas with $\gamma = 5/3$, except with the addition of a second pressure term, for self-interaction.}  This tool is especially important for modeling the TF regime, our primary focus here.
As we note, however, it is equally important at bringing insight regarding the CDM-like dynamics of SFDM beyond the de~Broglie scale, even in the free-field limit, without requiring either large, N-body simulations or solving the NLSE or QHD equations numerically-resolved well \emph{below} the de~Broglie scale. 

In \S\ref{sec:SI}, we discuss the characteristic length scale introduced by repulsive SI and its role in the gravitational instability, collapse, and virialization we study here, including the SI Jeans length $\lambda_\text{SI,J}$, 
the radius $R_\text{TF}$ of the $(n=1)$-polytrope that results when SFDM in the TF regime (henceforth abbreviated ``SFDM-TF''), has self-gravity balanced by SI pressure in hydrostatic equilibrium (HSE), and the expectation that such polytropes form the cores of larger virialized objects in SFDM-TF, supported outside their cores by an internal velocity dispersion analogous to that for CDM haloes. Constraints on the range of values allowed for $R_\text{TF}$
and on particle mass $m$ in order for SFDM to be viable and in the TF regime are also discussed there and in \S\ref{sec:mTF}.  

In \S\ref{sec:analytical}, we introduce an analytical approximation for SFDM-TF haloes as SI-modified isothermal spheres in HSE -- the $[(n=\infty) + (n=1)]$-\textit{double-polytrope} -- and solve the ODEs numerically. Our solutions anticipate the outcome of our dynamical collapse simulations to follow, showing how virialized objects larger than $R_\text{TF}$ can exist in SFDM-TF, with their pressure-support dominated by self-interaction inside  $R_\text{TF}$ but by velocity dispersion, instead, 
at larger radii.  We use this analytical model to predict a
core-halo mass relationship and its dependence on $R_\text{TF}$,
which we shall later compare with that from the simulations. 

In \S\ref{sec:methods}, we introduce our numerical methodology for simulating the gravitational instability and nonlinear collapse problem in SFDM-TF, leading to the formation of a virialized object we identify as a halo.  We solve the SFDM-TF fluid approximation equations described above, in
1D, spherical symmetry, by a Lagrangian hydrodynamics code
which is described there and
in Appendix~\ref{sec:avis}.
Our set-up and initial conditions are described there, as well, for a suite of simulation cases, for different input parameters.  

In \S\ref{sec:results}, we present the results of these simulations, 
showing for the first time how the fluid dynamics of gravitational 
instability and collapse in SFDM-TF leads to the formation of a virialized halo,
associated with the spherical post-shock region bounded by the strong 
accretion shock that arises during the collapse,
just as it does in the CDM model (i.e. where we have shown previously 
that CDM follows a similar fluid approximation derived directly from the CBE).  
However, in SFDM-TF haloes, a universal core-envelope structure develops in 
which the core follows the profile of the $(n=1)$-polytrope inside $R_\text{TF}$, 
while the envelope outside $R_\text{TF}$ follows the same profile that would 
have been produced had SFDM-TF been replaced, instead, by CDM, from the same 
initial conditions. In \S\ref{sec:Virialequilibrium}, we show that the post-shock 
region is, indeed, in approximate virial equilibrium, using our simulation results 
to evaluate the terms in the virial theorem for SFDM derived in 
Appendix~\ref{sec:virial}. 

In \S\ref{sec:observations}, we apply these results to model galactic rotation 
curves, to see if SFDM-TF can resolve the small-scale structure 
problems of CDM described above, namely, the too-big-to-fail and cusp-core problems,
by comparing the rotation curves of SFDM-TF haloes to observations of Local Group 
dwarf galaxies.  \cite*{VBB19} posed this question for FDM, finding a ``catch-22''
problem for that model. There is only one parameter that can be tuned for FDM, the particle mass $m$.  They found that, for $m$ small enough to make solitonic core radii large enough ($\sim 1$ kpc) and core densities small enough
to solve the too-big-to-fail problem in intermediate-mass dwarfs 
(halo masses $M_h\sim 10^{10} \text{ M}_\odot$),
the cores of more massive dwarfs ($M_h\sim 10^{11} \text{ M}_\odot$)
were too small and dense to
solve the cusp-core problem.
In fact, the latter problem for FDM haloes was made even worse than for CDM.  
For FDM solitonic cores, the core radius is inversely proportional to the mass in the core ($R_c \sim \lambda_\text{deB} \propto M_c^{-1}$), so their central density increases rapidly with increasing core mass ($\rho_c \propto M_c/R_c^3 \propto M_c^4$).
Since the core mass, in turn, increases with halo mass, thanks to the core-halo mass relationships discussed below, the central density increases, consequently, with increasing total halo mass, as well.
For SFDM-TF haloes, on the other hand, as we shall show, the core size is fixed by the polytropic radius ($R_c \sim R_\text{TF}$), which depends only on the constant quantity $g/m^2$, so their central density scales less steeply with core mass ($\rho_c \propto M_c$) and halo mass than for FDM. 
Therefore, as our analysis here will show, SFDM-TF can pass the [``too-big-to-fail'' +  ``cusp-core'']-test without suffering from the FDM ``catch-22'' problem identified by \cite{VBB19}, if the SI strength parameter $g/m^2$ is chosen to make $R_\text{TF}\gtrsim 1$ kpc.  

Finally, our conclusions are summarized in \S\ref{sec:Conclusions}.
In a companion paper \citep*[][``Paper II'']{paper2}, we build upon these results and apply the tools developed here
further, to consider the dynamics of structure formation in SFDM-TF in a broader cosmological context.

\section{Basic Equations and Modeling} 
\label{sec:model}

\subsection{The coupled Schr\"{o}dinger-Poisson equations for non-relativistic SFDM with self-interaction} \label{sec:NLSE}

    In the matter-dominated, structure formation era,
    the ultralight SFDM bosons form a non-relativistic Bose-Einstein condensate, described by a complex scalar field which
    obeys the time-dependent, nonlinear Schr\"{o}dinger equation (NLSE), also known as the Gross-Pitaevski equation:
    \begin{equation}
        i\hbar \frac{\partial \psi}{\partial t} = \Big( -\frac{\hbar^2}{2m}\nabla^2 + V \Big) \psi
        \label{eq:NLSE}
    \end{equation}
    where $m$ is the particle mass and $V$ is the potential energy term, which includes the field's self-gravity\footnote{We will neglect baryons throughout this work.} and repulsive self-interaction,
    \begin{equation}
        V = m\Phi + \frac{g}{m}|\psi|^2
        \label{eq:potential}
    \end{equation}
    Here, $g$ is the constant self-coupling strength, and $\Phi$ is the Newtonian gravitational potential, which obeys the Poisson equation
    \begin{equation}
        \nabla^2\Phi = 4\pi G \rho
    \end{equation}
    where the mass density is given by
    \begin{equation}
        \rho = |\psi|^2
        \label{eq:rho}
    \end{equation}
    which is the conserved charge density of the Schr\"{o}dinger field owing to its U(1) symmetry. That is, for the Noether 4-current density $j^\mu$, its time component is $j^0/c = \rho = |\psi|^2$.
    The corresponding 3-current density is given by
    \begin{equation}
        \bm{j} = \frac{i\hbar}{2m} (\psi \nabla\psi^* - \psi^* \nabla \psi )
        \label{eq:currentdensity}
    \end{equation}
    We can also relate this 3-current density $\bm{j}$ to the mass flux density, according to 
    \begin{equation}
        \bm{j} = \rho \bm{v}
        \label{eq:massflux}
    \end{equation}
   where $\bm{v}$ is the bulk velocity associated with mass transport, given in terms of $\psi$ by equating the expressions for $\bm{j}$ in equations (\ref{eq:currentdensity}) and (\ref{eq:massflux}).  
   
   \subsection{Quantum hydrodynamics equations} \label{sec:QHD}
   
  \subsubsection{Madelung-Bohm formulation} \label{sec:Madelung}

    It is also possible to use the NLSE to derive continuity and momentum equations in a form similar to classical hydrodynamics, which provides an exact, alternative formulation 
    of the NLSE, sometimes referred to as the equations of ``quantum hydrodynamics'' (QHD).
    There are two general approaches to obtaining these QHD equations that are often cited in the literature \citep[for a textbook presentation, we refer the reader to][and references therein]{Wyatt2005}.
    The first, pioneered by \citet{Madelung27}, is by substituting equation~(\ref{eq:polar}) below into the NLSE and separating the latter into a real equation and an imaginary one, as follows. 
    
    If we write the field in polar form (referred to as ``the Madelung transformation''), in terms of a real amplitude and phase,
    \begin{equation}
        \psi = |\psi|e^{iS}
        \label{eq:polar}
    \end{equation}
    and substitute this into equation (\ref{eq:currentdensity}), the 3-current density becomes
    \begin{equation}
        \bm{j} = \rho \frac{\hbar}{m} \nabla S
        \label{eq:Madelungcurrent}
    \end{equation}
    This represents a mass flux, for which the bulk flow velocity given by equating the $\bm{j}$ in equations (\ref{eq:massflux}) and (\ref{eq:Madelungcurrent}) is
    \begin{equation}
        \bm{v} = \frac{\hbar}{m}\nabla S
        \label{eq:v}
    \end{equation}
    Now, conservation of charge can be expressed as
    \begin{equation}
        \partial_\mu j^\mu = \frac{\partial \rho}{\partial t} + \nabla \cdot (\rho \bm{v}) = 0
        \label{eq:continuity}
    \end{equation}
    This is the continuity equation for a classical fluid with mass density $\rho$ and bulk velocity $\bm{v}$.
    
    In fact, just by identifying the mass density and bulk velocity as in equations~(\ref{eq:rho}) and (\ref{eq:v}), the imaginary part of the NLSE becomes the continuity equation (\ref{eq:continuity}). The gradient of the real equation, on the other hand, becomes a momentum conservation equation resembling the Euler equation of classical fluid mechanics, but instead of the usual term involving a gas pressure gradient in that case, there is a new, potential-like term
    (in addition to that involving $V$), called the ``quantum potential'', $Q$ \citep[sometimes also referred to as the ``Bohm potential'', after][]{Bohm52-1,Bohm52-2}:
    \begin{align}
        &\frac{\partial \bm{v}}{\partial t} + (\bm{v}\cdot \nabla)\bm{v} + \nabla Q + \frac{1}{m}\nabla V = 0 \label{eq:momentum-Madelung}\\
        &Q = -\frac{\hbar^2}{2m^2} \frac{\nabla^2 \sqrt{\rho}}{\sqrt \rho} \label{eq:Q}
    \end{align}
    For the potential energy $V$, given by equation~(\ref{eq:potential}), it is conventional to rewrite the self-interaction term as a pressure term, instead:
    \begin{equation}
        \frac{1}{m}\nabla V = \nabla\Phi + \frac{g}{m^2}\nabla \rho = \nabla\Phi + \frac{1}{\rho}\nabla P_\text{SI}
    \end{equation}
    where
    \begin{equation}
        P_\text{SI} = \frac{g}{2m^2} \rho^2
        \label{eq:PSI}
    \end{equation}
    So the momentum equation now explicitly contains two new terms, the quantum potential and the self-interaction pressure: 
    \begin{align}
        \frac{\partial \bm{v}}{\partial t} + (\bm{v}\cdot \nabla)\bm{v} + \nabla Q + \nabla\Phi + \frac{1}{\rho}\nabla P_\text{SI} = 0 \label{eq:momentum-Madelung-PSI}
    \end{align}
    
    In principle, with these exact QHD equations above, 
    we could now solve the problem we focus on here, of the formation of
    virialized objects by gravitational collapse and virialization, by familiar
    finite-difference methods for numerical hydrodynamics.  
    This has advantages over direct solution of
    the NLSE, such as are discussed, e.g., in \citet{Wyatt2005}.  For example, the 
    latter generally requires the introduction of a fictitious absorbing potential
    at the boundary. And there are great insights to be gained from 
    the study of hydrodynamical flows as initial value problems, including the identification of a bulk flow velocity, the 
    association of the kinetic energy term in the NLSE with the  
    quantum potential ($Q$-term) in the QHD Euler equation, and of the 
    conservation of current density with mass conservation via the 
    continuity equation. 
    There are significant numerical challenges for the QHD solution, too,
    however, including that posed by the nature of the nonlinear $Q$-term 
    involving high-order derivatives of the density.  The latter is referred to
    as $\emph{the node problem}$.  Near nodes, the quantum potential $Q$
    varies strongly on small spatial scales, and it becomes singular at 
    their exact locations.  This makes the equations stiff, as solutions involve
    modes with very different scales in time and space.  Even more serious,
    using standard algorithms, the evolution becomes unstable in these regions,
    and the solution fails.  And yet,
    this is precisely what we would need to overcome if we wanted to account
    for the important effects of small-scale structure on the 
    scale of $\lambda_\text{deB}$ and below, everywhere in our grid,
    as required to produce the large-scale CDM-like behavior described in our introduction.   
    And, to make matters worse, we are focused here on
    the TF regime, in which $\lambda_\text{deB}$ is far smaller than any
    scale of interest in our solution, so resolving this very small scale in
    detail everywhere would be a catastrophic computational burden. 
    Fortunately, as we shall see, we can develop an approximation that captures
    the large-scale effect of the small-scale behavior of the $Q$-term,
    but this approximation is best understood if we first relate this term to 
    an internal phase space velocity dispersion, as follows.

  \subsubsection{Phase space formulation} \label{sec:phasespace}  
    
    The second approach to obtaining the QHD equations, pioneered by \citet{Taka54}, is to take momentum moments of the equation of motion of a phase space distribution function constructed from $\psi$, known as the Wigner function \citep{Wigner32}:
    \begin{equation}
        W(\bm{x},\bm{p},t) = \frac{1}{(2\pi\hbar)^{3}} \int \psi^*(\bm{x} + \bm{y}/2, t) \psi(\bm{x} - \bm{y}/2, t) e^{i\bm{p}\cdot\bm{y}/\hbar} d^3\bm{y}
        \label{eq:wigner}
    \end{equation}
    The Wigner function defines the mass density structure of phase space, so the local mass density at a point in coordinate space is obtained by integrating $W$ over momentum space:
    \begin{equation}
        \rho(\bm{x},t) = \int W(\bm{x},\bm{p},t) d^3\!\bm{p} \label{eq:rho-W}
    \end{equation}
    Substituting equation (\ref{eq:wigner}) into (\ref{eq:rho-W}) and integrating
    then yields the same expression for density in terms of $\psi$ as before (equation~\ref{eq:rho}).
    The local average over momentum space of any quantity $A$, also known as its ``bulk'' value, is given by
    \begin{equation}
        \langle A \rangle (\bm{x},t) = \frac{1}{\rho(\bm{x},t)}\int A W(\bm{x},\bm{p},t) d^3\!\bm{p}
    \end{equation}
    For example, the bulk velocity is
    \begin{equation}
        \bm{v}(\bm{x},t) = \frac{\langle \bm{p} \rangle (\bm{x},t)}{m} = \frac{1}{\rho(\bm{x},t)}\int \frac{\bm{p}}{m} W(\bm{x},\bm{p},t) d^3\!\bm{p}
        \label{eq:v-W}
    \end{equation}
    Substituting equation~(\ref{eq:polar}) into equation (\ref{eq:v-W}) and integrating yields the same expression for bulk velocity in terms of $\nabla S$ as before (equation~\ref{eq:v}).
    We may also compute the velocity dispersion tensor (i.e. the momentum dispersion tensor, divided by $m^2$) as the mean-square average of deviations from the bulk velocity:
    \begin{align}
        \sigma^2_{ij}(\bm{x},t) &= \frac{1}{\rho(\bm{x},t)}\int \frac{(p_i - \langle p_i \rangle)(p_j - \langle p_j \rangle)}{m^2} W(\bm{x},\bm{p},t) d^3\!\bm{p} \nonumber \\
        &= (\langle p_i p_j \rangle - \langle p_i \rangle \langle p_j \rangle)/m^2 \label{eq:sigma-W}
    \end{align}
    
    The equation of motion for the Wigner function is found by computing its partial time derivative ($\partial W/\partial t$) and substituting in the NLSE (i.e. $\partial\psi/\partial t = i\hbar\nabla^2\psi/2m -iV\psi/\hbar$) as needed.
    The result, known as the Wigner-Moyal equation, is similar to the collisionless Boltzmann equation (CBE), but with additional terms that encode the effects of wave mechanics. 
    Momentum moments of the Wigner-Moyal equation are computed in the usual way, by operating on it with $\int\!d^3\!\bm{p}\, p_i^l \, p_j^m \, p_k^n$, corresponding to the $(l\!+\!m\!+\!n)^\text{th}$ moments.
    The 0$^\text{th}$ moment of the Wigner-Moyal equation produces the same continuity equation as before (equation~\ref{eq:continuity}).  The $1^\text{st}$ moment produces a momentum equation resembling the Euler equation of classical fluid mechanics,
    but instead of the usual term involving a gas pressure gradient in that case, there is a new, pressure-like term, called the ``quantum pressure'' tensor, $\bm{\Pi}$, which is sourced by the velocity dispersion tensor:
    \begin{align}
        &\frac{\partial v_i}{\partial t} + v_j \frac{\partial v_i}{\partial x_j} + \frac{1}{\rho}\frac{\partial \Pi_{ij}}{\partial x_j} + \frac{1}{m}\frac{\partial V}{\partial x_i} = 0 \label{eq:momentum} \\
        &\Pi_{ij} = \rho \sigma^2_{ij}
        \label{eq:PiRhoSigma}
    \end{align}
    The quantum pressure tensor acts as a momentum flux density tensor; its divergence accounts for the transport of momentum associated with the kinetic term in the NLSE.
    By direct integration of equation~(\ref{eq:sigma-W}), $\bm{\Pi}$ can be expressed in terms of the density $\rho$ and its spatial derivatives:
    \begin{equation}
        \Pi_{ij} = \Big(\frac{\hbar}{2m}\Big)^2 \Big( \frac{1}{\rho}\frac{\partial \rho}{\partial x_i}\frac{\partial \rho}{\partial x_j} - \frac{\partial^2 \rho}{\partial x_i \partial x_j} \Big)
        \label{eq:QPT}
    \end{equation}
    
    Despite the differences between the approaches in \S\ref{sec:Madelung} and \S\ref{sec:phasespace}, the difference between their results (equations \ref{eq:momentum-Madelung} and \ref{eq:momentum}, respectively) is purely cosmetic: in conjunction with equation~(\ref{eq:continuity}), both are exact reformulations of the NLSE, and are, indeed, equivalent to one another. By comparing equations (\ref{eq:momentum-Madelung}) and (\ref{eq:Q}) with equations (\ref{eq:momentum}) and (\ref{eq:QPT}), in fact, 
    we see that the quantum potential $Q$ and quantum pressure tensor 
    $\Pi_{ij}$ are related by  
    \begin{equation}
        \frac{\partial Q}{\partial x_i} = \frac{1}{\rho} \frac{\partial \Pi_{ij}}{\partial x_j}
        \label{eq:QeqivPi}
    \end{equation}
    As equations (\ref{eq:PiRhoSigma}) and (\ref{eq:QeqivPi}) now make clear, 
    however, the force contributed by the quantum potential
    term in the QHD momentum equation (\ref{eq:momentum-Madelung-PSI}),
    derived by substituting the Madelung transformation into the NLSE, 
    corresponds to the effective 
    ``pressure'' in the momentum flux density associated with the internal spread of
    momentum in the phase space derivation.  In both cases, these 
    ``quantum'' terms come from the kinetic term of the NLSE and encode the wave mechanical behavior of the scalar field, providing significant ``pressure'' to resist gravitational collapse, on scales comparable to the de Broglie wavelength, $\lambda_\text{deB}$. 
    As we shall see in the following section, in fact,
    this quantum pressure is even responsible for providing
    support against gravity on scales
    well beyond the de~Broglie wavelength, when the coupling of
    gravitational and quantum dynamics leads to conditions with a large-enough internal velocity dispersion. 
    
    \subsection{Dynamics on scales larger than the de~Broglie wavelength: phase space smoothing and the collisionless Boltzmann equation} \label{sec:classical}
    
    Both the NLSE and the QHD equations can be used to model SFDM dynamics, but doing so requires resolving the characteristic length scale, $\lambda_\text{deB}$. 
    However, for the purposes of this work, we are interested in SFDM in the TF regime, which considers the limit of relatively large particle mass and small de Broglie wavelength ($mc^2 \gg 10^{-21}$ eV $\Leftrightarrow \lambda_\text{deB} \ll 0.1$ kpc for a typical halo),\footnote{
    The SFDM model parameters, including the particle mass, must still fall within the limits of Bose-Einstein condensation, because we will always treat $\psi$ as a classical field.}
    so it is impractical to resolve a length scale that is, by design, negligibly small. 
    Therefore, we seek an alternative, albeit approximate, route to modeling SFDM dynamics -- one that captures the important emergent effects of the field's wave mechanics on large scales ($\gg \lambda_\text{deB}$) by smoothing over the complex details of small-scale ($\lesssim \lambda_\text{deB}$) dynamics governed by the NLSE.  As a bonus, the resulting approximation will also describe SFDM dynamics in the free-field limit (i.e. no SI), explaining why simulations of FDM halo formation report CDM-like behavior on scales larger than the de~Broglie wavelength inside haloes. 
    
    Following the work of \cite*{SKV89}, \cite{WK93}, and \cite{Mocz18}, we adopt a \textit{smoothed} phase space representation of $\psi$, known as the Husimi representation \citep{Husimi40}, by smoothing over $\psi$ with a Gaussian window of width $\eta$ and taking a Fourier transform:
    \begin{align}
        \Psi(\bm{x},\bm{p},t) &= \frac{1}{(2\pi\hbar)^{3/2}} \frac{1}{(\eta \sqrt{\pi})^{3/2}} \nonumber \\
        & \times \int e^{-(\bm{x}-\bm{y})^2/2\eta^2} \psi(\bm{y},t) e^{-i\bm{p}\cdot(\bm{y}-\bm{x}/2)/\hbar} d^3\bm{y}
        \label{eq:husimi}
    \end{align}
    The corresponding distribution function, which defines the (smoothed) mass density structure of phase space, amounts to a Gaussian-smoothed Wigner function, and is given by
    \begin{equation}
        \mathcal{F}(\bm{x},\bm{p},t) = |\Psi(\bm{x},\bm{p},t)|^2
        \label{eq:distfunc}
    \end{equation}
    The local mass density, bulk velocity, and velocity dispersion are defined in the same way as before in equations~(\ref{eq:rho-W}), (\ref{eq:v-W}),  and (\ref{eq:sigma-W}), respectively, except with the Wigner function now replaced by this smoothed phase space distribution function.
    The equation of motion for $\mathcal{F}$ can be found in the same manner as the Wigner-Moyal equation, as well, by computing $\partial\mathcal{F}/\partial t$ and substituting in the NLSE (i.e. $\partial\psi/\partial t = i\hbar\nabla^2\psi/2m -iV\psi/\hbar$) as needed.
    \cite{SKV89} and \cite{WK93} demonstrated that, when smoothing over scales much larger than the typical de~Broglie wavelength (i.e. $\eta \gg \lambda_\text{deB}$), this equation of motion reduces to the collisionless Boltzmann equation (CBE):
    \begin{equation}
        \frac{d\mathcal{F}}{dt} \equiv \frac{\partial \mathcal{F}}{\partial t} + \frac{p_i}{m}\frac{\partial \mathcal{F}}{\partial x_i} - \frac{\partial V}{\partial x_i}\frac{\partial \mathcal{F}}{\partial p_i}=0
        \label{eq:CBE}
    \end{equation}
    
    This is the same equation as that which governs the behavior of CDM, where $\mathcal{F}$ in that case would be replaced by the
    phase space distribution function of its collisionless N-body particles, and $V$ includes only the gravitational potential energy term (i.e. no SI, so $g=0$).  This explains why we expect FDM (the $g \rightarrow 0$ limit of SFDM) to be well-approximated by CDM dynamics on scales larger than $\lambda_\text{deB}$. Indeed, \cite{Mocz18} have shown that FDM simulations with a particle mass $mc^2 \gtrsim 10^{-21}$ eV bear a strong resemblance to analogous CDM simulations, 
    to the limit of their numerical resolution. Furthermore, many other FDM 
    simulations using smaller particle masses have demonstrated that FDM structures 
    resemble those of CDM on scales larger than their de~Broglie wavelengths. 
    This is, after all, one of the reasons SFDM is considered to be a viable 
    dark matter candidate -- the emergent large-scale behavior of SFDM wave mechanics 
    is CDM-like. When the repulsive SI term is included in the potential, 
    however, it will dominate over the collisionless kinematics below a certain length scale, 
    as we shall discuss below, causing the dynamics and equilibrium structures of our SFDM model 
    to deviate from those of standard CDM on that scale. The importance of this SI
    scale defines our regime of interest, the TF regime.
    
    With this formalism in place, we can proceed to derive fluid equations, similar to equations~(\ref{eq:continuity}) and (\ref{eq:momentum}), that approximate the dynamics 
    of SFDM in the TF regime on scales much larger than $\lambda_\text{deB}$ 
    \emph{while fully accounting for the effect of quantum pressure}.  
    This will involve taking momentum moments 
    of the equation of motion again, but this time the equation of motion will be the CBE.   
    As we shall see below, we must then assume something further, to close the moment hierarchy,
    in order to complete the set of fluid equations required to evolve the SFDM density and
    velocity fields dynamically from a given initial condition.

    \subsection{The fluid approximation}
    \label{sec:fluidapprox}
    
    \citet{AS05} constructed a set of fluid conservation equations for CDM dynamics
    by taking moments of the CBE for a gas of point-mass
    CDM particles that interact only gravitationally (in the limit that 
    their 2-body relaxation time is much longer than any time scales of interest).  
    As shown above, a similar CBE (equation~\ref{eq:CBE}) describes the
    evolution of the scalar field in the SFDM model when we smooth the results on scales which are large
    compared to the de~Broglie wavelength.  
    We can adopt the \citet{AS05} result for CDM to our case for SFDM, therefore.
    The potential term in their CBE, however, was just that for gravity, 
    so we must do something to add the effect of our repulsive self-interaction. 
    Fortunately, their analysis is valid for a general potential, so we can follow their work 
    and replace the purely gravitational potential by the one in equation~(\ref{eq:potential}), 
    which includes our repulsive SI potential, at the end.   
    We restate their main results here, and refer the reader to their paper for further details, 
    including justifications and tests for the approximations we make in this section. 
    
    As is standard procedure, we start by computing momentum moments of the CBE, where the $(l\!+\!m\!+\!n)^\text{th}$ moments are given by
    \begin{equation}
        \int p_i^l \, p_j^m \, p_k^n \, \frac{d\mathcal{F}}{dt} d^3\!\bm{p} = 0
    \end{equation}
    where $\mathcal{F}$ is given by equations~(\ref{eq:husimi}) and (\ref{eq:distfunc}).
    The 0$^\text{th}$ and 1$^\text{st}$ moments mimic the continuity and momentum equations of hydrodynamics \citep[see, e.g.,][]{BT1987}:
    \begin{align}
        \text{[0$^\text{th}$ moment]}& \quad \quad \frac{\partial \rho}{\partial t} + \frac{\partial (\rho v_j)}{\partial x_j} = 0 \label{eq:0mom}\\
        \text{[1$^\text{st}$ moment]}& \quad \quad \frac{\partial v_i}{\partial t} + v_j \frac{\partial v_i}{\partial x_j} + \frac{1}{\rho}\frac{\partial P_{ij}}{\partial x_j} + \frac{1}{m}\frac{\partial V}{\partial x_i} = 0 \label{eq:1mom}
    \end{align}
    where
    \begin{align}
        P_{ij} = \rho\sigma_{ij}^2 \label{eq:Pij}
    \end{align}
    This $P_{ij}$ is an effective ``pressure'' sourced by the phase space velocity dispersion, $\sigma_{ij}^2$, 
    directly analogous to the quantum pressure tensor, $\Pi_{ij}$ (equation~\ref{eq:PiRhoSigma}), in the exact QHD equations. 
    By comparing equations~(\ref{eq:momentum}) and (\ref{eq:PiRhoSigma}) with (\ref{eq:1mom}) and (\ref{eq:Pij}), 
    we see that the procedure of smoothing over scales larger than $\lambda_\text{deB}$ has 
    allowed us to approximate the ``quantum pressure'' tensor as this effective velocity-dispersion pressure tensor coming from the CBE.
    Therefore, $P_{ij}$ can rightly be thought of as an approximation of the large-scale behavior of $\Pi_{ij}$.
    We demonstrate the correspondence between $P_{ij}$ and $\Pi_{ij}$, and the validity of the smoothing procedure we employ here more generally, with worked examples of toy models in Appendix~\ref{sec:QPT-to-P}.
    
    To use the mass and momentum conservation equations (\ref{eq:0mom}) and (\ref{eq:1mom}) above,
    we must find a way to relate $\sigma_{ij}$ to the dependent variables in those equations.  
    Our ignorance of this relationship reflects the familiar conundrum of deriving continuity 
    and momentum equations from momentum moments of the CBE without closing the infinite hierarchy 
    of moments of higher order.  We are interested here in describing the dynamics of structure 
    formation in SFDM, as driven by a competition between gravitational and pressure forces.  
    This will include our describing the equilibrium structure of objects like galactic haloes 
    and their formation as the nonlinear outcome of gravitational instability and collapse.  
    For this purpose, we simplify the closure problem above by adopting spherical symmetry and 
    assuming the velocity dispersion to be isotropic, such that
    \begin{align}
        & v_\theta = v_\phi = 0 \\
        & \sigma^2_{\theta\theta} = \sigma^2_{\phi\phi} = \sigma^2_{rr} \equiv \sigma^2 \\
        & P_{\theta\theta} = P_{\phi\phi} = P_{rr} \equiv P_{\!\sigma} = \rho\sigma^2
    \end{align}
    where $r$, $\theta$, and $\phi$ are the usual spherical coordinates. This yields
    \begin{align}
        \text{[0$^\text{th}$ moment]}& \quad \quad \frac{\partial \rho}{\partial t} + \frac{1}{r^2}\frac{\partial (r^2 \rho v)}{\partial r} = 0 \label{eq:0momSpherical}\\
        \text{[1$^\text{st}$ moment]}& \quad \quad \frac{\partial v}{\partial t} + v \frac{\partial v}{\partial r} + \frac{1}{\rho}\frac{\partial P_{\!\sigma}}{\partial r} + \frac{1}{m}\frac{\partial V}{\partial r} = 0 \label{eq:1momSpherical}
    \end{align}
    where $v = v_r$, for short. Substituting our expression for the potential energy $V$ in  
    equation (\ref{eq:potential}) and using the definition of $P_\text{SI}$ in equation (\ref{eq:PSI}), equation~(\ref{eq:1momSpherical}) becomes
    \begin{align}
        \frac{\partial v}{\partial t} + v \frac{\partial v}{\partial r} + \frac{1}{\rho}\frac{\partial}{\partial r}(P_{\!\sigma}+P_\text{SI}) + \frac{\partial \Phi}{\partial r} = 0 \label{eq:1momPPSI}
    \end{align}
    This is the momentum equation for a self-gravitating fluid with two sources of pressure -- the repulsive self-interaction, and an isotropic velocity dispersion.  Now, there is one more step required to close
    the moment hierarchy. 
    
    \cite{AS05} further showed that, under the additional assumption that the velocity distribution is symmetric about its mean (i.e. $\mathcal{F}$ is symmetric in $\bm{p}$ about $\langle \bm{p} \rangle$), referred to as a skewless velocity distribution, 
    the 2$^\text{nd}$ momentum moment of the CBE can be written in the form of an energy equation for a $\gamma = 5/3$ ideal gas:
    \begin{align}
        \text{[2$^\text{nd}$ moment]}& \quad \quad\! \frac{\partial}{\partial t}\Big(\frac{3P_{\!\sigma}}{2\rho}\Big) + v \frac{\partial}{\partial r}\Big(\frac{3P_{\!\sigma}}{2\rho}\Big) + \frac{P_{\!\sigma}}{\rho r^2}\frac{\partial (r^2 v)}{\partial r} = 0 \label{eq:2momSpherical}
    \end{align}
    Therefore, the velocity dispersion pressure, $P_{\!\sigma}$, can be treated as monatomic ideal gas pressure, with a corresponding internal energy that increases when compression takes place, adiabatically,
    by the effect of $P_{\!\sigma}dV$-work. 

    In general, differential hydrodynamical conservation equations like these, 
    expressed in terms of continuous, differentiable dependent variables, 
    must be supplemented with additional relationships that ensure that 
    mass, momentum, and energy \emph{fluxes} are continuous even in the presence of 
    discontinuities, e.g. shocks, in the dependent variables.  As long as SFDM
    obeys equations~(\ref{eq:0momSpherical}), (\ref{eq:1momPPSI}), and 
    (\ref{eq:2momSpherical}), sound waves can occur, for example, which steepen
    into shocks, just as for any compressible ideal gas.  As pointed out by \cite{AS05}
    in their
    discussion of the fluid approximation for CDM,
    the gravitational instability and collapse that forms haloes
    involves supersonic infall that leads to strong accretion shocks, so we can expect the same
    to occur in SFDM.  In the adiabatic case which is relevant here 
    (i.e. no dissipation, viscosity, or heat conduction), the continuity of
    mass, momentum, and energy fluxes are expressed by the following modified adiabatic (``Rankine-Hugoniot'') shock
    jump conditions:
    \begin{align}
        & [ \rho v_\text{sf} ]_s = 0 \label{eq:massjump}\\
        & \big[ P_{\!\sigma} + P_\text{SI} + \rho v_\text{sf}^2 \big]_s = 0 \label{eq:momentumjump}\\
        &\!\left[ \rho v_\text{sf} \bigg( \frac{ 3P_{\!\sigma} }{ 2\rho } + \frac{ v_\text{sf}^2 }{ 2 } + \frac{ P_{\!\sigma} + P_\text{SI} }{ \rho } \bigg) \right]_s = 0 \label{eq:energyjump}
    \end{align}
    where the $s$-subscripted brackets refer to the difference between pre- and post-shock values, 
    i.e. $[A]_s \equiv A(\text{pre-shock}) - A(\text{post-shock})$.  
    The velocity $v_\text{sf}\equiv v - v_s$ is the bulk radial velocity as 
    measured in the rest frame of the shock (i.e. 
    $v_\text{sf}$ refers to 
    ``shock frame''), where $v_s$ is the radial velocity of the shock in the lab frame
    (i.e. the velocity of approach of the shock in the rest frame of pre-shock gas).  
    We note that, while these conditions resemble the usual adiabatic
    shock jump conditions for an ideal gas, 
    there is a distinction that results from the presence of SI. In particular, since it contributes a force
    per unit area that is discontinuous at the shock, 
    $P_\text{SI}$ must contribute to the momentum and energy flux 
    densities and their jump conditions just as does the ideal gas
    pressure $P_{\!\sigma}$.  One might be tempted to replace the pressure
    of the usual jump conditions everywhere by the sum of the two pressures,
    however this would not be correct for the first term in the brackets 
    of equation~(\ref{eq:energyjump}), since that term refers only to the
    internal energy of the ideal gas.
    
    These approximations allow us to use hydrodynamical equations similar to those for an
    ideal gas to model SFDM averaged on scales larger than the de~Broglie wavelength.   This will 
    account for the emergent, large-scale behavior of quantum pressure, as required of any meaningful approximation for
    SFDM dynamics, but without requiring us to follow the complicated 
    details of the de~Broglie-wavelength-scale variations that give rise to that behavior.
    This fluid approximation works to describe FDM (i.e. the limit of zero SI) if we
    do not care to resolve structure below the de~Broglie scale, but
    it also allows us to model SFDM \emph{with} SI, by
    adding a second pressure to the momentum equation,
    $P_\text{SI}$, that obeys the simple barotropic law of an ($n=1$)-polytrope. 
    This is particularly useful when 
    we want to study the limit of large particle mass, appropriate to SFDM-TF,
    which is the focus of this work. 
    In this limit, the smallest-scale structures we need to resolve are those
    characteristic of the repulsive SI, not the comparatively smaller scale of the de~Broglie
    wavelength, but we must still account for the full impact of quantum pressure on large scales. 
    In \S\ref{sec:analytical}, we shall use the fluid momentum equation to derive an approximate hydrostatic equilibrium solution for this model. In \S\ref{sec:methods}, we describe the method we will use to simulate the spherical collapse and formation of SFDM-TF haloes by a 1D, Lagrangian hydrodynamics scheme adapted to solve the fully time-dependent continuity, momentum, and energy equations.
    
    To summarize, our approximation scheme can be stated as follows. We want to model the behavior of SFDM in the limit where the de~Broglie wavelength is much smaller than the characteristic scale on which the
    self-interaction operates to oppose gravitational collapse (i.e. the TF regime).  
    While we are always talking about ultra-light bosons when we model SFDM, this TF regime is one in which particle mass and self-coupling strength are both large enough that the primary distinction between this model and standard CDM is the repulsive self-interaction. If the particle mass is large, then the typical de~Broglie wavelength is small, and, on scales much larger than $\lambda_\text{deB}$, the ``quantum pressure'' sourced by the kinetic term in the NLSE behaves like the effective pressure of a collisionless fluid sourced by its velocity dispersion, which is why FDM behaves like CDM on large scales. In spherical symmetry, this collisionless fluid ``pressure'' can be well-approximated by a monatomic ideal gas pressure law. Therefore, we can model the behavior of SFDM-TF in spherical symmetry by approximating the field's quantum pressure (given by equation~\ref{eq:QPT}) as the pressure of a $\gamma=5/3$ ideal gas (which obeys equation~\ref{eq:2momSpherical}), while treating the self-interaction pressure exactly (equation~\ref{eq:PSI}).
    
    The correspondence discussed by \cite{AS05} between solutions of the CBE and the analogous fluid approximation equations for the time-dependent formation of CDM haloes by gravitational instability and spherical collapse can serve as a guide to what we expect for the large-scale behavior of the solution for SFDM, as follows.  For CDM,
    as long as the initial random thermal velocities are small enough (i.e. the corresponding ideal gas is cold enough) to be highly Jeans unstable, gravitational collapse proceeds by spherical infall which, in the case of the CBE, has mass shells moving inward on purely ballistic, radial trajectories like cannonballs. In the fluid approximation, the same mass shells follow the same trajectories, falling supersonically, so the
    outward pressure force is negligible compared with the inward gravitational force. These fluid shells
    adiabatically compress as they fall.  In both cases, if that ballistic motion continued without interruption (i.e. as if the integrated mass interior to a shell was always the same, and the infall ``pressure-free''), each shell would reach the origin in a finite free-fall time which is later for shells that start at larger radii. 
    
    For the collisionless particles of the CBE solution, shells that reach the origin pass right through it 
    and reverse their radial velocities from infall to outflow.  On their way out,
    they move through other shells, including those which have not yet reached the origin, and the mass interior to a given shell in this region of shell-crossing is no longer constant in time.  Each outflowing shell reaches a maximum radius (or ``apoapse'') which is smaller than its starting radius (since there is then more mass interior to it then when it began its descent) at which it turns around and falls back again. The actual density distribution, which is smooth outside the region of shell-crossing, exhibits caustics of infinite density (but negligible mass) inside that region. 
    A plot of the phase space diagram of particle (or shell) velocities versus radius shows smooth
    infall outside the region of shell-crossing and a spiral of phase space windings around the origin within 
    the region of shell-crossing, where the
    turn-around radius of each winding (i.e. where $v = 0$) is the location of its density caustic. As time advances, more and more mass falls into the region of shell-crossing, which grows in radius as the phase space spiral within it ``winds up'', adding more and more windings to itself, and the
    shell-crossing region contains more and more caustics. 
    
    By contrast, in
    the solution of the fluid equations, infalling mass shells cannot cross each other and never reach the origin.  Instead, since the adiabatic compression of ideal gas as it approaches the origin eventually heats it enough to make the innermost velocity profile subsonic, a strong accretion shock forms at a small but finite radius, separating the subsonic gas within it from the supersonic, infalling gas outside it.
    Thereafter, as successive mass shells fall and reach the shock, their properties jump discontinuously across it, thermalizing their kinetic energy of infall into a mixture of thermal energy and a small amount of subsonic infall, shock-heating the gas to such high temperature that the sound speed is as high as the pre-shock infall velocity, while increasing its density by a factor of 4 (the strong-shock value for $\gamma = 5/3$).
    As time advances, the accretion shock overtakes more and more infalling mass and moves outward, creating a shock-bounded sphere in hydrostatic
    equilibrium, held up by its ideal gas pressure, whose radius and mass grow with time. 
    
    This post-shock sphere in the fluid solution corresponds
    closely to the region of shell-crossing and caustics
    in the CBE solution.  In fact, the mass profiles $M(<r)$ of the two solutions are almost identical at all
    radii, despite the presence of caustics in the CBE
    solution.  Moreover, the radius of the shock is close to that of the outermost caustic in the region of shell-crossing (the outermost winding of the phase space diagram, at its turn-around radius), 
    which, in recent years, has
    come to be referred to as ``the splashback radius''. 
    Finally, the velocity dispersion in the Maxwellian velocity distribution of the microscopic, random thermal motions of the ideal gas (interior to the shock) in the fluid case, as parameterized by its temperature, corresponds closely with the velocity dispersion in the rest frame of bulk motion in the phase space diagram of the particles in the CBE solution at the same radii (within the region of shell-crossing).  
    This is to be expected, since the ``temperature''\footnote{
    For SFDM, this effective velocity-dispersion ``temperature'' should not be confused with the temperature that appears in the Bose-Einstein distribution for bosons in thermodynamic equilibrium.} 
    associated with the ideal gas pressure in our energy equation~(\ref{eq:2momSpherical}), recall, corresponds directly to the velocity dispersion ($\sigma^2$) obtained from moments of the phase space distribution function and the CBE.
    
    We can identify the CDM haloes formed by gravitational instability and collapse with the ``virialized object'' in the fluid description, bounded by the accretion shock, and by the outermost caustic in the CBE description.   This is, in fact, a much more physical definition of the radial extent of a halo than the more conventional definition in the cosmological literature, which is based on the post-virialization radius predicted by a simple toy model of a homologously-expanding uniform sphere (i.e. the ``top-hat model'').  In the latter model, also known as the ``standard uniform sphere (SUS) approximation'', the initially-expanding sphere decelerates under its own gravity, reaches a maximum expansion radius from which its entire mass then collapses ballistically to the origin in a single free-fall time.  At this point, it is assumed that
    infinite collapse is avoided; initial inhomogeneities in the sphere, amplified by violent relaxation, are assumed to redistribute and ``thermalize'' its mass and energy instantaneously into a new, static, uniform sphere with the same total mass and energy, but now in virial equilibrium. In both models of virialized haloes, that of the post-shock region in the collapse calculation or of the post-collapse virialized sphere in the top-hat model, when haloes of different mass form from the same initial,  pre-collapse, mean density, the final halo mass $M_h$ and radius $R_h$ are related by $M_h \propto R_h^3$.  The only difference between them will be the constant of proportionality.
    When we compare the properties of SFDM haloes and their CDM counterparts with galaxy observations in \S\ref{sec:observations}, we will make it clear which convention we adopt for the outer radius of the halo.

    There is one more detail we note before moving on, which is that there were two versions of fluid approximation in \cite{AS05}, the one we adopt here in which we assume the velocities in the frame of bulk motion are isotropic, and another in which we assumed motions were strictly radial even on the microscopic level of the random internal motions of particles (but still skewless).   Strictly speaking, our summary above has neglected the detailed distinction between these two regimes regarding the correspondence between the
    CBE and fluid approximation
    solutions.  For more details regarding this and the correspondence between the CBE and fluid solutions
    for CDM, the reader is referred to \cite{AS05}.

    \subsection{The characteristic scale of repulsive self-interaction pressure support}
    \label{sec:SI}
    
    \subsubsection{Gravitational instability in the TF regime: the self-interaction Jeans length}
    \label{sec:SI_Jeans_Length}
    
    SI pressure introduces a characteristic length scale, analogous to the standard Jeans length, 
    below which gravitational instability is
    suppressed \citep[cf.][]{SC18}.
    Consider the QHD equations,
    equations~(\ref{eq:continuity}) and (\ref{eq:momentum-Madelung}), derived from the Madelung transformation, 
    and assume the quantum potential term in equation~(\ref{eq:momentum-Madelung})
    is everywhere subdominant and can be neglected in favor 
    of the SI term.  The term in equation~(\ref{eq:momentum-Madelung})
    involving the gradient of the SI potential
    introduces a term in that momentum equation identical to a 
    pressure force in a gas with the pressure, $P_\text{SI}$, given by equation~(\ref{eq:PSI}).
    In the TF regime, if we drop the Q-term in the momentum equation
    altogether, the continuity and momentum equations are identical
    to those for a Newtonian fluid with a barotropic equation of
    state.  As such, a uniform, infinite, static fluid which is linearly
    perturbed by a plane-wave density and velocity mode of wavelength
    $\lambda$ will, in the absence of gravity, result in 
    sound waves that travel with a sound speed $c_{s,\text{SI}}$ given by
    \begin{equation}
        c_{s,\text{SI}} = \sqrt{\frac{\partial P_\text{SI}}{\partial \rho}} = \sqrt{\frac{g \rho}{m^2}}\label{eq:SI_soundspeed}
    \end{equation}
    In the presence of self-gravity, the same plane-wave perturbation
    will oscillate like a sound wave for small wavelengths, while large-wavelength modes are gravitationally unstable, 
    with amplitudes that grow exponentially over time, instead.  
    The boundary between these
    regimes is the Jeans wavelength $\lambda_\text{SI,J}$ given by 
    \begin{equation}
         \lambda_\text{SI,J} = \sqrt{\frac{\pi c_{s,\text{SI}}^2}{G\rho}} = \sqrt{\frac{\pi g}{G m^2}}
    \end{equation}
    This Jeans length $\lambda_\text{SI,J}$ is a constant,
    independent of the density $\rho$ of the unperturbed
    fluid, and is apparently fixed entirely by SFDM particle 
    parameters ($m,g$) in their combination $g/m^2$.  For
    unstable modes with $\lambda \gg \lambda_\text{SI,J}$, however,
    exponential growth occurs on a gravitational free-fall time-scale,
    which does depend on $\rho$, proportional to $1/\sqrt{G\rho}$.
    In particular, for any mode of wavelength $\lambda$,
    the density perturbation amplitude evolves in
    proportion to $\exp{(-i\omega t)}$, with a dispersion relation
    given by
    \begin{equation}
         \omega \tau_\text{ff} = \sqrt{ \frac{\lambda_\text{SI,J}^2}{\lambda^2} - 1}
    \end{equation}
    where $\tau_\text{ff}$ is the free-fall time given by
    \begin{equation}
         \tau_\text{ff} = \frac{1}{\sqrt{4 \pi G \rho}}
    \end{equation}

    \subsubsection{Gravitational equilibrium in the TF regime:       (n~=~1)-polytropes}
    \label{sec:SI_Polytropes}
    
    The dependence of SI pressure on density in 
    equation~(\ref{eq:PSI}) is that of a polytrope ($P \propto \rho^{1+1/n}$)
    of index $n = 1$.  A self-gravitating fluid in hydrostatic equilibrium supported by 
    this pressure alone will take the shape of an $(n = 1)$-polytrope.
    According to the well-known analytical solution for such a polytrope
    in 1D spherical symmetry, 
    the density profile is given by
    \begin{equation}
        \rho = \rho_c \text{sinc}({\pi r/R_\text{TF}})
        \label{eq:polytropedensity}
    \end{equation}
    where $\rho_c$ is the central density, and $R_\text{TF}$ is a finite radius that bounds the polytrope, given by
    \begin{equation}
        R_\text{TF} = \pi\sqrt{\frac{g}{4 \pi G m^2}}
        \label{eq:RTF}
    \end{equation}
    which is just $\lambda_\text{SI,J}/2$. 
    
    The density profile above has a finite central density, drops rapidly to zero as $r\rightarrow R_\text{TF}$, and has 
    a ratio of its central-to-average density of $\pi^2/3$.  Hence,
    as an isolated object, this ($n = 1$)-polytrope would appear to have a
    flattened central profile of width $R_\text{TF}$.
    Like the Jeans length above,
    this $R_\text{TF}$ is fixed entirely by SFDM particle 
    parameters ($m,g$) in their combination $g/m^2$,
    so $R_\text{TF}$ is, itself, a physical constant of the SFDM
    model, independent of the total mass of the polytrope.
    
    \subsubsection{$R_\text{TF}$ and the polytropic cores of larger objects}
    \label{sec:Polytropic_Cores}
    
    \citet{RDS14} argued that objects larger than
    this polytrope radius can occur in SFDM in the TF regime, 
    when gravitational instability leads to its collapse and virialization,
    since quantum kinetic effects make it CDM-like on large scales.
    As we show above, in fact, when SFDM is averaged over scales larger 
    than the de Broglie wavelength, an effective ideal gas pressure
    is present which is just like that associated with 
    the velocity dispersion of random thermal motions in a 
    collisionless gas of CDM particles, orbiting inside a virialized 
    halo.  Recall that for CDM haloes, density $\rho$ diverges towards
    the center.  The effective ``gas pressure'' $\rho\sigma^2$ does not grow faster than one power of $\rho$ as radius approaches zero, 
    however, since velocity dispersion is close to isothermal 
    over most of the CDM halo, dropping in
    the very center. The polytropic SI pressure, on the other hand,
    increases toward the center as $\rho^2$, 
    so it must eventually come to dominate over
    the gas pressure at small radii.  
    In fact, in our detailed analysis to follow
    of gravitational instability, collapse and virialization in SFDM-TF, the equilibrium objects that result will 
    follow the $(n = 1)$-polytropic density profile inside their cores, 
    at $r \lesssim R_\text{TF}$ 
    (where $P_\text{SI}$ dominates over $P_{\!\sigma}$ in equation~\ref{eq:1momPPSI}), 
    while following a CDM-like profile for $r \gtrsim R_\text{TF}$. 
    If this generic core-envelope structure 
    is required to match observations of dwarf 
    galaxies in the Local Group, for example, which prefer kpc-sized cores
    to diverging CDM-like central profiles, 
    then particle mass and SI strength should be chosen to satisfy
    \begin{equation}
        R_\text{TF} \sim 1 \text{ kpc} \Longleftrightarrow \frac{g}{m^2c^4} \sim 2 \times 10^{-18} \text{ eV}^{-1} \text{ cm}^3
    \end{equation}

    \subsubsection{Cosmological constraints on $R_\text{TF}$ from the effect
        of SFDM on the expansion rate of the background Universe}
    \label{sec:SFDM_constraints} 
    
    \cite*{Li14} and \cite*{Li17}
    placed limits on the values of $m$ and $g$
    that are allowed based upon observational consequences if cosmic
    dark matter is SFDM. Unlike CDM, which was matter-like very early on,
    and only affected the expansion rate of the Universe roughly between the
    epochs of matter-radiation equality and matter-dark energy equality,
    SFDM has novel, additional consequences for the expansion rate 
    of the Universe at early times which depend on its particle parameters.
    For the case considered there, of a complex scalar
    field with a quartic SI,
    the NLSE and Poisson equation are valid only for the late-time, non-relativistic phase of
    evolution of the field as the Universe expands.  More generally,
    the field must obey the relativistic Klein-Gordon equation 
    coupled to the Einstein equations, and, 
    as the homogeneous Universe expands according to 
    the Friedmann equation, so must the homogeneous scalar field evolve, too,
    from relativistic at early times to non-relativistic at late times.
    For a given particle mass $m$, 
    the observed dark matter rest-mass density in the
    Universe today, inferred from cosmological observations, sets the
    particle number density, which is a conserved charge for this
    complex scalar field, and this allows us to trace its evolution back in 
    time, as a function of the two parameters, $m$ and $g/m^2$.
    
    According to this evolution, complex SFDM started in a stiff, 
    relativistic phase during which its
    equation of state had a ratio of pressure to energy density 
    $w \approx 1$.  This phase must have ended roughly by the time of 
    Big Bang nucleosynthesis (BBN)
    so as not to disturb the generally good 
    agreement between predictions of BBN 
    for a Universe dominated by radiation during that time and observations
    of the primordial abundance of the light elements. SFDM with SI
    then transitioned from this stiff phase to a radiation-like equation of
    state ($w \approx 1/3$), before finally transitioning to a matter-like equation of state ($ w \approx 0$), thereafter. This
    transition to matter-like must have preceded the epoch
    of matter-radiation equality in our observed Universe. In such a model,
    SFDM dominated the total energy density of the universe
    (and, therefore, its expansion rate) \textit{twice}, first during its stiff phase
    and again during its matter-like phase.  During its radiation-like
    phase, SFDM contributed to the total radiation-like energy density
    of the universe, a ``plateau'' of constant fraction of the critical
    density.   
    How late the transition from the radiation-like to the matter-like
    phase was, and how large the contribution of SFDM was to the total energy density of the universe during this radiation-like phase,
    depended on the value of $g/m^2$. If $g/m^2$ is too large, and the field remained radiation-like for too long, the universe would have
    transitioned from radiation-dominated to matter-dominated later than 
    in the $\Lambda$CDM model. However, the redshift at which this 
    transition occurs, $z_\text{eq}$, is determined observationally 
    from the Cosmic Microwave Background temperature anisotropy power spectrum, 
    so there is an upper limit for the value of $g/m^2$, 
    which \cite{Li14} determined to be
    \begin{equation}
        \frac{g}{m^2c^4} \leq 4 \times 10^{-17} \text{ eV}^{-1} \text{ cm}^3 \quad \quad \text{[for $w(z_\text{eq})\leq0.001$]}
        \label{eq:zeqlim}
    \end{equation}
    which corresponds to an upper bound on the polytrope radius of
    \begin{equation}
        R_\text{TF} \leq 5 \text{ kpc}
    \end{equation}
    The core radius needed to solve the cusp-core problem of observed
    galactic haloes ($R_\text{TF} \sim 1$ kpc), 
    is compatible with this constraint.  
    
    We noted in \cite{Li14, Li17} that $R_\text{TF}$ is further constrained by the dual requirements
    that the transition from stiff to radiation-like
    occur sufficiently early and, after transitioning,
    the radiation-like phase contributes an amount
    to the total energy density of the
    radiation-dominated Universe during BBN, so as to be
    consistent with BBN abundance determinations. 
    The stiff phase ends earlier if $m$ is higher,
    while the contribution to the energy density after transitioning
    to radiation-like is higher for higher values of $g/m^2$.  Hence,
    these dual requirements impose lower limits on these parameters. 
    For  
    \begin{align}
        \frac{g}{m^2c^4} &\geq 2 \times10^{-18} \text{ eV}^{-1} \text{ cm}^3 \label{eq:Nefflim-g}\\
        m c^2 &\geq 5 \times 10^{-21} \text{ eV} \label{eq:Nefflim-m}
    \end{align}
    the scalar field is radiation-like during BBN, as required,
    while adding to the effective number of relativistic degrees of freedom, $N_\text{eff}$ in such a way as to bring it 
    within 1$\sigma$ of the central value inferred from observations of the primordial element abundance.  Those
    observations favored a value of $N_\text{eff} = 3.56 \pm 0.23~(1\sigma)$  which is somewhat greater than the standard value of 3.046 
    for a universe which is radiation-dominated during BBN, 
    with 3 neutrino species and no extra relativistic degrees of freedom.\footnote{
    We note that, for a model like complex SFDM, the value
    of $N_\text{eff}$ during BBN can be 
    different, i.e. higher, than its value at much later times, after
    matter-radiation equality, as measured by observations
    of the CMB anisotropy, so there is no inconsistency if CMB
    observations favor a lower value of $N_\text{eff}$ than do
    those of BBN abundances.  Such a difference between 
    BBN- and CMB-era values of $N_\text{eff}$ cannot occur
    in the standard CDM model, however, so this BBN value
    of $N_\text{eff}$ is in mild tension with standard CDM,
    since CMB anisotropy measurements currently favor a value closer to
    the standard value of 3.046.}
    The lower bound on $g/m^2$ in equation~(\ref{eq:Nefflim-g}) corresponds to 
    \begin{equation}
        R_\text{TF} \geq 1 \text{ kpc}
    \end{equation}
    while, as we shall show below,
    the lower bound on $m$ in equation~(\ref{eq:Nefflim-m}) implies that $\lambda_\text{deB}$ 
    inside galactic haloes will typically be much smaller than this.
    In this regard, our assumption of the TF regime 
    with $R_\text{TF} \gtrsim 1$ kpc is self-consistent and
    occupies a particularly favored region of parameter space.  
    However, as also noted in \cite{Li14}, if
    we relax the 1$\sigma$ lower-bound above, on $N_\text{eff}$ inferred from BBN abundance
    measurements, then the lower bound on $g/m^2$ is also relaxed and, with it, the corresponding
    lower bound on $R_\text{TF}$. 
    For values of $R_\text{TF}$ below the bound in equation~(\ref{eq:Nefflim-g}), the lower bound on $m$ increases somewhat above that in equation~(\ref{eq:Nefflim-m}) as $R_\text{TF}$ decreases.   
    
    For the case of real SFDM with this same quartic SI, 
    the evolution of its equation of state
    differs at early times from that of the complex case and, as a result,
    the constraints described above do not all apply.  As we go back in time from the present, the transition between late matter-like phase 
    and the earlier phase of 
    radiation-like equation of state is subject to constraints for real SFDM, as well. 
    As such, the upper limit on $g/m^2$
    in equation~(\ref{eq:zeqlim}) is the same for both real and complex
    SFDM.   At earlier times, however, when the complex field described
    above made its transition between a radiation-like phase with 
    $w \approx 1/3$ and a stiff phase with $w \approx 1$, the real field
    makes a transition from radiation-like 
    to an early cosmological-constant-like phase with $w = -1$, instead.
    In this case, the universe was radiation-dominated at all times
    before the epoch of matter-radiation equality, and the contribution
    of an SFDM plateau to the critical density during the 
    radiation-dominated era was limited to its radiation-like phase,
    and absent during its earlier cosmological-constant-like phase.  As such, the SFDM plateau is also required to obey constraints from BBN.
    The lower limits on the values of $m$ and $g/m^2$ in equations~(\ref{eq:Nefflim-g})
    and (\ref{eq:Nefflim-m}) above, 
    required to keep the value of $N_\text{eff}$ during BBN 
    within 1$\sigma$ of the value
    inferred from primordial abundance measurements, should be
    similar for the real SFDM case, as well.  Furthermore,
    like the complex case, this lower limit on $g/m^2$ can also be relaxed 
    for the real field case if
    we relax the 1$\sigma$ lower-bound on $N_\text{eff}$.  However, the lower limit on $m$ in 
    equation~(\ref{eq:Nefflim-m}) would also be relaxed for the real field in this case, whereas
    it would remain for the complex field.

    \subsection{A lower bound on \emph{m} for the TF regime}
    \label{sec:mTF}

    We have imposed the small-$\lambda_\text{deB}$ limit on the SFDM phase space distribution function as a requirement for obtaining the CBE and establishing the TF regime. This is to ensure that $R_\text{TF} \gg \lambda_\text{deB}$, meaning that, for $r\lesssim R_\text{TF}$, quantum pressure is negligible compared to SI pressure, and for $r \gtrsim R_\text{TF} \gg \lambda_\text{deB}$, quantum pressure can be approximated by velocity-dispersion pressure. To get an estimate of the minimum particle mass that can be tolerated for these criteria to hold, we take $\lambda_\text{deB}$ for a virialized dark matter halo of mass $M_h$ and radius $R_h$ to be
    \begin{equation}
        \lambda_\text{deB} = \frac{h}{m v_\text{vir}} \propto \frac{1}{m}\sqrt{\frac{R_h}{M_h}} \propto \frac{1}{m}M_h^{-1/3}
    \end{equation}
    where $v_\text{vir} = \sqrt{G M_h/R_h}$ is approximately the virial velocity of the halo. Since $\lambda_\text{deB}$ depends on halo mass, the minimum allowed particle mass for the TF regime approximation to be valid will depend on the minimum halo mass, $M_{h,\text{min}}$, that needs to be accommodated by the model. The dependence can be expressed as
    
    \begin{equation}
        \frac{mc^2}{10^{-21}\text{ eV}} \gg \bigg( \frac{M_{h,\text{min}}}{10^{9} \text{ M}_\odot} \bigg)^{-1/3} \bigg(\frac{R_\text{TF}}{1 \text{ kpc}}\bigg)^{-1}
    \end{equation}

\section{Analytical Approximation}
\label{sec:analytical}

\subsection{SI-modified isothermal spheres: double-polytropes} 
    \label{sec:IsoPoly}

    We can construct an approximate SFDM-TF halo density profile by obtaining the hydrostatic equilibrium (HSE) solution of the momentum equation in 1D, spherical symmetry (equation~\ref{eq:1momPPSI}). As noted previously, the SI pressure term in equation~(\ref{eq:1momPPSI}) corresponds to that of an $(n=1)$-polytrope:
    $P_\text{SI} = K\rho^{1+1/n}$ with $K=g/2m^2$.
    As discussed above, there is a closed-form analytical solution for HSE in that case, if that is the only pressure term.  However, 
    to make this analytically tractable
    in the presence of the additional, velocity-dispersion ``pressure'' $P_{\!\sigma}$, we will need to make an assumption about the form of $P_{\!\sigma}$. In particular, we shall assume that the velocity dispersion is a constant, independent of radius:
    \begin{equation}
        \sigma^2 = \frac{P_{\!\sigma}}{\rho} = \text{ constant}
    \end{equation}
    which makes the associated pressure that of an isothermal sphere, i.e. the limiting case of a polytropic pressure law $P = K\rho^{1+1/n}$ for $n=\infty$, where $K=\sigma^2$ in this case. Virialized haloes formed in CDM simulations are usually close to isothermal, so it seems appropriate to assume that the CDM-like ``pressure'' (i.e. the term that should be responsible for forming a CDM-like envelope on large scales) in this SFDM-TF model is isothermal, as well. 
    
    Under this assumption, the combined effect of the two pressure terms makes this problem amount to a generalization of the well-known Lane-Emden equation for spherical polytropes, in which we replace the single polytropic pressure law by a sum of two polytropic laws, in this case corresponding to $n=1$ and $n=\infty$, respectively 
    -- a $\emph{double}$-polytrope.  In particular, we set the time-derivative and bulk velocity to zero in  equation~(\ref{eq:1momPPSI}), to
    express HSE as the balance between the inward force of gravity and the sum of the outward forces due to the isothermal and polytropic pressures:
    \begin{equation}
        -\frac{1}{\rho}\frac{d}{d r}\big(\sigma^2\rho+K\rho^{1+1/n}\big) = \frac{d \Phi}{d r}
    \end{equation}
    where we have written the SI pressure as a generic polytropic pressure law, $P_\text{SI}=K\rho^{1+1/n}$. For the model explored in this paper, $K=g/2m^2$ and $n=1$, but we will proceed with the generic form so that the result is valid for any polytropic pressure law added to a self-gravitating isothermal ideal gas. We can rewrite the equation above by eliminating the $1/\rho$ factor on the left, and taking the divergence of both sides so that the right side can be replaced with the Poisson equation:
    \begin{equation}
        -\frac{1}{r^2}\frac{d}{dr}\Big(r^2\frac{d}{dr}\big[ \sigma^2\ln\rho + (n+1)K\rho^{1/n} \big]\Big) = 4\pi G\rho
    \end{equation}
    Finally, we can non-dimensionalize using the conventional transformations adopted for the Emden-Chandrasekhar equation of a self-gravitating isothermal gas, with one additional non-dimensional variable to codify the strength of the polytropic pressure relative to the isothermal pressure:
    \begin{align}
        &\xi \equiv \frac{r}{\sqrt{\sigma^2/4\pi G \rho_c}} \equiv \frac{r}{r_0}\\
        &\zeta \equiv -\ln{(\rho/\rho_c)} \\
        &\chi \equiv (n+1) K \rho_c^{1/n}/\sigma^2
    \end{align}
    where $\rho_c$ is a parameter that, under the boundary conditions provided below, sets the central density of the HSE profile. The result is a combination of the Lane-Emden and Emden-Chandrasekhar equations, corresponding to the polytropic and isothermal pressure laws, respectively:\footnote{
    This modified isothermal Emden-Chandrasekhar equation 
    which adds the pressure term of an
    ($n=1$)-polytrope, was also explored by \cite{Chavanis19},
    although derived there from somewhat different considerations.}
    \begin{equation}
        \frac{1}{\xi^2}\frac{d}{d\xi}\Big(\xi^2\frac{d}{d\xi}\big[ \zeta - \chi e^{-\zeta/n} \big]\Big) = e^{-\zeta} \label{eq:emden}
    \end{equation}
    In anticipation of a density profile with a flat central core, we solve this differential equation with the following inner boundary conditions:
    \begin{align}
        \zeta(0) = 0 &\Longleftrightarrow \rho(0) = \rho_c \\
        \zeta'(0) = 0 &\Longleftrightarrow \rho'(0) = 0
    \label{eq:BC}    
    \end{align}
    
\begin{figure}
        \includegraphics[width=\columnwidth]{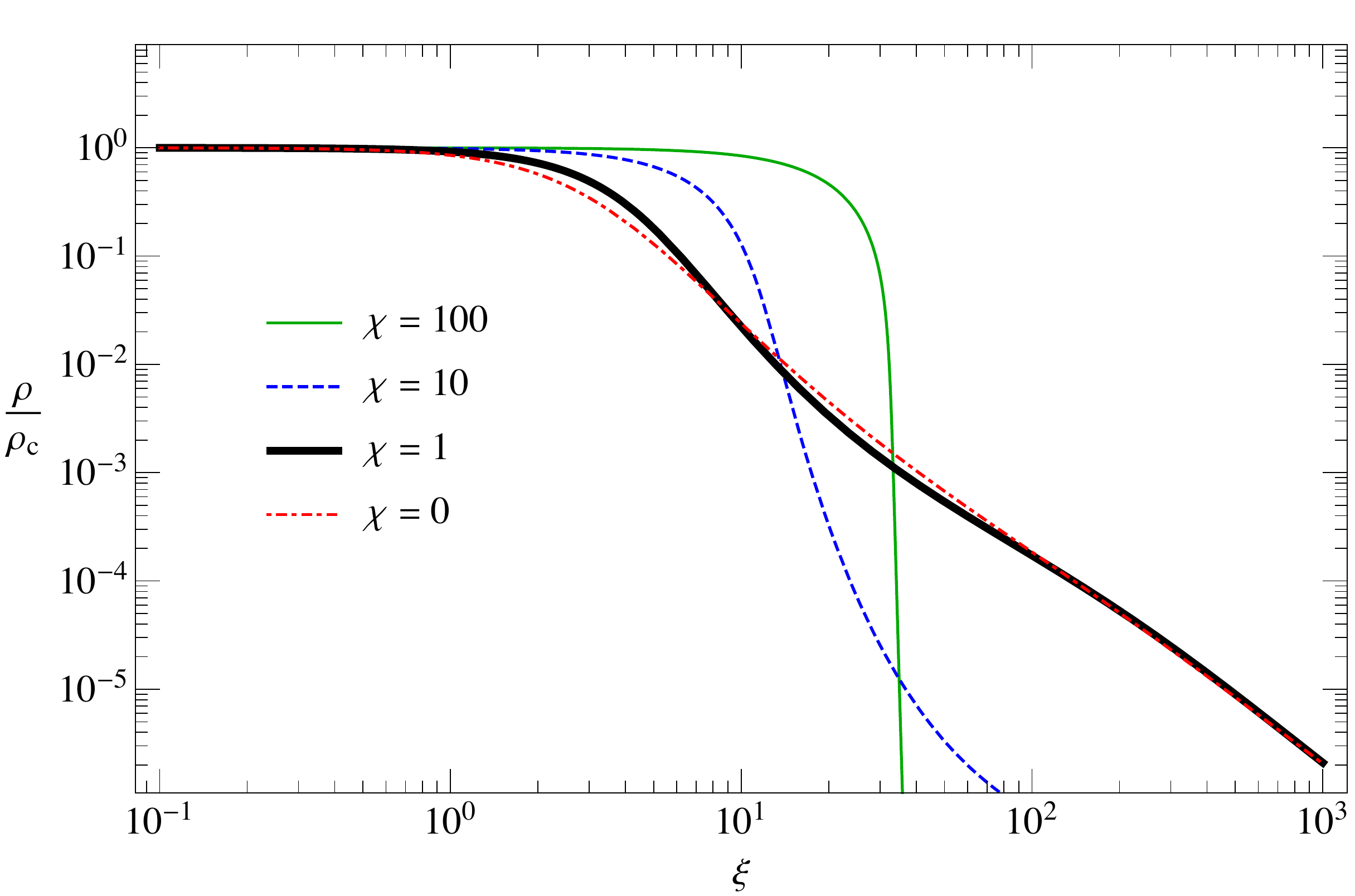}\\
        \includegraphics[width=\columnwidth]{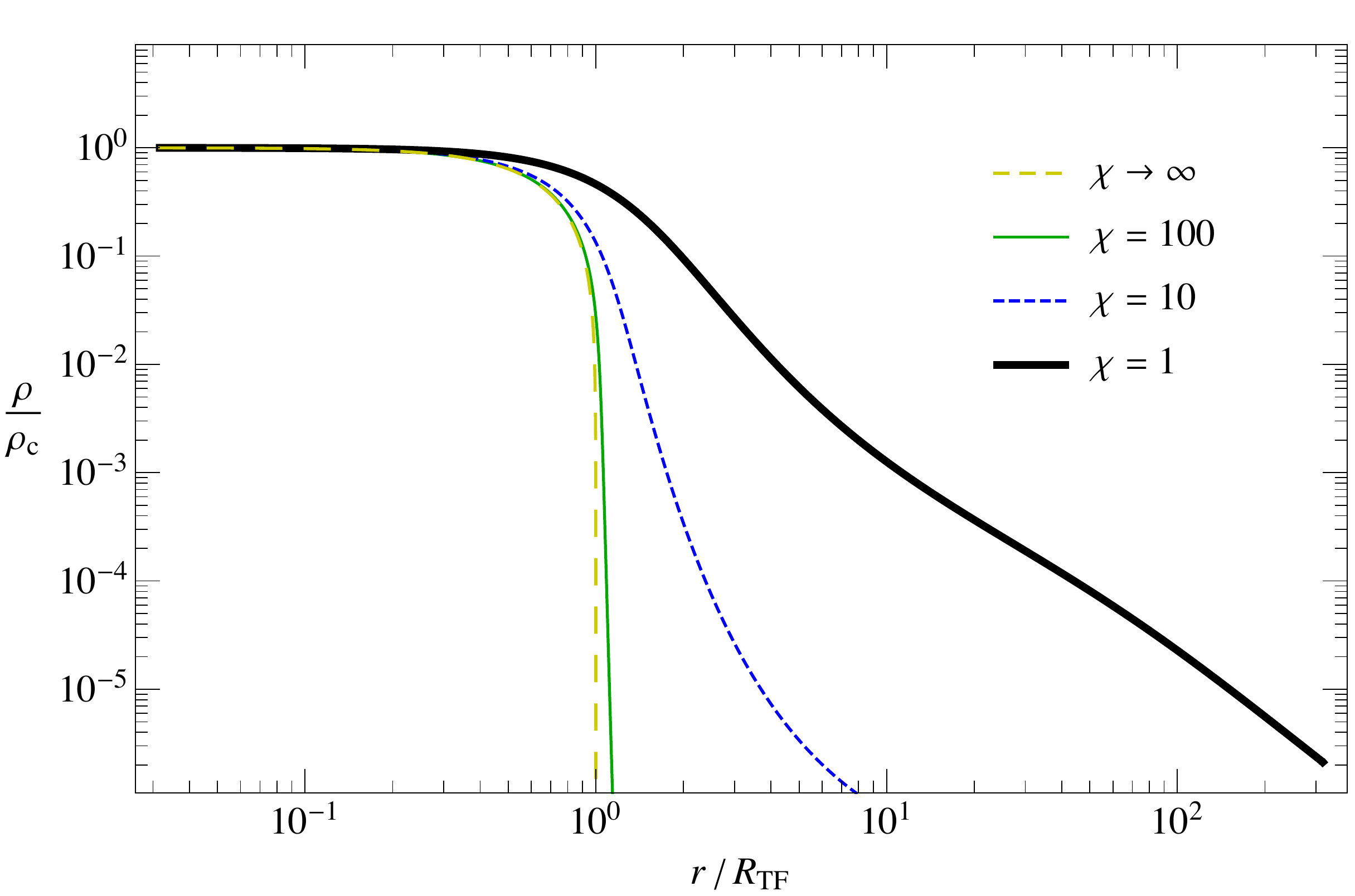}
        \caption{\textbf {SI-modified isothermal spheres: $\bm{[(n=\infty) + (n=1)]}$-polytropes.} Density profiles for SFDM-TF haloes in HSE, calculated by solving the double-polytrope equation~(\ref{eq:emden}) - (\ref{eq:BC}) for an isothermal sphere with added $(n=1)$-polytropic SI pressure, for different values of SI strength parameter, $\chi$, as labeled. The top panel has its radial coordinate normalized by the isothermal sphere radius ($\xi = r/r_0$), while the bottom panel is normalized by the polytrope radius ($R_\text{TF}$) and shows the ($n=1$)-polytrope profile recovered in the limit of $\chi \rightarrow \infty$.}
        \label{fig:isoHSE}
    \end{figure}

    The $\chi$ parameter encodes the strength of the polytropic pressure, which for this SFDM-TF model is sourced by the repulsive SI. When $\chi=0$, equation~(\ref{eq:emden}) reduces to the Emden-Chandrasekhar equation, whose solution (using the same boundary conditions given above) is a non-singular isothermal sphere which features a central core in its density profile that smoothly transitions to its envelope near $r= r_0 = \sqrt{\sigma^2/4\pi G \rho_c}$.
    For the SFDM-TF model, $\chi$ is close to the square of the ratio of the $(n=1)$-polytrope radius ($R_\text{TF}$) to this isothermal sphere radius:
    \begin{equation}
        \chi = \Big( \frac{R_\text{TF}}{\pi r_0} \Big)^2
    \end{equation}
    As $\chi$ increases, the central core in the density profile resulting from this HSE solution is extended out to $R_\text{TF}$, and the core-to-envelope transition is made sharper, with a larger jump in density. In the limit $\chi \rightarrow \infty$, equation~(\ref{eq:emden}) becomes equivalent to the Lane-Emden equation, and the ($n=1$)-polytrope profile is recovered. We show the HSE profile for different values of $\chi$ in Fig.~\ref{fig:isoHSE}.
    
   \subsection{Anticipating the equilibrium objects created by gravitational collapse}
    \label{sec:virialized_objects}

    Realistic values of $\chi$ for virialized SFDM-TF haloes with central polytropic cores and CDM-like envelopes can be approximated as follows. In the envelope ($r\gtrsim R_\text{TF}$), where the SI pressure should be subdominant, the velocity dispersion should be roughly equal to $GM/R$ in order for the CDM-like pressure to balance gravity there. Evaluated just outside the core, 
    \begin{equation}
        \sigma^2 \approx \frac{G M_c}{R_\text{TF}}
        \label{eq:sigma-core}
    \end{equation}
    where $M_c$ is the mass enclosed within $R_\text{TF}$, which we can approximate as the volume integral of the ($n=1$)-polytrope density profile in equation (\ref{eq:polytropedensity}),
    \begin{equation}
        M_c \approx \int_0^{R_\text{TF}} 4\pi r^2 \rho_c \text{sinc}(\pi r/R_\text{TF}) dr = \frac{4}{\pi}\rho_c R_\text{TF}^3
        \label{eq:Mc}
    \end{equation}
    In principle, if $\chi \ll 1$, this would not be a good way to approximate $M_c$, since $R_\text{TF}$ would be within the flattened central region of the non-singular isothermal sphere, so it would be better to set $M_c \approx \frac{4\pi}{3}\rho_c R_\text{TF}^3$. However, since we are interested in the core being set by the polytrope radius, $R_\text{TF}$, rather than the isothermal sphere radius, $r_0$, we use equation~(\ref{eq:Mc}), accordingly. Either way, the two approximations only differ by about a factor of 3, so to first-order, equation~(\ref{eq:Mc}) will suffice.
    In this case, $r_0$ and $\chi$ are found to be
    \begin{align}
        &r_0^2 = \frac{\sigma^2}{4\pi G \rho_c} \sim \frac{R_\text{TF}^2}{\pi^2} \\
        &\chi \sim 1
    \end{align}
    
    In its most general form, equation~(\ref{eq:emden}) requires 4 parameters to define its non-dimensional variables and provide a unique solution in physical units: \{$n, K, \rho_c, \sigma$\}. 
    For the SFDM-TF model explored in this work, we always take $n=1$ and use $R_\text{TF}$ in lieu of $K$, so the 3 remaining parameters to specify are \{$R_\text{TF}, \rho_c, \sigma$\}.
    In accordance with the reasoning outlined above, we will henceforth adopt $\chi = 1$, which specifies a relationship between $R_\text{TF}$, $\rho_c$, and $\sigma$, so the HSE solution is fixed with only 2 additional parameters.
    Since we will want to be able to vary $R_\text{TF}$ freely, the two-parameter model will consist of specifying either \{$R_\text{TF}, \rho_c$\} or \{$R_\text{TF}, \sigma$\}. 
    In some cases, it will be convenient to specify the central density and solve for $\sigma$ (i.e. use the former parameter pair),
    but if we want to obtain the HSE solution for a SFDM-TF halo with a given halo mass and radius, without imposing a central density, we will use the latter parameter pair with
    \begin{equation}
        \sigma^2 = \frac{G M_h}{R_h}
        \label{eq:sigma-env}
    \end{equation}
    and solve for $\rho_c$, instead.
    Again, this follows from the fact that the velocity-dispersion pressure alone must balance gravity in the halo envelope.
    Furthermore, the constancy of $\sigma^2$ allows us to equate equations~(\ref{eq:sigma-core}) and (\ref{eq:sigma-env}) to obtain an approximate relationship between the mass of the halo and the mass of its core, if we define the halo mass and radius with a fixed overdensity criterion such that $M_h \propto R_h^3$, as follows:\footnote{
    \citet{Padilla20} derived a more generalized core-halo mass relation for SFDM (i.e. not restricted to the TF regime), which reduces to equation~(\ref{eq:chTF}) in the TF limit. The equation for the TF regime was also found by \cite{Chav19}. }
    \begin{equation}
        \frac{M_c}{R_\text{TF}} \approx \frac{M_h}{R_h} \propto M_h^{2/3}
        \label{eq:chTF}
    \end{equation}
    
\begin{figure}
        \centering
        \includegraphics[width=\columnwidth]{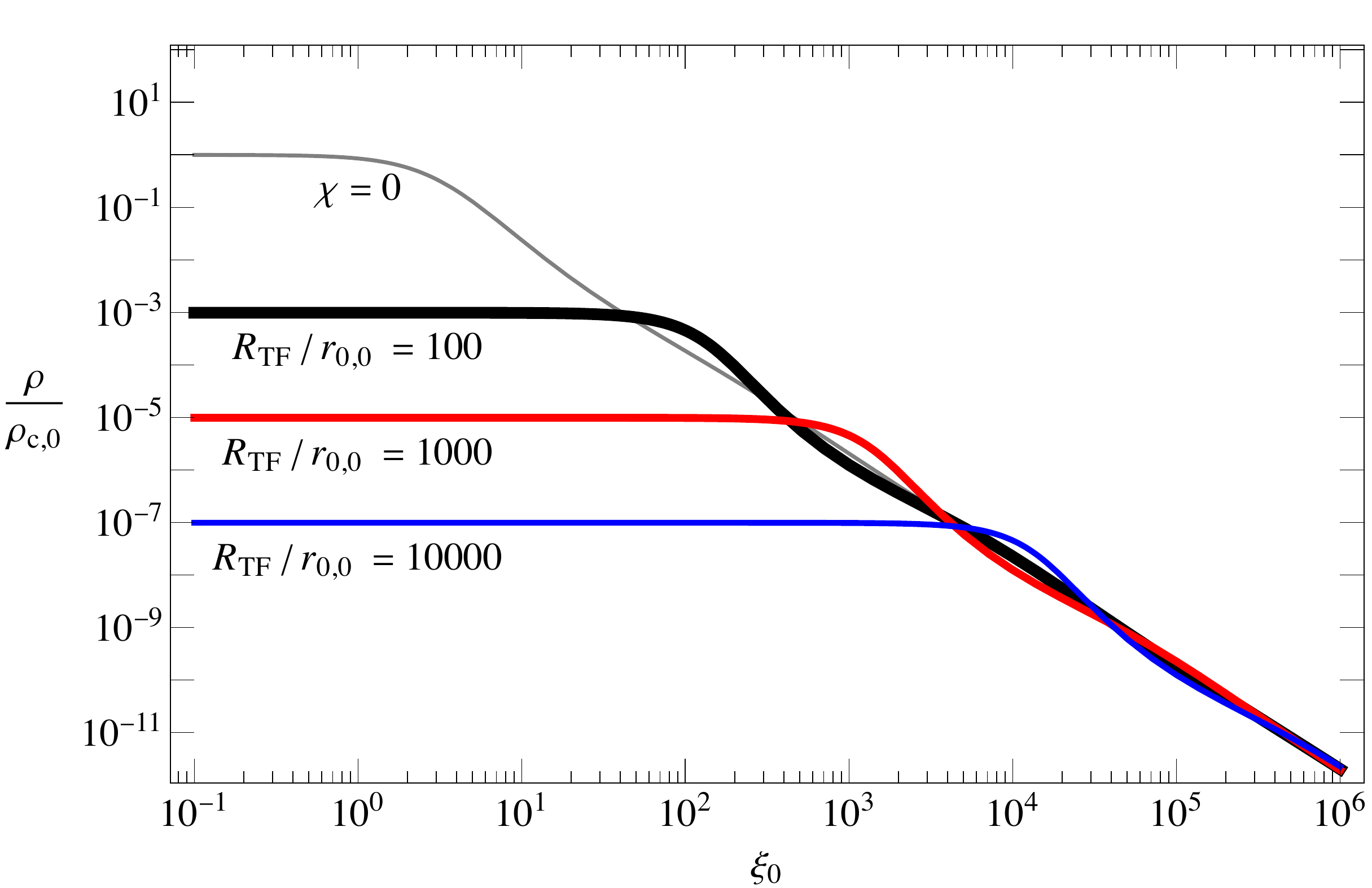}
        \caption{\textbf{SI-modified isothermal spheres, for the same halo if formed with different self-interaction strengths.} 
        We start with a standard non-singular isothermal sphere (i.e. equation~\ref{eq:emden} without SI, $\chi=0$; grey line) with core radius $r_{0,0}$ and central density $\rho_{c,0}$, corresponding to a velocity dispersion of $\sigma^2 = 4\pi G \rho_{c,0} \, r_{0,0}^2$. 
        We then ``add'' SI to this solution by setting $\chi=1$ and $R_\text{TF}/r_{0,0} = \{100,1000,10000\}$ (as labeled), but keeping the same value for $\sigma^2$,
        to show the effect of modifying the same underlying ``CDM'' halo with different amounts of SI.
        The density profiles (in units of $\rho_{c,0}$) are plotted against the dimensionless radius $\xi_\text{0}=r/r_\text{0,0}$. In each case, if we start at large radius and move inward, the profiles track the CDM-like profile ($\chi = 0$), down to $r\approx R_\text{TF}$, inside of which they flatten to the shape of an $(n=1)$-polytrope.}
        \label{fig:varyR0HSE}
    \end{figure}

    We can now anticipate the outcome
    of our simulations in the following sections, which follow the formation of equilibrium objects in SFDM-TF from gravitational instability, 
    leading to nonlinear collapse and virialization.  In particular, consider the virialized objects that
    form in the presence of SI of different strengths (i.e. different $R_\text{TF}$), for objects of the 
    same total mass and radius, $M_h$ and $R_h$. To approximate them by solutions of the double-polytrope presented above, we should take $\chi=1$ for
    every case.  As a reference case, however, we consider the case of $\chi=0$, the case with no SI.  
    Let us define the central density and core radius of this reference case as $\rho_{c,0}$ and $r_{0,0}$, respectively. 
    To model SFDM-TF haloes, we will require that $R_\text{TF} \gg r_{0,0}$.
    According to equation~(\ref{eq:sigma-env}), the velocity dispersion, $\sigma^2$, must be the same for all cases, 
    regardless of the value of $\chi$ or $R_\text{TF}$, because the choice of $M_h$ and $R_h$ fixes $\sigma^2$.
    In particular, given the reference $\chi=0$ case, we can write
    \begin{equation}
        \sigma^2 = 4\pi G \rho_{c,0} \, r_{0,0}^2
    \end{equation}
    In that case, the central densities of the $\chi=1$ solutions with different $R_\text{TF}$ scale relative to the $\chi=0$ solution as
    \begin{equation}
        \frac{\rho_c (R_\text{TF})}{\rho_{c,0}} = \frac{\sigma^2}{4\pi G \rho_{c,0} \, r_0^2(R_\text{TF})} = \Big( \frac{\pi r_{0,0}}{R_\text{TF}} \Big)^2
    \end{equation}
    where $r_0(R_\text{TF}) = R_\text{TF}/\pi$ since $\chi=1$.
    Therefore, increasing $R_\text{TF}$ results in progressively larger cores with lower central densities.
    The resulting density profiles for haloes of the same total mass and radius, but different values of $R_\text{TF}/r_{0,0}$, are plotted in
    Fig. \ref{fig:varyR0HSE}, in which densities and radii are all normalized by the same central density ($\rho_{c,0}$) and core
    radius ($r_{0,0}$) of the case with $\chi=0$ (i.e. no SI), for direct comparison.
    
    According to Fig. \ref{fig:varyR0HSE},
    we anticipate that a self-gravitating sphere of SFDM-TF
    in hydrostatic equilibrium that results from gravitational instability and collapse will follow
    a very simple description.  There will be a flattened core in the center, about the size of $R_\text{TF}$, that follows the profile of the $(n=1)$-polytrope
    for a sphere with only $P_\text{SI}$ to support it, surrounded by an envelope which roughly traces the same profile as 
    the case with no SI (i.e. the standard CDM case).
    In our analytical approximation here, the profile of the envelope follows that of an isothermal sphere in which the
    pressure support is provided by the velocity dispersion pressure $P_{\!\sigma}$, which is the same for all values of $R_\text{TF}$. 
    In the detailed numerical simulations to follow, we will find that the objects which correspond to the equilibrium spheres described above 
    are actually the virialized regions inside an accretion shock that forms when gravitational instability leads to collapse in SFDM-TF.  
    Remarkably enough, however, we will find a similar characteristic behavior for the density profile of that virialized region, in which
    there is an $(n=1)$-polytropic core of size $\sim R_\text{TF}$, surrounded by the envelope profile that would have been found in the absence of SI (i.e. the standard CDM case). 
    The envelope shall differ somewhat from that of our double-polytrope approximation, since the latter imposed a strict
    condition of isothermality for that envelope, while the detailed numerical solution has no such restriction built-in.

\section{Numerical Methodology} \label{sec:methods}

To study the dynamical formation of an SFDM-TF halo by gravitational collapse, 
we limit our attention here to 1D, spherical symmetry and take full advantage of the 
fluid approximation derived above in \S\ref{sec:fluidapprox} for this case. 
Even in spherical symmetry, these partial differential equations 
in time and space, equations~(\ref{eq:0momSpherical}) - (\ref{eq:2momSpherical}),  
require a numerical solution. Fortunately, 
the close resemblance of these SFDM-TF fluid 
equations to those of an ideal gas, with 
ideal gas pressure sourced by an effective
SFDM velocity dispersion, makes it possible to apply standard methods for 
solving the conservation equations of compressible gas dynamics, 
modified only by adding a second pressure force to the momentum equation,
to account for SI by its polytropic pressure law.  
We shall adopt the 1D-spherical, Lagrangian hydrodynamics code of \citet{AS07}, therefore,
modified to include the SI pressure of the SFDM-TF model.  
The reader is referred to the appendix of \citet{AS07} for additional details, including several code tests.

We apply this code to study the time-dependent, dynamical formation of a 
virialized halo as the nonlinear outcome of Jeans instability in SFDM-TF, 
starting from a spherically-symmetric, linear-amplitude 
density perturbation in a cold, static background density field, resulting in gravitational collapse and spherical infall.  As long as the initial density profile is centrally
concentrated, spherical shells in free-fall would reach the origin in a finite time which is larger
for shells of larger initial radius. If there are pressure forces present, however, they might decelerate
this infall before that can happen.  In fact, for the case of interest, the initial conditions 
will be chosen so that the perturbation is highly Jeans-unstable, both with respect to the Jeans
instability condition derived in \S\ref{sec:SI_Jeans_Length} for the case with only $P_\text{SI}$ and
no $P_{\!\sigma}$, and with respect to that in the other limit, with only $P_{\!\sigma}$ but no $P_\text{SI}$.  In that case,
shells infall with highly supersonic velocity, with trajectories that initially
do follow those of free-fall collapse, but the initially unimportant pressure forces eventually rise to
become important before any shell hits the origin.  In particular, $P_\text{SI}$ will grow as density increases during
infall, as $P_\text{SI} \propto {\rho}^2$, while the ideal-gas-like pressure $P_{\!\sigma}$ of the SFDM from its effective 
velocity dispersion also increases with density,
according to an adiabatic gas law, $P_{\!\sigma} \propto {\rho}^{\gamma}$, with
$\gamma = 5/3$, as dictated by the energy equation~(\ref{eq:2momSpherical}).  Since the former has a stronger density dependence than the latter, the innermost
shells are decelerated by $P_\text{SI}$, which dominates
over $P_{\!\sigma}$ when the shells reach a radius within $\sim R_\text{TF}$.
In the absence of this SI pressure, however, the ideal-gas-like pressure causes a strong shock-front to form, 
which decelerates the infalling shells and thermalizes
their bulk kinetic energy into the internal energy associated with velocity dispersion on the post-shock (inner radius) side, 
bringing their infall almost to rest (i.e. reduced to subsonic).  When both
kinds of pressure are present, 
a combination of these phenomena occurs: SI pressure halts infall of the
innermost shells, supporting a central core, while an accretion shock forms at the outer edge of
this core and propagates outward, surrounding that core
with post-shock SFDM in which the total pressure is
dominated by the high $P_{\!\sigma}$ that results from shock ``heating''.  As we shall see, this core-envelope structure, interior to the accretion shock,
is the virialized region we associate with the
halo that forms from gravitational instability and collapse, with mass and radius that grow with time
as the shock moves outward to overtake
additional infalling mass.

To follow this evolution for different SI interaction strengths, we ran a suite of spherical collapse simulations, each with 2000 mass shells,\footnote{
We determined that 2000 shells are sufficient for our work here by convergence testing.}
and track their radial position, density, velocity, and internal energy over time. The Lagrangian code trivially satisfies the continuity equation, because the mass of each shell is fixed.
The Poisson equation is also directly satisfied, in that
case, by taking the gravitational acceleration at shell radius $r$ to be $-GM(<r)/r^2$, since $M(<r)$, the total mass interior to that Lagrangian mass-shell, is also exactly conserved.
The time-integration uses a symplectic ``leap-frog'' integration scheme to evolve the position and velocity of each shell, given the pressure forces and gravitational acceleration specified by the momentum equation, including the SI pressure. The energy equation for the effective ideal-gas-like pressure is enforced by calculating the $P_{\!\sigma}dV$ work done on each shell (i.e. a change in volume results in a change in
the internal energy associated with the ideal-gas-like pressure
$P_{\!\sigma}$). In addition, artificial viscosity is added to
the equations of momentum and energy, to stabilize the calculation in the presence of shock formation, as described in more detail below and in Appendix~\ref{sec:avis}.    

For direct comparison, we also use this code to perform runs with $P_\text{SI} = 0$ in the momentum equation.   This serves to model the CDM limit of
SFDM, i.e. FDM in the limit of small
de~Broglie wavelength.  In that case, as shown
originally by \citet{AS05}, 
the fluid approximation of \S\ref{sec:fluidapprox} is a valid description of the dynamics of standard, collisionless CDM.  
As such, we shall
refer to these runs
with $P_\text{SI} = 0$ as \textit{CDM}, or \textit{CDM-like}.

\subsection{Initial conditions}
\label{sec:ICs}
The initial condition used in these simulations was that of a small-amplitude, linear density perturbation ($\Delta\rho \ll \rho$), which grows over time through gravitational instability, leading to highly nonlinear collapse and the formation of a virialized dark matter halo. 
The spherical shells are logarithmically-spaced in the initial condition,
such that their radial widths $\Delta r$ are proportional to their outer boundary radii $r$
(i.e. $\Delta r/r = $ constant).
They start out
cold ($P_{\!\sigma}\simeq0$) and static ($v=0$), with their initial acceleration dominated by gravity.  This initial condition is chosen to be highly Jeans unstable, not
only with respect to the velocity-dispersion pressure $P_{\!\sigma}$, alone, but also in the sense described in \S\ref{sec:SI_Jeans_Length} for the TF regime when $P_\text{SI} = g\rho^2/2m^2$ dominates over $P_{\!\sigma}$, instead.  
The initial density profile follows that of a non-singular isothermal sphere in the absence of SI (i.e. the solution to the Emden-Chandrasekhar equation; equation~\ref{eq:emden} with $\chi=0$), for which the central core is chosen to have a radius $r_0$ that
greatly exceeds the value of $R_\text{TF}$ in each case (in particular, we use $r_0 = 10^5 R_\text{TF}$).  This is equivalent to saying that the 
SI-Jeans number of the
initial condition $N_\text{J} = r_0/\lambda_\text{SI,J} \gg 1$. 
In addition, by taking $P_{\!\sigma}\simeq0$ initially, (i.e. by ``resetting'' the velocity dispersion of that initial isothermal sphere to close to zero), 
the conventional Jeans number, which results if we replace $P_\text{SI}$ in our analysis of the Jeans wavelength in \S\ref{sec:SI_Jeans_Length} by velocity-dispersion pressure $P_{\!\sigma}$, is also very large.  
Since we are interested in starting from a ``linear perturbation'', we initialize the shells to fit the shape of the far interior of this isothermal sphere's central core. 
This provides the desired linearity while
ensuring that the mass which collapses and 
virializes by the time our simulation ends does so much earlier than the outer regions of the sphere,
i.e. outside the initial core radius.
In practice, we initialize the outermost shell in each simulation at a radius $R_\text{sim}$ that is only 1\% of the isothermal sphere's core radius (i.e. $R_\text{sim} = 10^{-2} r_0 = 10^3 R_\text{TF}$), so the entire simulation domain is well within the asymptotically flat central region of the isothermal sphere profile. 
This ensures that the density gradient is very small everywhere in the initial profile and that the particular shape of the isothermal sphere profile has minimal impact on the ensuing dynamics. The central density of this initial profile ($\rho_{c,i}$) is tuned, along with the polytrope radius ($R_\text{TF}$), to produce haloes of various core sizes and densities.

\subsection{Dynamics and time-stepping}

 For the ``leap-frog'' integrator, we
 follow \cite{AS07} in calculating gravitational acceleration, shell-crossing, and hydrodynamical sound-crossing time-steps for each shell at each
 time-slice. We also consider an additional ``sound-crossing'' time for $P_\text{SI}$:
\begin{equation}
    \Delta t_\text{SI} = \frac{\Delta r}{c_{s,\text{SI}}}
\end{equation}
where $\Delta r$ is the width of the shell, and
$c_{s,\text{SI}}$ is the speed of sound waves in the SI-pressure-supported limit, given by equation (\ref{eq:SI_soundspeed}).  A new, global time-step for all
shells is then chosen at each time-slice, equal to the
minimum value of the time-steps of all shells at that
time-slice.

We handle shocks with numerical artificial viscosity, as detailed by \citet{AS07}, with features described in Appendix~\ref{sec:avis} which were added to avoid spurious heating during high-Mach-number gravitational collapse in regions far outside the strong accretion shock that forms near the center 
during such collapse.
That shock is essential to forming the CDM-like envelope of the halo, because it converts the infall kinetic energy of the collapsing shells into ``thermal'' velocity-dispersion energy, bringing them to rest by providing them with the velocity-dispersion pressure that is needed for a CDM-like profile to be supported against gravity. This is how the shells reach virial equilibrium once shocked. 
In this sense, the shock acts as a physical radius of the halo -- the boundary between the virialized halo and the infalling matter outside of it. As more shells collapse, the shock travels outward, accreting more matter, and the equilibrium profile of the halo is left in its wake. The correspondence between shock-induced ideal gas pressure and collisionless CDM dynamics was discussed in \citet{AS05}.

\begin{figure}
\centering
\includegraphics[width=\columnwidth]{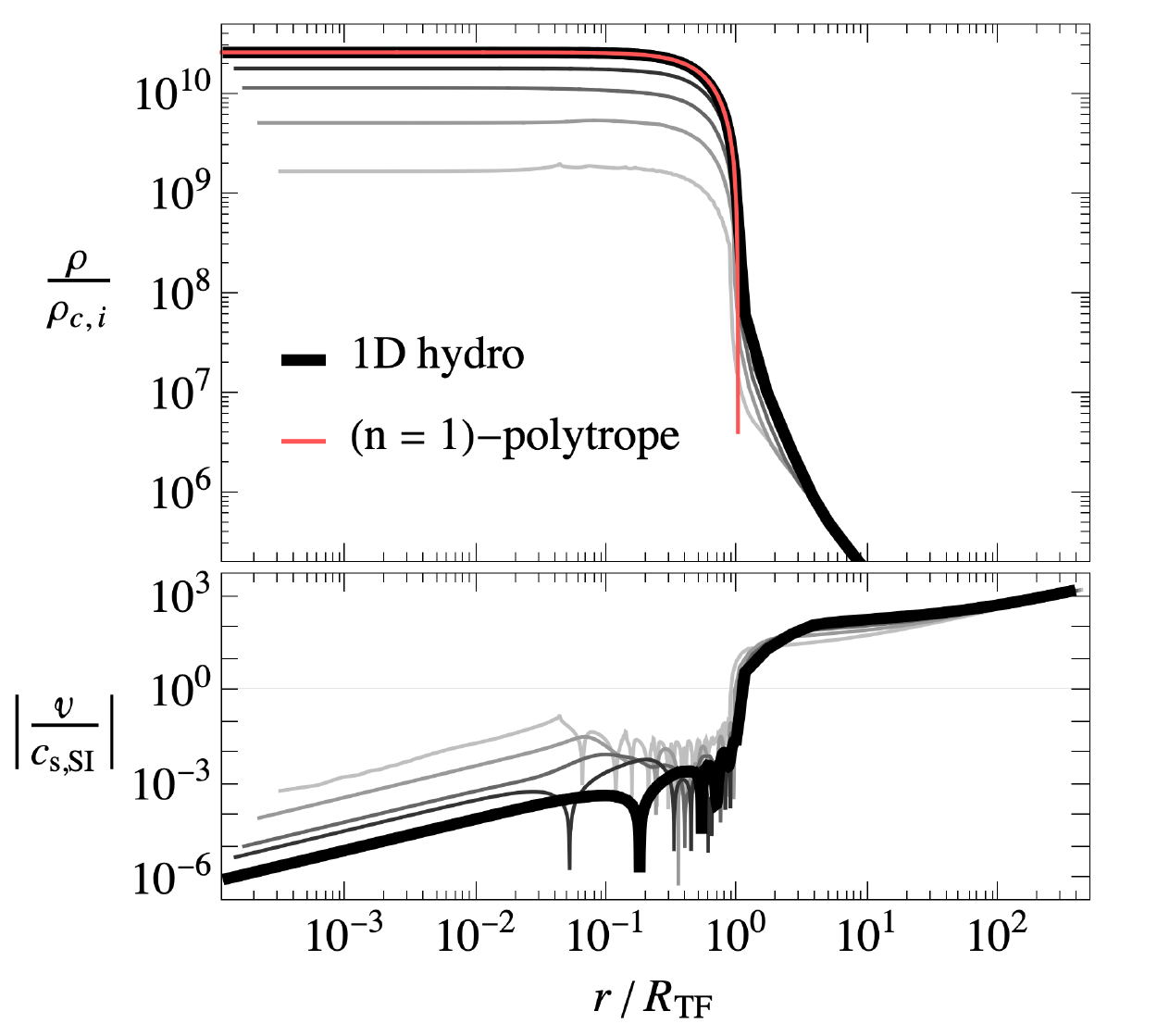}
\caption{\textbf{Gravitational collapse with only self-interaction pressure: filling the polytropic core.} 
Lighter lines represent earlier times in the simulation, with time progressing between them in equal increments of $\Delta t = 0.0025 \tau_{\text{ff},i}$, 
from the lightest grey line at $t = 1.925 \tau_{\text{ff},i}$ to the thick black line at $t = 1.935 \tau_{\text{ff},i}$, 
where $\tau_{\text{ff},i} = 1/\sqrt{4\pi G \rho_{c,i}}$ and $\rho_{c,i}$ is the central density of the initial condition.
Top: Evolution of the density profile from the 1D hydro simulation without velocity-dispersion pressure ($P_\text{SI}$ only), starting from initial conditions described in \S\ref{sec:ICs}. 
The central density increases over time as more matter falls within the polytrope radius ($R_\text{TF}$).
The red line shows the density profile of an $(n=1)$-polytrope, the expected hydrostatic equilibrium profile, with its central density matched to that of the thick black line. 
Bottom: The ``Mach number'' profile of the $P_\text{SI}$-only fluid given by the ratio of its bulk velocity to the local SI ``sound speed'', $c_{s,\text{SI}} = \sqrt{g\rho/m^2}$.
Matter outside $R_\text{TF}$ is infalling supersonically ($|v/c_{s,\text{SI}}| \gg 1$), while matter inside $R_\text{TF}$ is near hydrostatic equilibrium ($|v/c_{s,\text{SI}}| \ll 1$).
Acoustic oscillations are apparent at $r \lesssim R_\text{TF}$, driven by homologous expansion and contraction ($v \propto \pm r$) at $r \ll R_\text{TF}$.
The secular trend for even these innermost shells, however, is to fall further inward, as is necessary for the density profile to grow.}
\label{fig:polytrope}
\end{figure}

\subsection{A code test with only SI pressure: filling the polytropic core}

The new component added to the 1D-spherical hydrodynamics code we employ is the repulsive SI-term, $P_\text{SI}$. We demonstrate the ability of our code to handle this term correctly. If we perform a spherical collapse calculation with only this SI pressure, having dropped the effective velocity-dispersion pressure, we expect the infalling shells to relax to an $(n=1)$-polytrope (see \S\ref{sec:SI}). Indeed, we find this is the case, as is shown in Fig.~\ref{fig:polytrope}. Starting from the same initial conditions described in \S\ref{sec:ICs}, infalling matter that falls to within $r\simeq R_\text{TF}$ settles into the shape of the polytropic core through acoustic motion. As more matter falls into the core, its density increases, but its radius stays fixed. We also find that a polytropic core in hydrostatic equilibrium, without any infalling matter outside of it, with shells initialized to reproduce the analytical profile in equations (\ref{eq:polytropedensity}) and (\ref{eq:RTF}), remains static and retains this exact equilibrium profile shape over time, as expected.

\section{Simulation Results} \label{sec:results}

\subsection{SFDM-TF halo profiles from spherical collapse} \label{sec:1Dsim}
\begin{figure}
        \centering
        \includegraphics[width=\columnwidth]{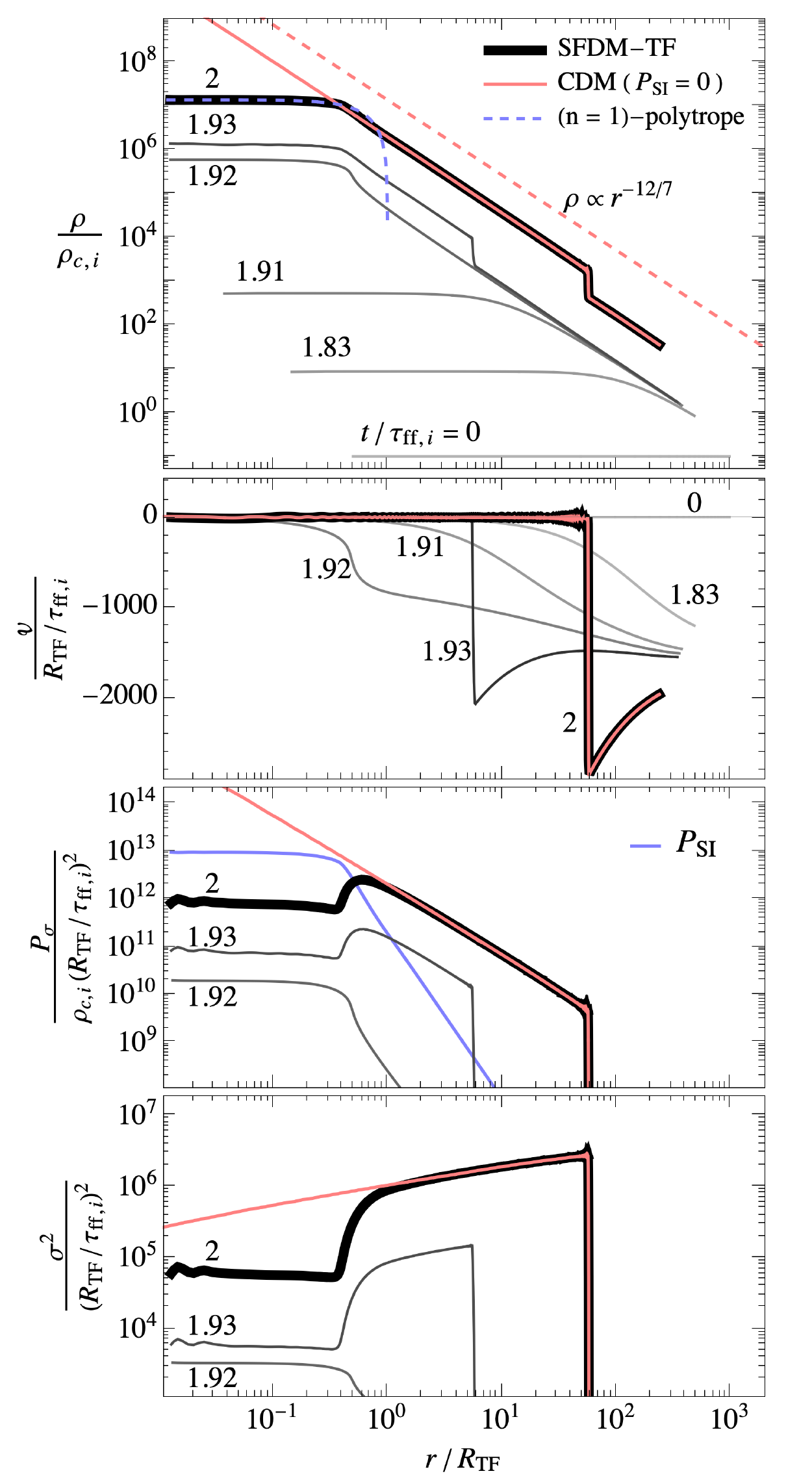}
        \caption{\textbf{Simulation results for the nonlinear gravitational collapse of a Jeans-unstable SFDM-TF perturbation.} From top to bottom, non-dimensionalized density, velocity, velocity-dispersion pressure, and velocity-dispersion ``temperature'' profiles for a SFDM-TF halo produced by a spherical collapse simulation.
        The thick black lines show the fiducial late-time profile at $t = 2\tau_{\text{ff},i}$, where $\tau_{\text{ff},i} = 1/\sqrt{4\pi G \rho_{c,i}}$ is the initial free-fall time, and $\rho_{c,i}$ is the central density of the initial condition.
        The thin grey lines show the profiles at earlier times in the simulation, vertically offset by an order of magnitude in the $\rho$, $P_{\!\sigma}$, and $\sigma^2$ panels for visual clarity. Each of these profiles is labeled with a time-stamp, denoting the time elapsed since the beginning of the simulation, in units of $\tau_{\text{ff},i}$.
        The discontinuity near $r=60R_\text{TF}$ is the accretion shock, which sets the virial radius of the halo. The thin red lines are profiles for a corresponding CDM run using the same initial conditions. The CDM density profile fits the shape of a power law with a logarithmic slope of $-12/7$. The SFDM-TF density profile closely tracks the CDM profile outside $r\simeq R_\text{TF}$, but takes the shape of a polytropic core (blue short-dashed line) near the center.
        The solid blue line in the $P_{\!\sigma}$ panel shows the profile for $P_\text{SI} = g\rho^2/2m^2$ at the fiducial time with the same normalization.}
        \label{fig:numprof}
    \end{figure}

\begin{figure}
        \centering
        \includegraphics[width=\columnwidth]{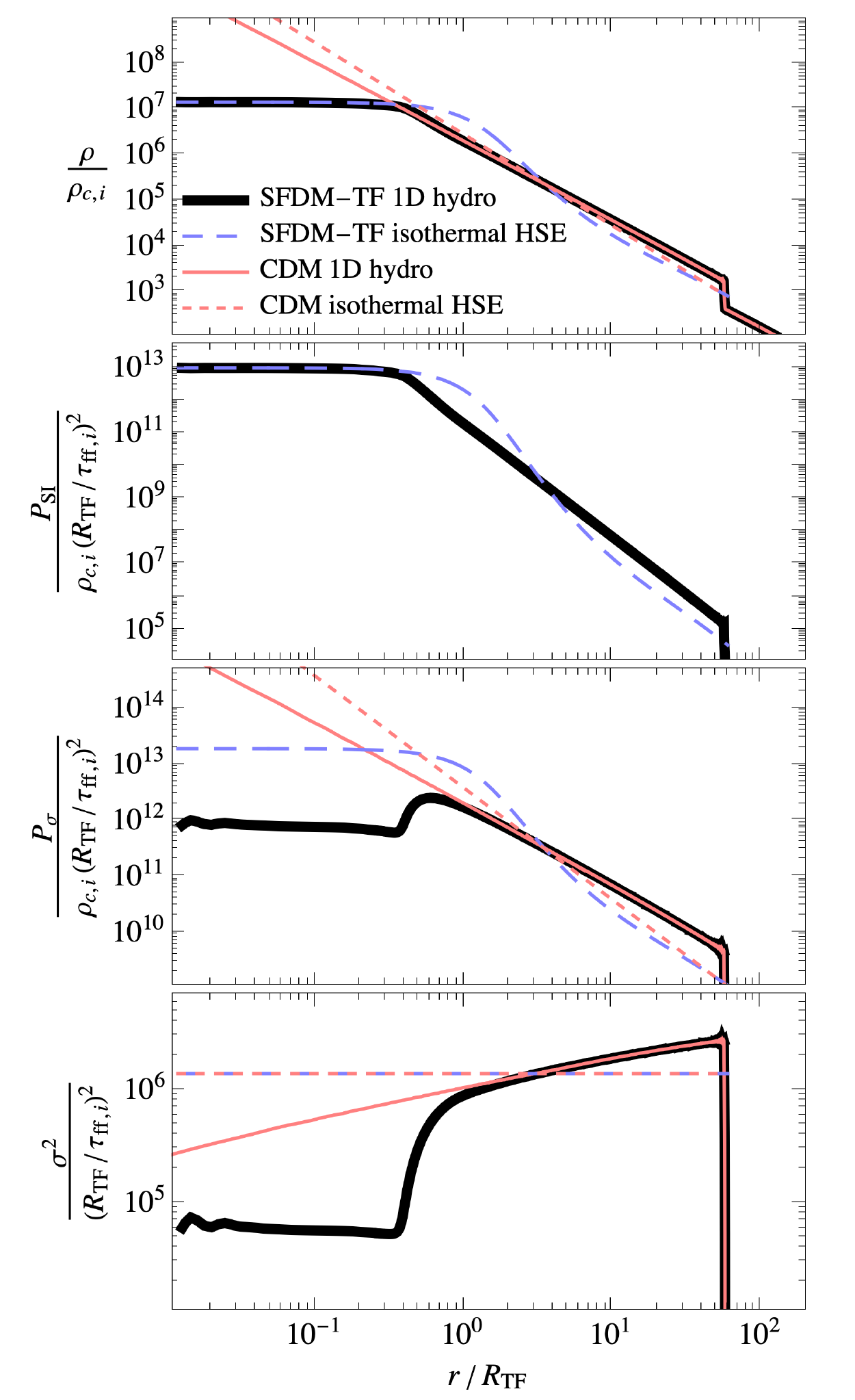}
        \caption{\textbf{Comparison of the post-shock profiles from the simulation of gravitational collapse of a Jeans-unstable SFDM-TF perturbation with the SI-modified isothermal sphere solution for $\bm{\chi=1}$.} Same quantities
        plotted as in Fig.~\ref{fig:numprof}, as labeled.  
        Simulation results for the same
        fiducial time-slice ($t=2\tau_{\text{ff},i}$) as the thick black lines 
        in Fig.~\ref{fig:numprof} are compared with the SI-modified isothermal
        sphere with the same $R_\text{TF}$ and central density, for $\chi=1$
        (long-dashed blue lines).  In addition, the simulation results for the
        CDM-like limiting case with no SI at that time-slice (thin red lines)
        are compared with the HSE profiles for the CDM-like limit of the isothermal sphere without SI 
        (i.e. $\chi=0$; short-dashed red lines).}
        \label{fig:numprofwHSE}
    \end{figure}

\begin{figure}
        \centering
        \includegraphics[width=\columnwidth]{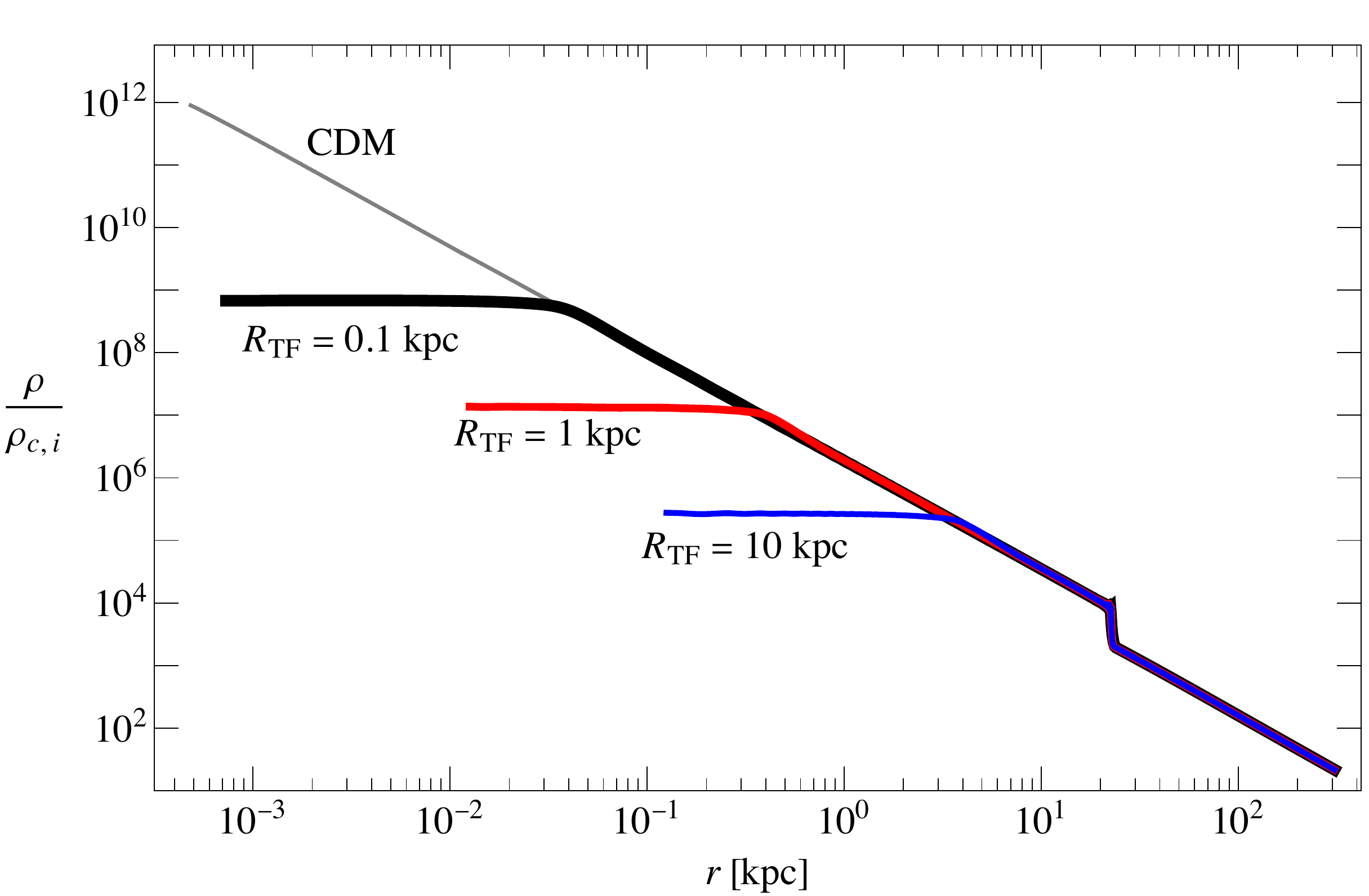}
        \caption{\textbf{SFDM-TF haloes formed by gravitational collapse for different self-interaction strengths.} Comparison of the SFDM-TF halo density profiles resulting from 3 runs of the 1D spherical collapse calculation, all starting from the same initial conditions, but with different values of $R_\text{TF}$, as labeled. The profile for a CDM run from the same initial conditions is shown, as well. In each case, the SFDM-TF profile tracks the CDM profile down to around $r\simeq 0.8R_\text{TF}$, and flattens to the shape of a polytropic core inside $r\simeq 0.4R_\text{TF}$. Compare this behavior inside the accretion shock with that of the SI-modified isothermal spheres in HSE, for different values of $R_\text{TF}$,
        in Fig. \ref{fig:varyR0HSE}.}
        \label{fig:varyR0}
    \end{figure}

\begin{figure}
        \centering
        \includegraphics[width=\columnwidth]{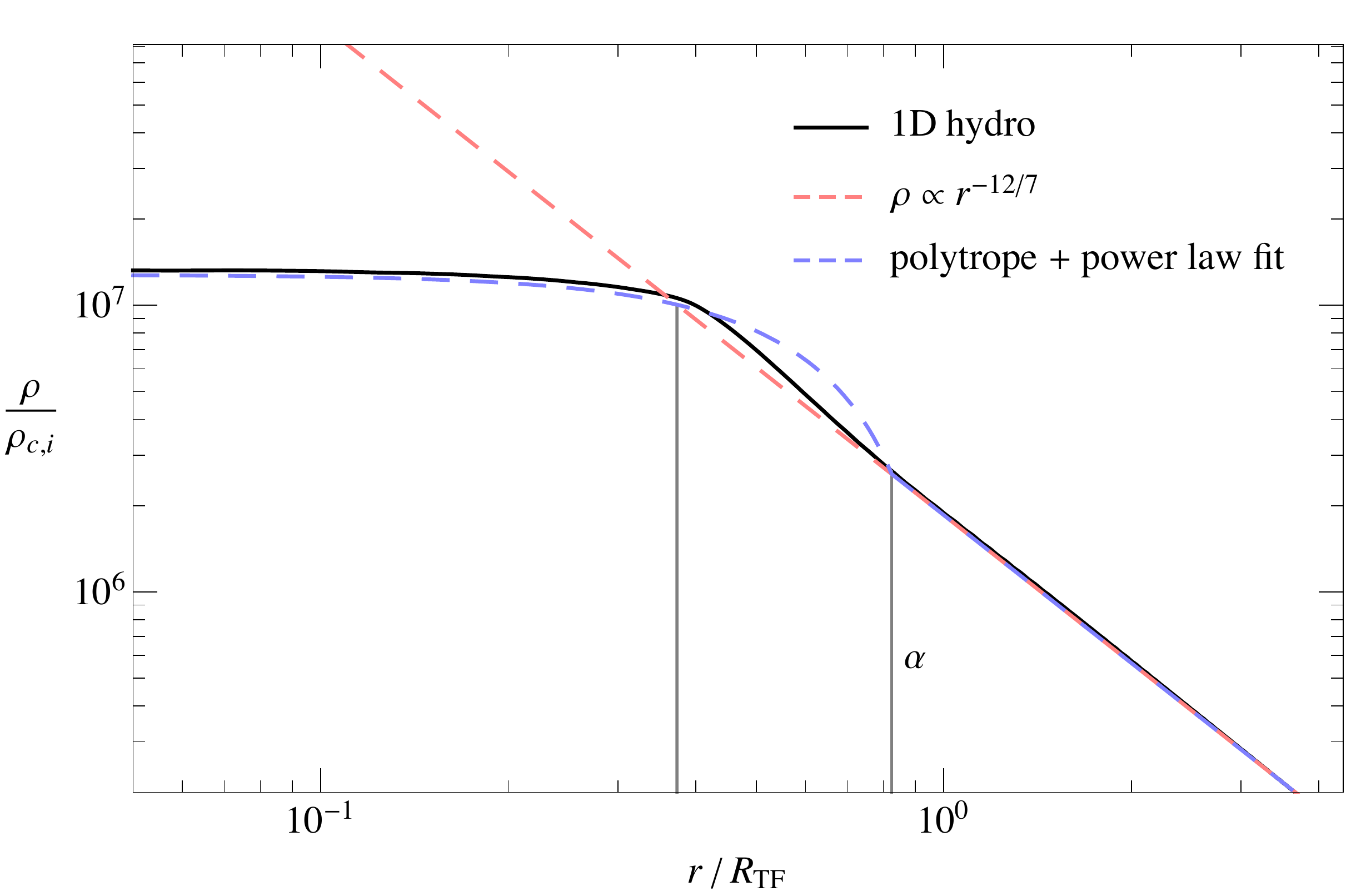}
        \caption{\textbf{A fit to the simulated SFDM-TF halo profile: polytropic core + CDM-like envelope.}
        The density profile resulting from our 1D hydro simulation (black), also plotted in Fig.~\ref{fig:numprof}, 
        is re-plotted here for comparison with the analytical fit from equation~(\ref{eq:numfit}) (blue dashed), which combines an 
        SI-pressure-supported ($n=1$)-polytrope inside a radius $r=\alpha R_\text{TF}$ with a 
        CDM-like envelope supported by accretion-shock-heated velocity-dispersion pressure.
        Vertical lines indicate points of intersection between the fitted polytropic core and the 
        CDM-like profile, which takes the shape of a power law with a logarithmic slope of $-12/7$ (red dashed). 
        The outer intersection point at $r/R_\text{TF} = \alpha = 0.825$ is set by requiring density continuity (equation~\ref{eq:denscont}) 
        and mass conservation (equation~\ref{eq:masscons}), meaning that the polytropic core and power law profiles both contain the same
        mass within this radius.}  
        \label{fig:numfit}
    \end{figure}

    We simulated the dynamical evolution of the spherical, Jeans-unstable perturbation described in \S\ref{sec:methods} for the SFDM-TF model, using our 1D-spherical, Lagrangian hydrodynamics code to solve the fluid equations derived in \S\ref{sec:fluidapprox}, for a range of SI strengths. The results, shown in Figs.~\ref{fig:numprof} - \ref{fig:virial}, 
    demonstrate that the spherical collapse of a linear SFDM-TF perturbation produces a virialized halo with a polytropic core in the center of a CDM-like envelope.
    
    The density, velocity, velocity-dispersion pressure ($P_{\!\sigma}$), and velocity-dispersion ``temperature'' ($\sigma^2 = P_{\!\sigma}/\rho$) profiles resulting from one of the runs of our SFDM-TF simulation suite are shown in Fig.~\ref{fig:numprof}. These variables are non-dimensionalized in terms of the central density of the initial condition ($\rho_{c,i}$), the initial free-fall time \big($\tau_{\text{ff},i} = 1/\sqrt{4\pi G \rho_{c,i}}$\big), and the polytrope radius ($R_\text{TF}$):
    \begin{align}
        &r \rightarrow \frac{r}{R_\text{TF}} \\
        &t \rightarrow \frac{t}{\tau_{\text{ff},i}} \\
        &\rho \rightarrow \frac{\rho}{\rho_{c,i}} \\
        &v \rightarrow \frac{v}{R_\text{TF}/\tau_{\text{ff},i}} \\
        &P_{\text{SI} } \rightarrow \frac{P_{\text{SI} }}{\rho_{c,i}(R_\text{TF}/\tau_{\text{ff},i})^2} \\
        &P_{\!\sigma } \rightarrow \frac{P_{\!\sigma} }{\rho_{c,i}(R_\text{TF}/\tau_{\text{ff},i})^2} \\
        &\sigma^2 \rightarrow \frac{\sigma^2}{(R_\text{TF}/\tau_{\text{ff},i})^2} 
    \end{align}
    In these non-dimensional variables, the resultant profiles are self-similar under changes to $\rho_{c,i}$: different simulation runs that start with different $\rho_{c,i}$ (and all else being equal) produce the same profiles at the same times as long as the variables (including time) are non-dimensionalized with respect to each run's own $\rho_{c,i}$.
    Therefore, for a given $R_\text{TF}$, a single simulation run can be used to model an infinite family of SFDM-TF haloes, e.g. with different central core densities, simply by re-scaling $\rho_{c,i}$.
    Similarly, for different values of $R_\text{TF}$, the profiles of all quantities
    plotted in these nondimensionalized variables are indistinguishable if
    plotted for the time-slice at which 
    their shock is located at the same value of $r/R_\text{TF}$.
    
    The thick black curves in Fig.~\ref{fig:numprof} show the numerical profiles at a fiducial late time $t = 2\tau_{\text{ff},i}$, while
    the thin grey curves show earlier times to illustrate the profiles' evolution, with time progressing as labeled in units of $\tau_{\text{ff},i}$. These grey curves have been shifted downward in the density, pressure, and velocity dispersion plots by an order of magnitude for visual clarity. The profiles for $t= 1.92 \tau_{\text{ff},i}$ show the accretion shock starting to form at the edge of the flat central core. This shock then propagates outward, leaving the envelope's equilibrium profile in its wake, and converting the kinetic energy of infalling shells into ``dispersion'' energy, to provide them with pressure support and bring them to rest.
    
    For comparison with these SFDM-TF results in Fig.~\ref{fig:numprof}, we overlay the 
    results of a CDM-like run (i.e. FDM in the limit of small $\lambda_\text{deB}$) from the 
    same initial conditions, produced by excluding the SI pressure in the momentum equation. 
    The SFDM-TF and CDM profiles track each other outside of the polytrope radius, as expected, 
    illustrating the CDM-like behavior of the SFDM-TF model on larger scales.
    The density profile of this CDM-like envelope fits the shape of a power law with a logarithmic slope of $-12/7$, as expected for the gravitational collapse of a spherical matter distribution from an initial
    condition that is cold, static, centrally concentrated, linearly perturbed, and smooth (i.e. differentiable everywhere, including at the center) \citep[see, e.g.,][]{Penston69}. The fact that pre- and post-shock profiles both follow the same power law reflects this CDM-like behavior as 
    a $\gamma = 5/3$ ideal gas, for which the density jumps across a strong shock
    by a constant factor of 4, while the post-shock gas is approximately hydrostatic.
    For $r\lesssim 0.4R_\text{TF}$, the SFDM-TF density profile is well-fit by an ($n=1$)-polytropic core profile, as indicated by the short-dashed blue curve in the top panel of the figure. 
    The CDM run, on the other hand, continues to follow the power law at small radii, 
    creating a diverging density cusp at its center. In addition, the velocity-dispersion pressure and ``temperature'' for the SFDM-TF case are low near the center compared to the CDM case, because the system is supported there by SI pressure, instead. 
    In the region $0.4 \lesssim r/R_\text{TF} \lesssim 0.8$, the SFDM-TF density profile smoothly morphs from the shape of the polytrope to the CDM-like power law, while the pressure and temperature profiles increase sharply, reflecting the shock ``heating'' that took place at that location when the strong accretion 
    shock formed there during collapse. 
    
    In Fig.~\ref{fig:numprofwHSE}, we compare our 1D hydro simulation results with the corresponding isothermal sphere HSE solutions (equation~\ref{eq:emden} with $n=1$).
    The SFDM-TF HSE solution ($\chi = 1$; long-dashed blue curves) is chosen to have the same central density and polytrope radius as the simulation profile (i.e. specifying \{$R_\text{TF}, \rho_c$\} and calculating $\sigma^2 = 4\pi G \rho_c R_\text{TF}^2/\pi^2$, as described in \S\ref{sec:virialized_objects}), 
    while the CDM-like isothermal sphere HSE solution ($\chi=0$; short-dashed red curves) uses the same value for $\sigma^2$ and is in the limit where its core radius is small compared to the polytrope radius of the corresponding SFDM-TF profile (i.e. $r_{0,0} \ll R_\text{TF}$; see \S\ref{sec:virialized_objects}).
    Consequently, Fig.~\ref{fig:numprofwHSE} shows only the asymptotic CDM-like HSE profile, far outside its isothermal sphere core radius (i.e. $r \gg r_{0,0}$), where its density profile follows $\rho \propto r^{-2}$, which differs slightly from the spherical-collapse result of $\rho \propto r^{-12/7}$.
    The SFDM-TF HSE density profile has a somewhat larger core, extending out past $R_\text{TF}$, and features a sharper transition to the envelope than the simulation result, but is otherwise similar to it. 
    The larger core is due to the fact that, by assumption, the HSE approximation is isothermal everywhere, including in the core, whereas the ``temperature'' profile produced by the 1D-spherical dynamics drops considerably near the center, where the SI pressure dominates. 
    As a result, the corresponding velocity-dispersion pressure is much larger near the center of the HSE profile than it is in the simulation results, leading to a more puffed-out central core. 
    
    The numerical profile's core size and shape are robust to changes in the simulation parameters. 
    For example, in Fig.~\ref{fig:varyR0}, we show the resulting density profiles of 4 runs, each using a different value for the polytrope radius ($R_\text{TF} = \{0 \text{ [i.e. CDM]}, 0.1, 1, 10\}$ kpc). 
    Each of the profiles for the cases with SI tracks the CDM case for $r\gtrsim 0.8R_\text{TF}$, and flattens into the shape of a polytropic core inside $r\lesssim 0.4R_\text{TF}$.
    As a reasonable approximation, we can understand the structure of our numerical profiles as resulting from a redistribution of the mass interior to some fraction of $R_\text{TF}$ from the CDM-like power law profile into the shape of an ($n=1$)-polytrope. That is, the density profile should fit something close to
    \begin{equation}
        \rho(r) = 
        \begin{cases}
            \rho_c \text{sinc}(\pi r/R_\text{TF}) & r\leq\alpha R_\text{TF} \\
            \rho_0 (r/R_\text{TF})^{-12/7} & r\geq\alpha R_\text{TF}
        \end{cases}
        \label{eq:numfit}
    \end{equation}
    where $\rho_0$ is the density of the CDM-like envelope at $r=R_\text{TF}$, and is related to the central density ($\rho_c$) by two constraints:
    (i) that the density profile be continuous at the switch point between the polytropic core and CDM-like envelope at $r=\alpha R_\text{TF}$,
    \begin{equation}
        \rho_c \text{sinc}(\pi \alpha) = \rho_0 \alpha^{-12/7}
        \label{eq:denscont}
    \end{equation}
    and (ii) that the mass interior to $r=\alpha R_\text{TF}$ in the SFDM-TF profile be the same as that of the corresponding CDM profile (i.e. the mass that would have been there had SI been absent),
    \begin{align}
        \int_0^{\alpha R_\text{TF}} \!\!\!\rho_c \text{sinc}\Big(\frac{\pi r}{R_\text{TF}}\Big) 4\pi r^2 dr = \int_0^{\alpha R_\text{TF}} \!\!\!\rho_0 \Big(\frac{r}{R_\text{TF}}\Big)^{-12/7} 4\pi r^2 dr
        \label{eq:masscons}
    \end{align}
    The latter constraint allows us to think of the SFDM-TF profile as a transformation of the CDM-like profile in which the mass interior to $r=\alpha R_\text{TF}$ is conservatively reconfigured into the polytropic core.
    Satisfying these two constraints simultaneously requires
    \begin{equation}
        \alpha \simeq 0.825
    \end{equation}
    We can then fit equation~(\ref{eq:numfit}) to our simulation results using this value of $\alpha$ by matching $\rho_0$ to the corresponding numerical CDM profile, as shown in Fig.~\ref{fig:numfit}.
    Here, the black curve shows our numerical SFDM-TF density profile, the red dashed curve shows the matching CDM-like power law profile, and the blue dashed curve shows the result from equation~(\ref{eq:numfit}) under the constraints given above. 
    The two vertical lines mark the radii at which the polytrope profile in equation~(\ref{eq:numfit}) intersects the power law, the outer of which is, by construction, at $r/R_\text{TF} = \alpha \simeq 0.825$, while the inner is at $r/R_\text{TF} \simeq 0.373$.
    As can be seen in the figure, these lines also demarcate the intermediate region in which the numerical profile transitions from the core to the envelope, which is where the simple polytrope + power law fitting function deviates most noticeably from the numerical result.
    At larger radii ($r/R_\text{TF} > 0.825$), the SFDM-TF numerical profile tracks the CDM-like power law envelope, and at smaller radii ($r/R_\text{TF} < 0.373$), it tracks the ($n=1$)-polytropic core (although with a slightly higher central density than the relationship between $\rho_c$ and $\rho_0$ would suggest, according to the considerations outlined above).
    At fixed $R_\text{TF}$, and for a fixed overdensity criterion defining the halo mass and radius (i.e. $M_h \propto R_h^3$), it is straightforward to show that this core-envelope halo structure, where the envelope follows $\rho \propto r^{-12/7}$, admits the mass relation
    \begin{equation}
        M_c \propto M_h^{12/21} \simeq M_h^{0.57}
    \end{equation}
    which is close to the ``isothermal'' assumption of $M_c \propto M_h^{2/3}$ (equation~\ref{eq:chTF}) that corresponds to an envelope profile of $\rho \propto r^{-2}$.

\subsection{Virial equilibrium}
\label{sec:Virialequilibrium}
  
\begin{figure}
        \centering
        \includegraphics[width=\columnwidth]{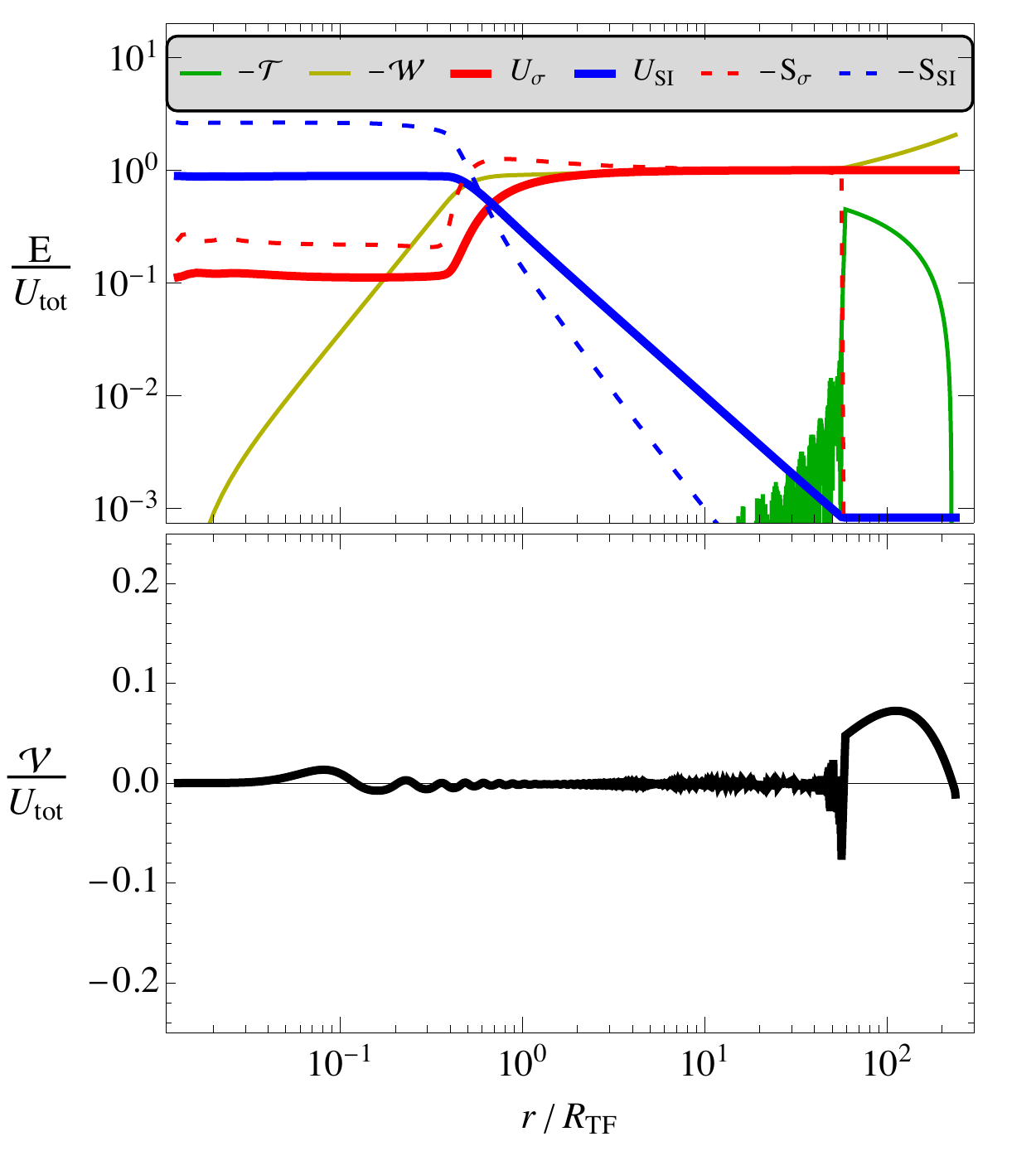}
        \caption{\textbf{Virial equilibrium results for simulated SFDM-TF halo formation.} Top: Cumulative energy and surface pressure profiles given by equations~(\ref{eq:T}) - (\ref{eq:surfPSI}) for the same SFDM-TF halo shown in Fig.~\ref{fig:numprof}, produced by our 1D spherical collapse calculation. The curves are normalized by the total cumulative internal energy, $U_\text{tot} = U_{\!\sigma} + U_\text{SI}$, at each radius. This total is dominated by the self-interaction internal energy inside the central core, and the internal energy due to dispersion pressure outside the core. Bottom: The total cumulative virial (equation~\ref{eq:virial}) for the same SFDM-TF halo profile. The figure demonstrates that, inside of the accretion shock (marked by the discontinuity near $r = 60R_\text{TF}$), the system is in virial equilibrium ($\mathcal{V} \simeq 0$).}
        \label{fig:virial}
    \end{figure}

    The accretion shock serves to virialize the collapsing mass, and so acts as a physical outer radius of the region we characterize as the 
    halo, discontinuously separating it from the cold, infalling material outside of it. To demonstrate that our numerical SFDM-TF haloes are, in fact, in virial equilibrium, we plot the cumulative profiles of all the terms that contribute to the virial $\mathcal{V}$ (top panel), along with their sums in the cumulative virial (bottom panel) for our fiducial run, in Fig.~\ref{fig:virial}.
    The terms plotted are those associated with bulk kinetic energy, gravitational potential energy, and internal energies associated with the velocity-dispersion and SI pressures, integrated over volume from the center out to a given radius:
    \begin{align}
        &\mathcal{T} = \frac{1}{2}\int v^2dM - \frac{4\pi r^2}{2} \left(\rho r v^2 + \frac{1}{2} r^2 \frac{\partial}{\partial t}(\rho v)\right) \label{eq:T}\\
        &\mathcal{W} = -\int \frac{GM}{r}dM \label{eq:W}\\
        &U_{\!\sigma} = \frac{3}{2}\int 4\pi r^2 P_{\!\sigma} dr = \frac{3}{2}\int \frac{P_{\!\sigma}}{\rho} dM \label{eq:Ugas}\\
        &U_\text{SI} = \int 4\pi r^2 P_\text{SI} dr = \int \frac{P_\text{SI}}{\rho} dM \label{eq:Usi}
    \end{align}
    In addition, we plot the velocity-dispersion and SI surface pressure terms given by
    \begin{align}
        &S_{\!\sigma} = -4\pi r^3 P_{\!\sigma} \label{eq:surfPgas} \\
        &S_\text{SI} = -4\pi r^3 P_\text{SI} \label{eq:surfPSI} 
    \end{align}
    The curves are normalized at each radius by the total cumulative internal energy within this radius, $U_\text{tot} = U_{\!\sigma} + U_\text{SI}$. In the central
    region where pressure support is dominated by $P_\text{SI}$, this total is dominated by the SI term. Near $r\simeq 0.4R_\text{TF}$, however, the velocity-dispersion pressure term starts to increase in importance, becoming dominant by $r=R_\text{TF}$, as the halo profile makes the transition from the polytropic core to the CDM-like envelope. The total virial is given by (see Appendix~\ref{sec:virial}):
    \begin{equation}
        \mathcal{V} = 2\mathcal{T} + \mathcal{W} + 2U_{\!\sigma} + 3U_\text{SI} + S_{\!\sigma} + S_\text{SI} \label{eq:virial}
    \end{equation}
    The bottom panel of the figure shows that the halo is, indeed, in approximate virial equilibrium ($\mathcal{V} \simeq 0$) inside of the accretion shock (the discontinuity at about $60R_\text{TF}$); the shock virializes infalling matter as it accretes onto the halo. 
    
\section{Comparisons to Observations}
\label{sec:observations}

    Following the work of \cite{VBB19}, we shall now assess the ability of the SFDM-TF model to solve the small-scale structure problems of CDM, 
    namely the too-big-to-fail and cusp-core problems, by comparing the rotation curves of SFDM-TF haloes to observations of Local Group dwarf galaxies.  
    In analyzing the FDM model, \cite{VBB19} found that if they tuned the particle mass $m$ (the only free parameter for FDM) 
    so as to reduce the central densities of $\lesssim 10^{10} \text{ M}_\odot$ haloes 
    enough to bring them into agreement with observed dwarf galaxies and resolve the too-big-to-fail problem, then
    higher-mass ($\sim 10^{11}\text{ M}_\odot$) FDM haloes would fail to match dwarf galaxy observations, 
    faring even worse than corresponding CDM haloes with regards to the cusp-core problem.
    This represents a kind of ``catch-22'' problem for FDM,
    in which the internal structure of FDM haloes depends upon halo mass in an unfavorable way.
    As we shall demonstrate in what follows, this is not the case for SFDM-TF haloes of the same total mass and size as the
    CDM (and FDM) counterparts studied by \cite{VBB19} for their comparison.
    
    \subsection{Models for SFDM-TF halo rotation curves}
    
    We use 3 approximations to model the mass profiles of SFDM-TF haloes for this purpose: the profile produced by the 1D dynamics of our spherical collapse simulation (\S\ref{sec:1Dsim}), the HSE solution for SI-modified isothermal spheres  (\S\ref{sec:analytical}), and a fitting function,
    similar to equation~(\ref{eq:numfit}), but in which the CDM-like envelope
    that surrounds the SI-dominated polytropic core follows the standard NFW profile of CDM haloes from N-body simulations, rather than the power law profile
    for the 1D hydro results
    of our noncosmological 
    (i.e. static) initial conditions.
    This fitting function is analogous to that for FDM halo profiles used by \cite{VBB19}, which they based upon the results of numerical simulations of FDM halo formation \citep[e.g.][]{SCB14} that found solitonic cores of size of order the de~Broglie wavelength inside the halo, surrounded by CDM-like envelopes, as the generic outcome. 
    We demonstrated above that halo formation in SFDM-TF leads to the same kind of core-envelope halo structure, but with
    a polytropic core of size $\sim R_\text{TF}$, instead of the solitonic cores of size $\sim \lambda_\text{deB}$.
    Hence, we construct our fitting function to
    represent the cosmological dark matter halo profiles for SFDM-TF by
    replacing the CDM-like envelope of our simulated profile from noncosmological initial conditions with an NFW density profile outside some fraction of $R_\text{TF}$,
    but with the same $(n=1)$-polytropic core as the simulations, within this radius, as follows:
    \begin{align}
        &\rho(r) = 
        \begin{cases}
            \rho_c \text{sinc}(\pi r/R_\text{TF}) & r\leq\alpha R_\text{TF} \\
            \rho_\text{NFW}(r) & r\geq\alpha R_\text{TF}
        \end{cases}
        \label{eq:fitfn} \\
        &\rho_\text{NFW}(r) = \frac{\delta_\text{NFW} \, \rho_\text{crit}}{(c_\text{NFW} \, r/R_h)(1 + c_\text{NFW} \, r/R_h)^2} \\
        &\delta_\text{NFW} = \frac{\Delta_\text{crit}}{3} \frac{c_\text{NFW}^3}{ \ln(1+c_\text{NFW}) - c_\text{NFW}/(1+c_\text{NFW}) }
    \end{align}
    where $c_\text{NFW}$ is the halo concentration parameter, $\rho_\text{crit} = 3H_0^2/8\pi G$ is the present-day critical background density, and $\Delta_\text{crit}$ is the mean overdensity of the halo relative to $\rho_\text{crit}$.
    We take $H_0 = 70$ km/s/Mpc and adopt the convention that the NFW halo radius is $R_h = R_{200}$ -- the radius within which the mean density is $200$ times the critical density -- and define the halo mass accordingly (i.e. $M_h = \frac{4\pi}{3} R_h^3 \Delta_\text{crit} \rho_\text{crit}$, $\Delta_\text{crit} = 200$).\footnote{
    Our approach, while similar to that of \citet{VBB19}, 
    differs from theirs in certain details. Their paper states that their assumed halo overdensity corresponded to that for the standard top-hat model in our observed Universe
    \citep[e.g. as fit by][]{BN98}. However, from discussions with the authors, we were informed that the value of overdensity they actually assumed is about 
    3 times higher than that.
    The resulting higher overdensity is then inconsistent with their statement that their plotted CDM haloes were based upon the median values of the concentration parameter for the halo mass-concentration relation derived from CDM N-body simulations for haloes defined by the standard overdensity criterion.   Nonetheless, as we find that their general conclusions remain valid after correcting for these errors, we continue to refer and compare our results to theirs, while constructing our own CDM and FDM profiles, but with halo masses and concentration parameters self-consistently related by the mass-concentration relation in the literature for haloes defined by overdensity $\Delta_\text{crit}=200$.}
    
\begin{figure}
    \centering
    \includegraphics[width=\columnwidth]{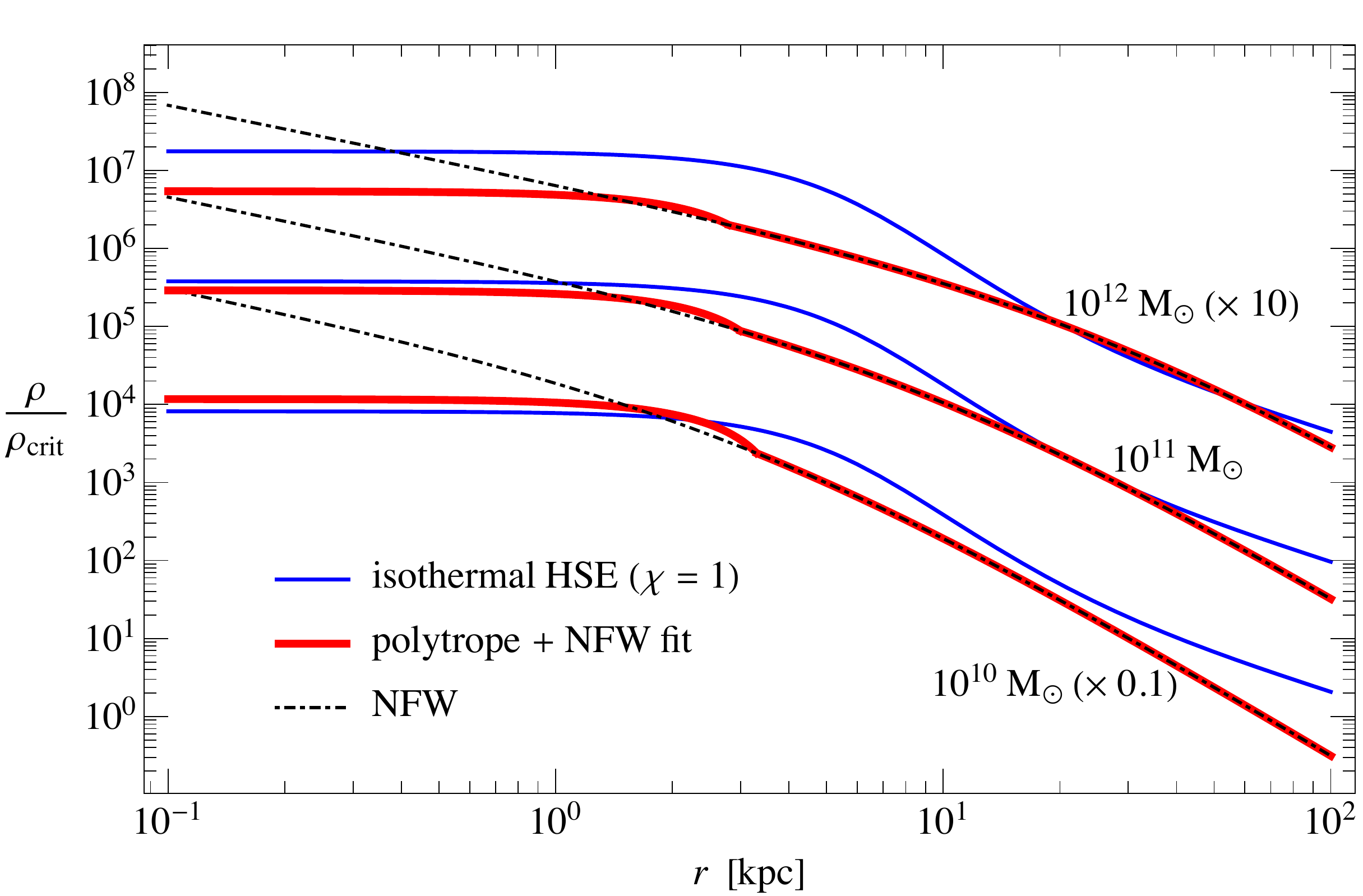}
    \caption{\textbf{Polytropic core + NFW envelope fitting function for SFDM-TF halo profiles.} Comparison of the approximate SFDM-TF halo density profiles for $R_\text{TF} = 4$ kpc given by the core+NFW fitting function (equation~\ref{eq:fitfn}; red) and the SI-modified isothermal hydrostatic equilibrium solution (equation~\ref{eq:emden}; blue) for 3 different halo masses, as labeled.
    Density is in units of the present-day critical density ($\rho_\text{crit}$).
    The NFW profiles are at median concentration for each mass, as per the mass-concentration relation found by \citet{Klypin16}.
    The $10^{10}$ and $10^{12} \text{ M}_\odot$ halo profiles are vertically offset by an order of magnitude for visual clarity.
    As determined by the constraints of density continuity and mass conservation (equations~~\ref{eq:denscontNFW} and \ref{eq:massconsNFW}),
    the switch point between the polytropic core and NFW envelope occurs at $\alpha \simeq \{0.824, 0.748, 0.699\}$ for the $10^{10}, 10^{11}$, and $10^{12}\text{ M}_\odot$ cases, respectively.}
    \label{fig:fitfn}
    \end{figure}

    Apart from the difference in the envelope profile, we set the parameters of this fitting function ($\alpha$ and $\rho_c$) in the same manner as what was done for the fit to our 1D hydro results with equation~(\ref{eq:numfit}). Conceptually, we start with an NFW profile for a given halo mass and concentration parameter,
    and conservatively redistribute the mass interior to $r = \alpha R_\text{TF}$ into the shape of the polytropic core. 
    Continuity in the density profile at the switch point requires
    \begin{equation}
        \rho_c \text{sinc}(\pi\alpha) = \rho_\text{NFW}(\alpha R_\text{TF})
        \label{eq:denscontNFW}
    \end{equation}
    while mass conservation requires
    \begin{equation}
        \int_0^{\alpha R_\text{TF}} \!\!\!\rho_c \text{sinc}\Big(\frac{\pi r}{R_\text{TF}}\Big) 4\pi r^2 dr = \int_0^{\alpha R_\text{TF}} \!\!\!\rho_\text{NFW}(r) 4\pi r^2 dr
        \label{eq:massconsNFW}
    \end{equation}
    These two conditions are sufficient to determine $\alpha$ and $\rho_c$ for a given halo, however unlike the simple power law envelope profile used in equation~(\ref{eq:numfit}), since the shape of the NFW profile depends on halo mass and concentration, the value of $\alpha$ will vary from halo to halo and with different choices of $R_\text{TF}$.
    In Fig.~\ref{fig:fitfn}, we compare the resulting fitting function profiles for $R_\text{TF} = 4$ kpc and 3 different halo masses ($10^{10}$, $10^{11}$, $10^{12}$ M$_\odot$) -- at median concentration as per the mass-concentration relation found by \citet{Klypin16} -- to the isothermal hydrostatic equilibrium solutions for those masses (i.e. specifying \{$R_\text{TF}, \sigma = \sqrt{GM_h/R_h}$\} and using the procedure outlined in \S\ref{sec:analytical}).
    The values of $\alpha$ corresponding to these 3 cases are found to be $\simeq$ 0.824, 0.748, and 0.699, respectively.
    
     \subsection{SFDM-TF vs. FDM haloes: solving the CDM too-big-to-fail problem?}
    
\begin{figure}
        \centering
        \includegraphics[width=\columnwidth]{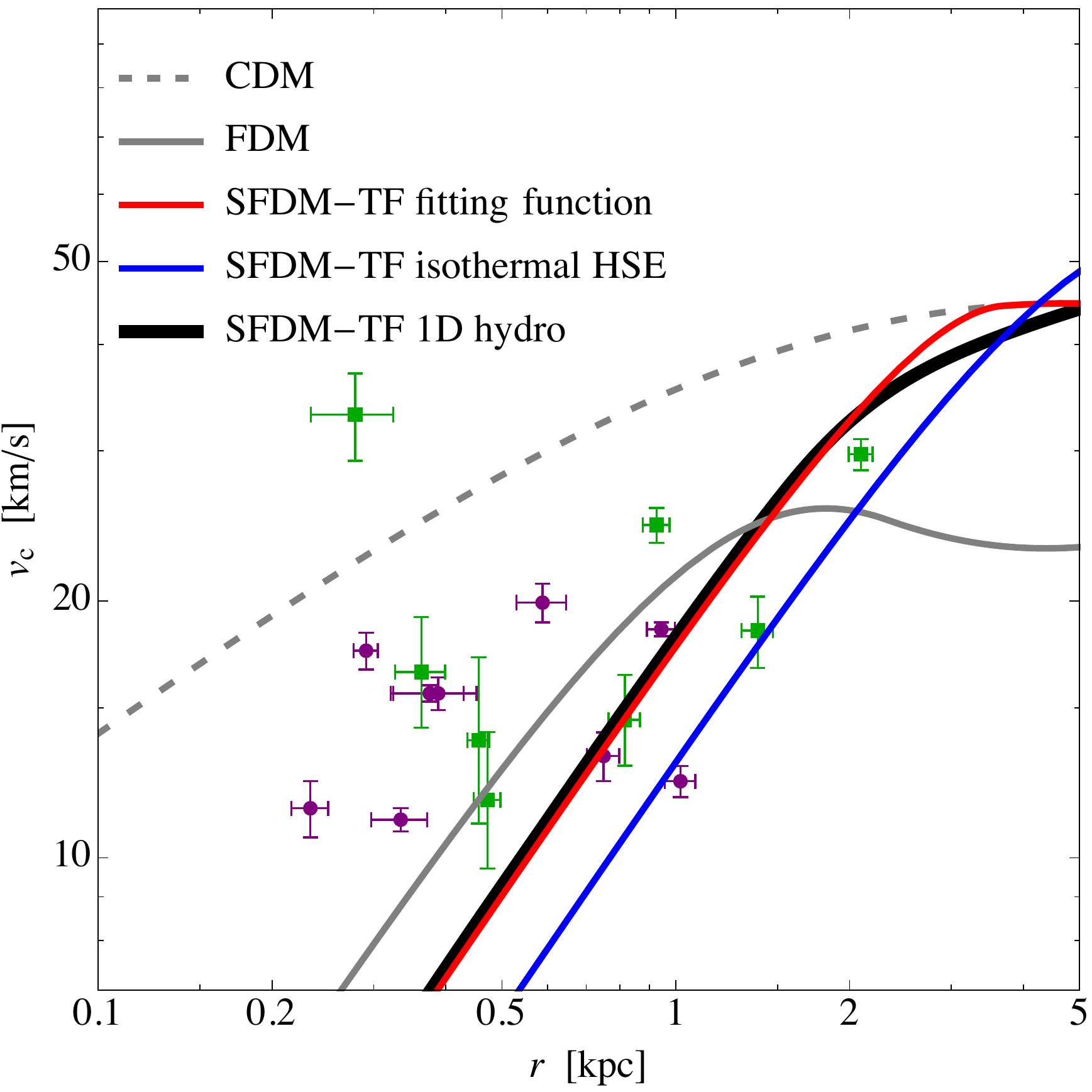}
        \caption{\textbf{The SFDM-TF model solution to the too-big-to-fail problem.} Comparison of predicted SFDM-TF halo rotation curves with dwarf galaxy observations, similar to that presented in \citet{VBB19} for FDM haloes. The grey dashed curve is the rotation curve of a $10^{10} \text{ M}_\odot$ NFW halo with a concentration parameter of $c_\text{NFW}=20$. This is compared to the rotation curves for an FDM halo, and our 3 approximations for an SFDM-TF halo -- the core+NFW fitting function (equation~\ref{eq:fitfn}), the SI-modified, isothermal sphere HSE solution (equation~\ref{eq:emden}), and the numerical profile produced by our 1D spherical collapse calculation. We use a particle mass of $mc^2=0.8\times10^{-22}$~eV for the FDM model, and a polytrope radius of $R_\text{TF} = 4$~kpc for the SFDM-TF model. These non-CDM models solve the too-big-to-fail problem, because their rotation curves are consistent with measurements of Milky Way satellites \citep[purple circles; from][]{Wolf10} and field dwarfs \citep[green squares; from][]{Kirby14}, whereas the CDM model is not.}
        \label{fig:tbtf}
    \end{figure}

    One way to state the too-big-to-fail problem is that many of the most massive subhaloes that are formed in CDM simulations of the Local Group are too centrally dense to be consistent with observations of the bright Milky Way satellites and field dwarfs that the subhaloes should host.
    \cite{VBB19} illustrate this by plotting the NFW rotation curves for $10^{9.5}$ and $10^{10}$ M$_\odot$ haloes, along with the circular velocities at the half-light radii of observed dwarf galaxies, to show that the CDM haloes fail to be consistent with any of the galaxies, having too high a rotation curve at small radii.
    On the other hand, the rotation curve for an FDM halo of the same mass -- generated by the fitting function procedure outlined in their work -- matches many of the galaxy data points, due to its low-density central core.
    We show a similar result for the SFDM-TF model in Fig.~\ref{fig:tbtf}, comparing to rotation curves for CDM and FDM, and data for the half-light circular velocities of bright Milky Way satellites \citep[purple circles; from][]{Wolf10} and Local Group field dwarfs \citep[green squares; from][]{Kirby14}.
    For the CDM halo, we use a $10^{10}$ M$_\odot$ NFW profile with a 
    concentration parameter of $c_\text{NFW}=20$,
    corresponding to a maximum circular velocity of about $v_\text{c,max}=45$ km/s at a radius of about $r_\text{max}=5$ kpc, which is typical for the too-big-to-fail problem \citep[see, e.g.][]{BKBK11}. The CDM rotation curve fails to match the half-light circular velocities of observed galaxies.
    We use the same halo mass to generate the rotation curve for FDM \citep[according to the prescription in][with a particle mass of $mc^2=0.8\times10^{-22}$ eV]{VBB19}, 
    as well as those for the SI-modified isothermal HSE approximation and the SFDM-TF polytrope + NFW fitting function.
    Finally, we tune the central density of the initial condition of our 1D spherical collapse calculation to produce a numerical SFDM-TF halo with the same circular velocity at the same radius as $v_\text{c,max}$ and $r_\text{max}$ for the CDM halo (i.e. a circular velocity of 45 km/s at a radius of 5 kpc). 
    For all 3 SFDM-TF rotation curve approximations, we adopt a polytrope radius of $R_\text{TF} = 4$ kpc.
    As can be seen in the figure, the SFDM-TF halo profiles also avoid the too-big-to-fail problem, just as FDM does, due to their low-density central cores.  

    \begin{figure}
        \centering
        \includegraphics[width=\columnwidth]{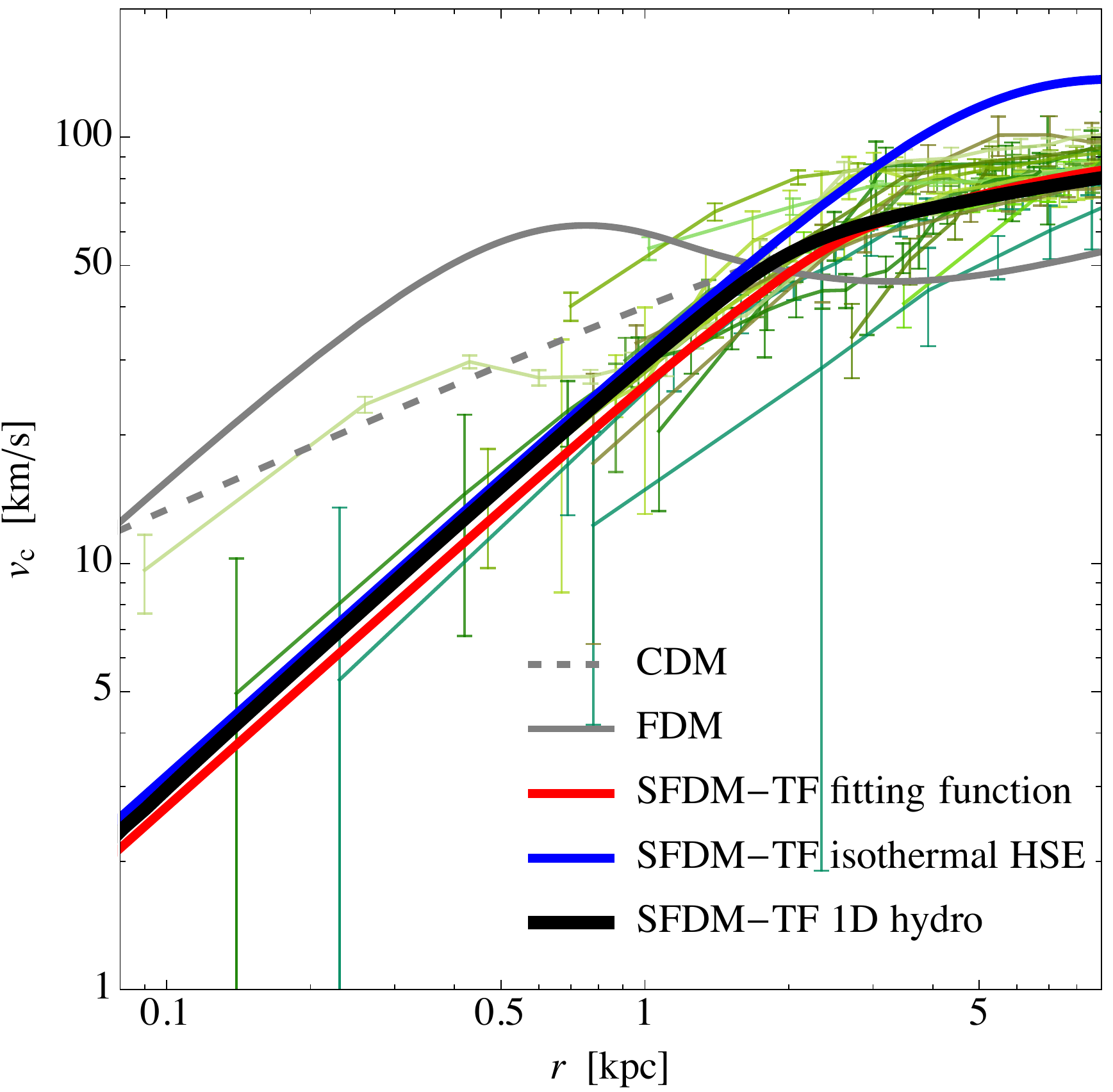}
        \caption{\textbf{The SFDM-TF model solution to the cusp-core problem.} 
        The grey dashed curve is the rotation curve of a $10^{11.15} \text{ M}_\odot$ NFW halo at median concentration. 
        This is compared to the rotation curves for an FDM halo, 
        and our 3 approximations for an SFDM-TF halo -- the core+NFW fitting function (equation~\ref{eq:fitfn}), 
        the SI-modified, isothermal sphere HSE solution (equation~\ref{eq:emden}), and the numerical profile produced by our 1D spherical collapse calculation. 
        We use a particle mass of $mc^2=0.8\times10^{-22}$ eV for the FDM model, 
        and a polytrope radius of $R_\text{TF} = 4$ kpc for the SFDM-TF model. 
        The SFDM-TF model is a better fit to the dwarf galaxy rotation curve data \citep[thin green lines; from][]{sparc} than the CDM and FDM models.}
        \label{fig:cuspcore}
    \end{figure}
    
    \subsection{SFDM-TF vs. FDM haloes: solving the CDM cusp-core problem, too?}
    
    While FDM successfully avoids the too-big-to-fail problem, \cite{VBB19} also demonstrated that the model fails to avoid the cusp-core problem in higher-mass, $\sim 10^{11} \text{ M}_\odot$ haloes, due to the fact that the characteristic size of the FDM halo core (the de~Broglie wavelength) shrinks with increasing core mass ($\lambda_\text{deB} \propto 1/M_c$), causing the central density to increase rapidly with mass. Heuristically, under the assumption of isothermality (i.e. $GM/R = $ constant), the ratio of FDM solitonic core mass to core radius should be 
    roughly equal to the ratio of halo mass to halo radius:
    \begin{align}
        &\frac{M_c}{\lambda_\text{deB}} \approx \frac{M_h}{R_h} \propto M_h^{2/3} \\
        &M_c \sim M_h^{1/3}
    \end{align}
    Therefore, the central density scales with halo mass as
    \begin{equation}
        \rho_c \propto \frac{M_c}{\lambda_\text{deB}^3} \propto M_c^4 \sim M_h^{4/3}
    \end{equation}
    As a result, although FDM does produce $\sim 10^{11} \text{ M}_\odot$ haloes with central cores rather than cusps, 
    such that their density profiles/rotation curves have the right \textit{slope} 
    in comparison to observed high-mass dwarf galaxies, the \textit{amplitudes} of those profiles
    are too high -- even higher than for CDM for a significant range of radii.
    
    We ask the same question now for the SFDM-TF model.
    First, following \cite{VBB19}, we plot CDM and FDM (with $mc^2=0.8\times10^{-22}$ eV) 
    rotation curves for a $10^{11.15}$~M$_\odot$ halo, along with observed dwarf galaxy rotation curves from the SPARC catalog in \cite*{sparc}, to illustrate the discrepancy between these models and the data at this mass scale, in Fig.~\ref{fig:cuspcore}.
    The CDM curve is an NFW profile with concentration parameter $c_\text{NFW}\simeq 10$,
    the median value for haloes of that mass today
    according to the mass-concentration relation reported by
    \cite{Klypin16}, 
    which corresponds to a maximum circular velocity of around $v_\text{c,max}=90$ km/s 
    at a radius of about $r_\text{max}=23$ kpc. 
    We extract the approximate contribution of dark matter to the 
    observed galaxy rotation curves by subtracting the baryonic components from the data, 
    assuming a constant stellar mass-to-light ratio of 0.2 M$_\odot$/L$_\odot$ at 3.6~$\mu$m, 
    as was done by \citet{VBB19}. 
    The resulting dark matter-only rotation curve is what is shown in the figure. 
    In addition, only those galaxies with (dark matter-only) asymptotic circular 
    velocities in the range 80~-~100 km/s are plotted.
    
    In the SFDM-TF model, by contrast with FDM,
    the core size is constant. Under the same assumption of isothermality, we can see that the central density grows more slowly with increasing halo mass than it does for FDM:
    \begin{align}
        &\frac{M_c}{R_\text{TF}} \approx \frac{M_h}{R_h} \propto M_h^{2/3} \\
        &M_c \sim M_h^{2/3} \\
        &\rho_c \propto \frac{M_c}{R_\text{TF}^3} \propto M_c \sim M_h^{2/3}
    \end{align}
    Consequently, the SFDM-TF halo profile can have both, the right slope \textit{and} the right amplitude, and so is a better fit to the data at small radii than CDM and FDM. We show the SFDM-TF halo rotation curves for our 3 approximation schemes with $R_\text{TF} = 4$ kpc in Fig.~\ref{fig:cuspcore}. Again, we use the same halo mass that is adopted for the CDM and FDM profiles to generate the isothermal HSE solution and fitting function curves, and tune the central density of the initial condition of our 1D spherical collapse calculation to produce a halo with the same circular velocity at the same radius as $v_\text{c,max}$ and $r_\text{max}$ for the NFW profile (i.e. a circular velocity of 90 km/s at a radius of 23 kpc). 
    All 3 SFDM-TF curves have central profiles that are much more consistent with the data than the CDM or FDM curves, demonstrating the ability of the SFDM-TF model to avoid the cusp-core problem in high-mass dwarf galaxies.

\section{Summary and Conclusions}
\label{sec:Conclusions}

We have explored an alternative to the standard cold dark matter (CDM) paradigm, scalar field dark matter (SFDM) in the Thomas-Fermi (TF) regime, referred to as SFDM-TF, which features a strong repulsive self-interaction (SI) that dominates the small-scale structure and dynamics of dark matter haloes.
The more-commonly studied free-field limit of SFDM (i.e. no SI), known as fuzzy dark matter (FDM), differentiates itself from CDM on the scale of the de~Broglie wavelength ($\lambda_\text{deB}$), suppressing structure and producing centrally cored density profiles below that scale (i.e. the FDM halo profile features a $\lambda_\text{deB}$-scale core surrounded by a CDM-like envelope).
In the TF regime, $\lambda_\text{deB}$ is tuned to be much smaller than the characteristic length scale of the repulsive SI, known as the polytrope radius ($R_\text{TF}$), so the latter is thought to be the relevant scale below which SFDM-TF dynamics deviates from CDM, and we have shown in this work that this is indeed the case.

We built on previous work related to the FDM model from \citet{SKV89}, \citet{WK93}, and \citet{Mocz18} which demonstrated that by constructing a smoothed phase space representation of the scalar field and taking $\lambda_\text{deB}$ to be much smaller than the smoothing scale, the phase space distribution function satisfies the collisionless Boltzmann equation (CBE) -- the same equation that governs CDM dynamics -- illustrating that FDM is CDM-like on scales larger than $\lambda_\text{deB}$.
By adding in the repulsive SI potential and taking momentum moments of the CBE, we derived the continuity and momentum equations for the SFDM-TF model, complete with SI pressure and a CDM-like velocity-dispersion pressure -- which approximates the large-scale behavior of quantum pressure -- to provide support against gravity.
Furthermore, \citet{AS05} showed that under the assumptions of spherical symmetry and an isotropic and skew-free velocity distribution, the CDM-like velocity-dispersion pressure obeys the energy equation of a $\gamma = 5/3$ ideal gas.
Therefore, the SFDM-TF fluid equations can be solved numerically with a standard 1D-spherical hydrodynamics code, needing only slight modification to accommodate the repulsive self-interaction.

We employed such a code to run a suite of simulations, for various $R_\text{TF}$ values, of the gravitational collapse of Jeans-unstable spherical Lagrangian mass shells from a linear, cold and static initial condition.  
The collapse results in the formation of virialized dark matter haloes that feature flattened central cores dominated by SI pressure below the scale of $R_\text{TF}$, and CDM-like envelopes dominated by velocity-dispersion pressure above that radius, as expected; SFDM-TF is CDM-like on scales larger than $R_\text{TF}$.
The density profiles of these haloes can be fit by equation~(\ref{eq:numfit}), which attaches an ($n=1$)-polytrope profile to the center of the CDM-like envelope (which, for our setup, is a power law with a logarithmic slope of $-12/7$) at a given fraction ($\simeq 0.825$) of the polytrope radius determined by enforcing density continuity and mass conservation at the connection point.
For the purposes of comparing to observations, we also constructed an analogous analytical ``fitting function'', given by equation~(\ref{eq:fitfn}), which uses an NFW profile for the CDM-like envelope instead of the power law profile that results from our noncosmological 1D simulations, by applying the same criteria of density continuity and mass conservation for attaching an ($n=1$)-polytrope to the center.
For a third point of comparison, we solved the equation of hydrostatic equilibrium (HSE) in spherical symmetry for the SFDM-TF model (equation~\ref{eq:emden} with $n=1$) under the assumption of isothermality (i.e. constant velocity-dispersion ``temperature'', $P_{\!\sigma}/\rho$).
The solution is a combined $[(n=\infty) + (n=1)]$-polytrope and, for well-motivated values of the $\chi$ parameter ($\chi \sim 1$), again features a central core below the scale of $R_\text{TF}$ and an envelope above that scale, and so is in reasonably good agreement with our simulation results, modulo the fact that the latter are not exactly isothermal and feature a sharp drop-off in the velocity-dispersion ``temperature'' inside $R_\text{TF}$.

Finally, following the work of \citet{VBB19} on the FDM model, we compared our 3 SFDM-TF halo profile approximations -- the isothermal HSE solution, the polytropic core + NFW envelope fitting function, and the results of our 1D spherical collapse simulations -- to half-light circular velocities and rotation curves of observed Local Group dwarf galaxies to assess the ability of the SFDM-TF model to avoid the too-big-to-fail and cusp-core problems of standard CDM.
\citet{VBB19} found that if the particle mass is tuned to allow the FDM model to avoid the too-big-to-fail problem in $\sim 10^{10} \text{ M}_\odot$ haloes by creating kpc-scale, low-density central cores, then higher-mass ($\sim 10^{11} \text{ M}_\odot$) FDM haloes would fail to avoid the cusp-core problem, because their central densities are significantly higher -- even higher than their CDM counterparts for a considerable range of radii -- so their rotation curves are inconsistent with observations of high-mass dwarf galaxies.
However, unlike FDM for which the characteristic size of cores in haloes ($\lambda_\text{deB}$) shrinks with increasing halo mass, the characteristic core size of SFDM-TF ($R_\text{TF}$) is a fixed physical constant, so the central density of an SFDM-TF core increases more slowly with increasing halo mass than the central density of an FDM core does.
Consequently, we find that SFDM-TF haloes can match observations on both the lower-mass and higher-mass ends, and so are a better fit to the data than both CDM and FDM.
The dwarf galaxy data are well-fit if $R_\text{TF}$ is in the range $1 \text{ kpc} \lesssim R_\text{TF} \lesssim 5$~kpc, consistent with  the cosmological constraints of \citet{Li14,Li17}. 
This corresponds to self-coupling strengths in the range $1.6\times10^{-18}~{\rm eV^{-1}~cm^3} \lesssim g/m^2 c^4 \lesssim 4\times10^{-17}~{\rm eV^{-1}~cm^3}$, and requires a particle mass larger than $\sim$ (2 - 10)$\times 10^{-22}$ eV$/c^2$ to be in the TF regime. 
Like \citet{VBB19}, our SFDM haloes were made to correspond with the CDM haloes used in the comparison with galaxy data, replacing the internal structure of each CDM halo by that given by our SFDM prescriptions.  To go beyond this, we must consider the abundance and structure of SFDM haloes self-consistently, as a product of large-scale structure formation from cosmological initial conditions for the SFDM model.  That is something we will consider in the future.

In this work, the first in a series devoted to the study of structure formation in the SFDM-TF model, we developed the fluid dynamics approximation as a useful tool, anticipating its further application, and focused on elucidating the fundamental behavior of gravitational instability and collapse to form virialized objects, from noncosmological (i.e. static) initial conditions.  The latter amounts to the classic Jeans instability problem and its nonlinear outcome, in the novel context of SFDM with repulsive self-interaction.  In a 
companion paper \citep[][``Paper II'']{paper2}, we extend the scope to place structure formation in the SFDM-TF model in the broader cosmological context of the expanding Universe.  This will apply the tools developed here again, but with 
fully cosmological boundary conditions, and will consider the implications for placing further constraints on particle parameters in comparison with astronomical observations.    

Some other limitations of our work here include its neglect of the coupling of baryonic dynamics to the dark matter, including dissipation and feedback effects in the former, and the restriction of our simulations here to 1D spherical symmetry. The latter neglects the punctuated effects on galaxy formation from merger events, as well as the angular momentum from tidal torquing by distant matter, during halo mass assembly.  Angular momentum, to first order, is a small effect on the halo formation studied here, since 3D cosmological N-body simulations of CDM tell us that the degree of rotational support is typically quite small (e.g. as measured by halo spin parameter,  $\lambda_\text{spin}\lesssim0.05$).  However, in \cite{RDS12}, we showed that even such small amounts of halo angular momentum, for SFDM-TF with a sufficiently strong self-interaction, make it possible for haloes to form a quantum vortex at the center of their polytropic cores, which might drop the density near the center.  
We will explore these issues and their consequences for the SFDM-TF model in the cosmological context, including full 3D SFDM-TF plus baryon simulations, in future work.  Finally,
even in the full cosmological context described above, structure formation is assumed to be
non-relativistic in this era (i.e. ``the Newtonian
approximation'').  However, as discussed in \citet{Padilla20}, for some parameters, halo cores are sufficiently dense that effects of general relativity become important, making them unstable to gravitational collapse, and leading to formation of supermassive black holes at the center.  While our Newtonian approximation cannot follow this process in detail, we will be able to use it to refine our understanding of the circumstances under which the threshold for instability might be reached.

\section*{Acknowledgements}

We are grateful for helpful discussions with Luis Padilla regarding the core-halo mass relation for SFDM, with Bohua Li regarding cosmological constraints from the homogeneous evolution of SFDM, and with Victor Robles and Michael Boylan-Kolchin regarding their previous work.  We thank Kyungjin Ahn and Hyunbae Park for their collaboration involving the original 1D hydro code which we modified for this work.   
This material is based upon work supported by the National Science Foundation Graduate Research Fellowship Program under Grant No. DGE-1610403. Any opinions, findings, and conclusions or recommendations expressed in this material are those of the authors and do not necessarily reflect the views of the National Science Foundation. 
T.R.-D. is supported by the Austrian Science Fund FWF through an Elise Richter fellowship, Grant No. V 656-N28. 
Simulations presented here were conducted on the Texas Advanced Computing Center's Stampede2 supercomputer under accounts asoz-630 and NSF XSEDE account TG-AST090005.

\section*{Data Availability Statement}

The data underlying this article are available in the article and in its online supplementary material.



\bibliographystyle{mnras}
\bibliography{ref} 



\appendix

\section{The correspondence between quantum pressure and CDM-like velocity-dispersion pressure
-- analytical examples} \label{sec:QPT-to-P}

In this section, we give some simple examples to confirm that the approximate momentum flux density in SFDM on large scales
associated with its internal velocity dispersion -- as computed explicitly 
for a given scalar field configuration, $\psi(x,t)$, by taking momentum moments of a smoothed phase space distribution function -- 
corresponds to the quantum pressure computed from the
density and its spatial derivatives according to the QHD equations, when 
smoothing on scales larger than $\lambda_\text{deB}$.
In particular, as we shall show, in the presence of a suitable 
spatial variation in the density at a given location, 
there is an internal momentum spread with nonzero variance, 
even when the bulk velocity is everywhere zero.  This will help us understand how 
the wave-mechanical nature of SFDM provides a means of supporting it
against gravity on large scales with an effective ``pressure'' force, 
a force which we can 
describe accurately even without resolving the structure of the field
on scales as small as $\lambda_\text{deB}$. 

As discussed in \S\ref{sec:model}, the exact QHD equations (equations~\ref{eq:continuity} and \ref{eq:momentum}) can be derived from moments of the Wigner function, which is an exact phase space distribution function for $\psi$. 
Analogously, in \S\ref{sec:classical}, we adopt a smoothed phase space distribution function, the Husimi function $\mathcal{F}$, to derive the approximate hydrodynamical equations for SFDM on scales larger than $\lambda_\text{deB}$ (equations~\ref{eq:0mom} and \ref{eq:1mom}).
The derivations of the exact and smoothed fluid equations follow the same basic procedure, and the exact and smoothed fluid quantities, such as mass density, bulk velocity, and velocity dispersion, are defined in the same way with respect to the appropriate distribution functions.
Importantly, the quantum pressure tensor is defined in terms of the exact fluid quantities by
\begin{equation}
    \Pi_{ij} = \rho \sigma_{ij}^2
\end{equation}
(see equation \ref{eq:PiRhoSigma}),
which is the same as the definition of the CDM-like velocity dispersion pressure tensor ($P_{ij}$; intended to approximate the effects of quantum pressure on scales larger than $\lambda_\text{deB}$) given in equation~(\ref{eq:Pij}), except that,
in the case of equation~(\ref{eq:Pij}), the mass density and velocity dispersion there are smoothed fluid quantities, not exact ones.
In order to avoid confusion, for the remainder of this section we will
distinguish the smoothed fluid quantities obtained through the procedure outlined in \S\ref{sec:classical} by denoting them with an overhead hat (e.g. $\hat{\rho}$),
in contrast to the exact fluid quantities related directly to $\psi$ through equations~(\ref{eq:rho}, \ref{eq:v}, \ref{eq:QPT}) etc., without a hat (e.g. $\rho$).

\subsection{Static plane wave}
\label{sec:static}

Throughout this work, we rely on the smoothing procedure given in \S\ref{sec:classical} to provide us with hydrodynamical equations that are a good approximation of the large-scale behavior of SFDM when $\lambda_\text{deB}$ is small. 
In particular, comparing equations~(\ref{eq:momentum}) and (\ref{eq:1mom}), we require that the CDM-like velocity-dispersion pressure ($P_{ij}$; since this quantity is, by definition, a ``smoothed'' one, we do not need to disambiguate it with an overhead hat), obtained from moments of the smoothed phase space distribution function ($\mathcal{F}$) and the CBE (equation~\ref{eq:CBE}), capture the approximate large-scale behavior of the quantum pressure tensor ($\Pi_{ij}$), i.e. $P_{ij} \equiv \hat{\Pi}_{ij} \approx {\Pi}_{ij}$ on large scales.

To illustrate this, we explore a simple toy model of a 1D plane wave to calculate and compare the exact and smoothed expressions for its fluid quantities:
\begin{equation}
    \psi = \sqrt{\rho_0} \sin{(2\pi x/\lambda)}
\end{equation}
\begin{equation}
    \rho = \rho_0 \sin^2(2\pi x/\lambda)
    \label{eq:rhosinewave}
\end{equation}
As we shall see, the de Broglie wavelength $\lambda_\text{deB}$ in this case is just the wavelength $\lambda$ of the plane wave, 
so we can think of the oscillations of this plane wave as the small-scale variations over which we intend to smooth when
we calculate the smoothed quantities.
The smoothing should then amount to a kind of spatial average over the oscillations, 
and provide oscillation-free fluid quantities that are good 
large-scale approximations to the exact ones.
For example, we should expect that the smoothed density be $\hat{\rho} \approx \rho_0/2$, since the average of the $\sin^2$ function is 1/2.

Furthermore, using equations~(\ref{eq:v}) and (\ref{eq:QPT}), respectively, 
the exact bulk velocity and quantum pressure of the plane wave are given by
\begin{align}
    &v = 0 \\
    &\Pi = \frac{\rho_0}{2}\Big(\frac{2\pi \hbar}{m\lambda}\Big)^2 \label{eq:Pisinewave}
\end{align}
Since these are both constants, we should expect the smoothed quantities to be  
the same as the exact ones (i.e. $\hat{v} \approx v$, $\hat{\Pi} \approx \Pi$).
The exact velocity dispersion is then given by
\begin{align}
    &\sigma^2 = \frac{\Pi}{\rho} = \frac{1}{2 \sin^2(2\pi x/\lambda)} \Big(\frac{2\pi \hbar}{m\lambda}\Big)^2 \label{eq:sigmasinewave}
\end{align}
We can see from these expressions that $\lambda$ does, indeed, act as a de~Broglie wavelength.
Since the de~Broglie wavelength should be the characteristic length scale of quantum pressure,
we can take the velocity dispersion associated with quantum pressure, $\sigma$, to be the characteristic velocity that enters the usual
de~Broglie wavelength formula, and average over the sinusoidal oscillations to get the typical value (the overhead bar denotes a spatial average):
\begin{equation}
    \lambda_\text{deB} = \overline{\frac{h}{m \sigma}} = \overline{\frac{h}{m\sqrt{\Pi/\rho}}} = \overline{2 \lambda \sin^2(2\pi x/\lambda)} = \lambda
    \label{eq:deBdef}
\end{equation}
In virialized dark matter haloes, $\sigma \sim v_\text{vir}$, so equation~(\ref{eq:deBdef}) serves as a useful operational definition of $\lambda_\text{deB}$ for SFDM.

Following the prescription in \S\ref{sec:classical}, we can now calculate the smoothed phase space distribution function and use it to derive the smoothed fluid quantities; the simplicity of the plane wave toy model makes this analytically tractable.
From equations~(\ref{eq:husimi}) and (\ref{eq:distfunc}), we obtain
\begin{align}
    \mathcal{F} = \frac{\rho_0 \eta}{4\sqrt{\pi}\hbar} &\exp\Big[ - \frac{\eta^2}{\hbar^2} \big( p^2 + (2\pi \hbar/\lambda)^2 \big) \Big] \nonumber\\
    &\times \Big( e^{4\pi \eta^2 p/\hbar\lambda} + e^{-4\pi \eta^2 p/\hbar\lambda} - 2\cos(4\pi x/\lambda) \Big)
\end{align}
which, when the smoothing scale $\eta$ is much larger than $\lambda$, yields the following fluid quantities:
\begin{align}
    \hat{\rho} &= \int \mathcal{F} dp = \frac{\rho_0}{2}\Big(1 - e^{-(2\pi\eta/\lambda)^2}\cos(4\pi x/\lambda)\Big) \approx \frac{\rho_0}{2}
\end{align}
\begin{align}
    \hat{v} &= \frac{\langle p \rangle}{m} = \frac{1}{\hat{\rho}}\int \frac{p}{m}\mathcal{F} dp = 0
\end{align}
\begin{align}
    \langle p^2 \rangle &= \frac{1}{\hat{\rho}}\int p^2 \mathcal{F} dp \nonumber \\
    &= \frac{\rho_0/2}{\hat{\rho}} \Big(\frac{2\pi \hbar}{\lambda}\Big)^2 \left[1 + \frac{\lambda^2}{8\pi^2\eta^2}\Big(1 - e^{-(2\pi\eta/\lambda)^2}\cos(4\pi x/\lambda)\Big)\right] \nonumber\\
    &\approx \Big(\frac{2\pi \hbar}{ \lambda}\Big)^2
\end{align}
\begin{align}
    \hat{\sigma}^2 &= \frac{\langle p^2 \rangle - \langle p \rangle^2}{m^2} \approx \Big(\frac{2\pi \hbar}{m \lambda}\Big)^2 
\end{align}
\begin{align}
    P &\equiv \hat{\Pi} = \hat{\rho} \hat{\sigma}^2 \approx \frac{\rho_0}{2}\Big(\frac{2\pi \hbar}{m \lambda}\Big)^2 = \Pi
\end{align}
Thus, we find that the momentum flux density ($\hat{\Pi}$) which results
from the smoothing procedure, after smoothing on scales larger than $\lambda_\text{deB}$, 
agrees with that calculated exactly, without smoothing ($\Pi$). This
demonstrates that our smoothing procedure successfully captured
the crucial effect of small-scale density variations on the de~Broglie scale,
by their contribution to the internal velocity dispersion, while still providing a smoothed density profile for which those variations are absent.

\subsection{Moving plane wave}

We can also consider a ``moving'' plane wave with a linear phase, which is more complicated but still analytically tractable:
\begin{equation}
    \psi = \sqrt{\rho_0} \sin{(2\pi x/\lambda)} e^{i p_0 x/\hbar}
\end{equation}
In this case, the exact density and quantum pressure are the same as before, but the bulk velocity is now a non-zero constant:
\begin{align}
    \rho &= \rho_0 \sin^2(2\pi x/\lambda) 
\end{align}
\begin{align}
    v &= \frac{p_0}{m}
\end{align}
\begin{align}
    \Pi &= \frac{\rho_0}{2}\Big(\frac{2\pi \hbar}{m\lambda}\Big)^2
\end{align}
The corresponding smoothed fluid quantities are found to be
\begin{align}
    \hat{\rho} &= \frac{\rho_0}{2} \Big(1 - e^{-(2\pi\eta/\lambda)^2}\cos(4\pi x/\lambda)\Big) \approx \frac{\rho_0}{2}
\end{align}
\begin{align}
    \hat{v} &= \frac{\rho_0/2}{\hat{\rho}} \frac{p_0}{m} \Big(1 - e^{-(2\pi\eta/\lambda)^2}\cos(4\pi x/\lambda)\Big) \approx \frac{p_0}{m}
\end{align}
\begin{align}
    \langle p^2 \rangle &= \frac{\rho_0/2}{\hat{\rho}} \Big(\frac{2\pi \hbar}{\lambda}\Big)^2 \nonumber\\
    &\quad\quad \times \left[1 + \left( \frac{p_0^2 \lambda^2}{4\pi^2\hbar^2} + \frac{\lambda^2}{8\pi^2\eta^2} \right) \Big(1 - e^{-(2\pi\eta/\lambda)^2}\cos(4\pi x/\lambda)\Big)\right] \nonumber \\
    &\approx \Big(\frac{2\pi \hbar}{ \lambda}\Big)^2 + p_0^2
\end{align}
\begin{align}
    \hat{\sigma}^2 &= \frac{\langle p^2 \rangle - \langle p \rangle^2}{m^2} \approx \Big(\frac{2\pi \hbar}{m \lambda}\Big)^2
\end{align}
\begin{align}
    P &= \hat{\rho}\hat{\sigma}^2 \approx \frac{\rho_0}{2}\Big(\frac{2\pi \hbar}{m \lambda}\Big)^2 = \Pi
\end{align}
As expected, the smoothed density is the spatial average of the exact density, while the smoothed bulk velocity and pressure agree with their corresponding exact expressions, since the latter are constants. 
The important difference with this ``moving'' plane wave case is that the $\langle p^2 \rangle$ term in the velocity dispersion picks up a contribution from the bulk velocity.
However, this contribution is subtracted off by the $\langle p \rangle^2$ term, so the velocity dispersion and its associated pressure are the same as they were in the previous case without bulk motion.
This is as it should be, since the smoothed pressure is supposed to approximate quantum pressure, which depends only on the density distribution, not the bulk velocity.

\subsection{Gaussian-modulated plane wave}
\label{sec:Gaussianmodulated}

Finally, we consider an example that features a separation of scales: the usual small-scale variations to be smoothed over amidst large-scale variations that should be resolved, not smoothed over. 
This kind of example is relevant for the CDM-like envelopes modeled in this work. 
FDM simulations of virialized dark matter haloes show that the density distributions in halo envelopes feature fluctuations on the scale of $\lambda_\text{deB}$, while the large-scale-averaged density profile fits an NFW-like shape, which has a typical scale of variation that is much larger than $\lambda_\text{deB}$.
In our work, we will need to recover similar large-scale density variations in halo envelopes while smoothing over the small-scale fluctuations.
We can see how the smoothing procedure accomplishes this by replacing 
the plane wave in \S\ref{sec:static} by one with the same
(small-scale) sinusoidal oscillations, but modulated by a Gaussian 
that varies more slowly with position:
\begin{align}
    &\psi = \sqrt{\rho_0} e^{-x^2/4L^2} \sin(2\pi x/\lambda) \\
    &\rho = \rho_0 e^{-x^2/2L^2} \sin^2(2\pi x/\lambda)
\end{align}
Smoothing over only the small-scale ($\lambda$) but resolving the large-scale ($L$) requires $L \gg \eta \gg \lambda$.
The exact quantum pressure corresponding to this density distribution is
\begin{align}
    \Pi &= \frac{\rho_0}{2} e^{-x^2/2L^2} \Big(\frac{2\pi \hbar}{m\lambda}\Big)^2 \left[ 1 + \Big(\frac{\lambda}{4\pi L}\Big)^2 \Big( 1 - \cos(4\pi x/\lambda) \Big) \right]\nonumber\\
    &\approx \frac{\rho_0}{2} e^{-x^2/2L^2} \Big(\frac{2\pi \hbar}{m\lambda}\Big)^2
\end{align}
This demonstrates that the characteristic length scale of quantum pressure (i.e. the de~Broglie wavelength) is that of the small-scale fluctuations, not the large-scale variation:
\begin{equation}
    \lambda_\text{deB} = \overline{\frac{h}{m\sqrt{\Pi/\rho}}} = \lambda
\end{equation}
Since the quantum pressure is not constant in this case, we can also calculate the acceleration it produces:
\begin{align}
    &a_\Pi = -\frac{1}{\rho} \frac{\partial \Pi}{\partial x} \approx \frac{1}{2 \sin^2(2\pi x/\lambda)} \Big(\frac{2\pi \hbar}{m\lambda}\Big)^2 \frac{x}{L^2} \label{eq:aPi-avg}
\end{align}

Following the same smoothing procedure as before and taking the limit $L \gg \eta \gg \lambda$, the smoothed fluid quantities are found to be
\begin{equation}
    \hat{\rho} = \frac{\rho_0}{2} e^{-\frac{x^2}{2L^2+\eta^2}} \sqrt{\frac{2L^2}{2L^2 + \eta^2}} \approx \frac{\rho_0}{2} e^{-x^2/2L^2} 
\end{equation}
\begin{equation}
    \hat{v} = 0 
\end{equation}
\begin{align}
    \langle p^2 \rangle &= \frac{\rho_0/2}{\hat{\rho}} e^{-\frac{x^2}{2L^2+\eta^2}} \sqrt{\frac{2L^2}{2L^2 + \eta^2}} \Big(\frac{2\pi \hbar}{\lambda}\Big)^2 \nonumber \\
    &\quad\quad\quad\times\left[ 1 + \frac{\lambda^2}{8\pi^2\eta^2}\frac{2L^2 + \eta^2}{2L^2} \left( 1 - \cos\Big(\frac{2L^2}{2L^2 + \eta^2} \frac{4\pi x}{\lambda}\Big) \right) \right] \nonumber \\
    &\approx \Big(\frac{2\pi \hbar}{\lambda}\Big)^2
\end{align}
\begin{align}
    \hat{\sigma}^2 &= \frac{\langle p^2 \rangle - \langle p \rangle^2}{m^2} \approx \Big(\frac{2\pi \hbar}{m \lambda}\Big)^2 
\end{align}
\begin{align}
    P &= \hat{\rho}\hat{\sigma}^2 \approx \frac{\rho_0}{2} e^{-x^2/2L^2} \Big(\frac{2\pi \hbar}{m \lambda}\Big)^2 \approx \Pi
\end{align}
\begin{equation}
    \hat{a}_\Pi = -\frac{1}{\hat{\rho}} \frac{\partial \hat{\Pi}}{\partial x} \approx \Big(\frac{2\pi \hbar}{m\lambda}\Big)^2 \frac{x}{L^2} 
\end{equation}
These results exhibit smoothing over the small-scale oscillations, effectively replacing $\sin^2(2\pi x/\lambda) \rightarrow 1/2$ in the exact expressions, 
while leaving the large-scale Gaussian profile in tact.
Crucially, however, the small-scale oscillations are ``gone, but not forgotten''; their impact is reflected in the quantum pressure and acceleration.
These terms are larger by a factor of $(4\pi L/\lambda)^2$ than they would be if derived from a density distribution in which the small-scale fluctuations were absent.
In other words, if we resolved only the smoothed, large-scale Gaussian density profile ($\hat{\rho}$), and erroneously used that to calculate the quantum pressure term for the purpose of modeling SFDM dynamics, the result would be
\begin{align}
    \Pi_\text{err} &= \Big(\frac{\hbar}{2m}\Big)^2 \left[ \frac{1}{\hat{\rho}} \Big(\frac{\partial \hat{\rho}}{\partial x}\Big)^2 - \frac{\partial^2 \hat{\rho}}{\partial x^2} \right]  \nonumber \\
    &= \frac{\rho_0}{2} e^{-x^2/2L^2} \Big(\frac{\hbar}{2 m L}\Big)^2 \ll \Pi
\end{align}
This is why it is necessary to account for the effects of de~Broglie-scale variations on SFDM dynamics, as the smoothing procedure of \S\ref{sec:classical} does, even if that scale seems too small to be relevant, astrophysically; failing to do so could dramatically underestimate the quantum pressure.

\section{Artificial viscosity in the 1D-spherical hydrodynamics code} \label{sec:avis}

The gravitational collapse we model here, starting from cold initial conditions that lead to highly supersonic infall and the formation of a strong accretion shock at small radius, requires some extra care with regard to the numerical treatment of shocks. If we were solving this problem with physical viscosity included, as in the Navier-Stokes equations, we would expect the thickness of the shock jump transition layer to be several times the mean-free-path of the microscopic process that mediates the shock.  In fact, this scale is assumed to be uninterestingly small and unresolved by our numerical method, with numerical diffusion that would, ideally, smooth the shock transition over the thickness of a few
mass shells.  In practice, however, numerical diffusion is not sufficient to stabilize the code in the presence of shocks, so we employ a standard von Neumann-Richtmeyer artificial viscosity to handle their formation and propagation accurately and stably, spreading the transition over a few shell widths. In the standard prescription, an artificial viscous pressure term, $P_\text{vis}$, is added to the momentum and energy equations (i.e. $P_{\!\sigma} \rightarrow P_{\!\sigma} + P_\text{vis}$ in equations~\ref{eq:1momPPSI} and \ref{eq:2momSpherical}) so as to provide an additional effective pressure to help decelerate and heat the gas as it is about to encounter a shock, thereby boosting the dispersion pressure of the fluid enough to help bring it to rest at the right time, while satisfying the Rankine-Hugoniot shock jump conditions. At each time-step, and for each shell with local mass density $\rho$, the viscous pressure is given by
\begin{equation}
    P_\text{vis} = 
    \begin{cases}
        \frac{2 c_\text{vis} (\Delta v)^2}{1/\rho - 1/\rho_\text{old}} & \Delta v < 0 \\
        0 & \Delta v > 0
    \end{cases}
\end{equation}
where $\rho_\text{old}$ is the local mass density of the shell from the previous time-step, $\Delta v = v_\text{outer} - v_\text{inner}$ is the difference in radial velocity between the outer and inner boundaries of the shell, and $c_\text{vis}$ is a tuning parameter that sets the magnitude of the viscous pressure and determines the number of shells over which the shock is spread out. We use $c_\text{vis}=4$, corresponding to a shock width of about 4 shells. In order to prevent $P_\text{vis}$ from spuriously heating the shells before they need to be shocked, the artificial viscosity is only active when $\Delta v < 0$, meaning the shell boundaries are approaching each other, which is a necessary condition for shock formation.

In our simulations, we noticed that this standard prescription still produced some spurious heating at small radii prior to shock formation. This was most noticeable in the CDM runs because it resulted in the CDM density profile flattening out into the shape of a core at very small radii due to the excess pressure there.
In previous work by \citet{Shapiro96}, the problem of spurious heating was dealt with by disabling artificial viscosity in the energy equation until the formation of a shock was imminent, while leaving the prescription for the momentum equation unaltered. 
For their adaptive smoothed particle hydrodynamics (ASPH) code, they set a criterion that predicted when a particle crossing was about to occur, which, without artificial viscosity, would result in a caustic.
Viscous heating in the energy equation was enabled for a particle only once it satisfied this predictive criterion, allowing it to be shocked at the right time while minimizing the chance of spurious heating prior to shock formation.
Analogously, we added a criterion for enabling artificial viscosity in the energy equation of our 1D-spherical hydrodynamics code, which predicts the occurrence of \textit{shell} crossing.
In order for $P_\text{vis}$ to be active in the energy equation for a given shell, we require (in addition to $\Delta v < 0$) that the shell-crossing time for that shell be less than some multiple of the global time-step, $\Delta t$:
\begin{equation}
    \Delta t_\text{cross} = c_\text{cross}\Big|\frac{\Delta r}{\Delta v}\Big| < N\Delta t
    \label{eq:cross-crit}
\end{equation}
where $\Delta r$ is the width of the shell, and $c_\text{cross}$ is a tuning parameter which we set to 0.05, as was done in \citet{AS07}. 
This criterion further ensures that only the shells for which a shock is imminent have viscous heating turned on in their energy equation, which helps prevent the excess heating near the center of the CDM profile prior to shock formation, thereby hindering the formation of the artificial core. In practice, we find $N=3$ to be a suitable choice, as higher values do not have a strong enough effect, and lower values prevent the shock from ever forming. 
It is worth emphasizing that this additional criterion only affects the CDM profiles at very small radii; outside of the artificial core, the CDM results are the same with or without this alteration.

In the SFDM-TF runs, the innermost shells are brought to rest by SI pressure, rather than dispersion pressure (artificial or otherwise). However, since these shells still collapse and contract on their way to equilibrium, they satisfy the artificial viscosity criteria and trigger spurious heating, even with the additional requirement of equation~(\ref{eq:cross-crit}) (although it does reduce the amount of spurious heating slightly). This can be seen as the flat central region of the SFDM-TF dispersion pressure and ``temperature'' profiles in Fig.~\ref{fig:numprof}.
Regardless, even with this spurious heating, the dispersion pressure of these shells is negligible compared to their SI pressure (see, e.g., the $U_\text{gas}$ and $U_\text{SI}$ curves in Fig.~\ref{fig:virial}), so their dynamics are unaffected. As such, the addition of equation~(\ref{eq:cross-crit}) has no effect on the SFDM-TF density profile.

\section{Scalar virial theorem with boundary terms for SFDM haloes} \label{sec:virial}

We derive the virial theorem for a self-gravitating 
SFDM halo in the Newtonian regime. For isolated objects in static
equilibrium, the virial
theorem for BECs was derived from the NLSE, assuming a harmonic
external trap potential, by \citet{Dalfovo99}, 
based on a variational principle involving the total energy.  This 
approach was extended to the case of a self-gravitating object in \citet{Wang01}.  
In \citet{RDS12}, we presented the virial theorem for such objects, as derived from 
the hydrodynamical formulation that follows from the Madelung 
transformation. 
Realistic haloes that form from gravitational instability and collapse, however,
are not isolated, but are rather
embedded in a background density field
and subject to infall, as studied in this paper. 
We shall here derive the virial theorem more generally, therefore,
by including surface terms that were neglected by the previous treatments, 
which limited them to isolated objects. We will again follow
the hydrodynamical formulation as in \citet{RDS12}, but now 
accounting for infall by our inclusion of 
boundary terms in the virial, as in \cite*{SIR99}. 
In the main body of this paper, we restrict ourselves to spherical symmetry and focus on the TF regime, where the de Broglie wavelength is small, allowing us to approximate the large-scale effects of quantum pressure with a velocity-dispersion pressure that comes from the collisionless Boltzmann equation (CBE).
Here, for the sake of completeness, we will first derive the virial theorem in full 3D with an exact formulation for quantum pressure, and then provide the appropriate form for the 1D-spherical TF regime approximation explored in this work, at the end.

We start from the exact 3D momentum equation for the SFDM model in Madelung form (equation~\ref{eq:momentum-Madelung}):
\begin{equation}\label{eq:newmom}
 \rho \frac{\partial \bm{v}}{\partial t} + \rho (\bm{v}\cdot \nabla)\bm{v} = -\rho \nabla Q - \rho \nabla \Phi - \nabla P_\text{SI}
\end{equation}
Now, the usual procedure with the momentum equation (\ref{eq:newmom}) can be adopted by which we
take the scalar product with the position vector $\bm{r}$, and
integrate the resulting equation over the volume $V$ interior to
some closed surface $S$, which surrounds the system:
\begin{equation} \label{eq:vir}
    \int_V \rho \frac{D\bm{v}}{D t} \cdot \bm{r}~ d^3\bm{r} = -\int_V \big( \rho
    \nabla Q  + \rho \nabla \Phi + \nabla P_\text{SI} \big) \cdot \bm{r} ~ d^3\bm{r}
\end{equation}
where $D/Dt$ is the Lagrangian derivative given by
\begin{equation}
    \frac{D}{Dt} = \frac{\partial }{\partial t} + \bm{v} \cdot \nabla
\end{equation}
Let us calculate the integrals term by term: the left-hand side in
equation~(\ref{eq:vir}) can be written as
\begin{align}
    \int_V \rho \frac{D\bm{v}}{D t} \cdot \bm{r} ~d^3\bm{r} &= \mathcal{V} - 2T + S_L = \mathcal{V} - 2\mathcal{T}
\end{align}
where the virial ($\mathcal{V}$) and inertia ($I$) are given by
\begin{align}
    &\mathcal{V} \equiv \frac{1}{2}\frac{\partial^2 I}{\partial t^2} \\
    &I \equiv \int_V \rho r^2 ~d^3\bm{r}
\end{align}
and the kinetic energy due to bulk motion ($T$) and its corresponding surface term ($S_L$) are given by
\begin{equation}
    T \equiv \frac{1}{2}\int_V \rho v^2 ~d^3\bm{r}
\end{equation}
\begin{equation}
    S_L \equiv \int_S \left(\rho x_j v_j \bm{v} + \frac{1}{2}x_jx_j\frac{\partial}{\partial t}(\rho \bm{v})\right) \cdot d\bm{S}
\end{equation}
We also define the combined volume+surface kinetic term
\begin{equation}
    \mathcal{T} \equiv T - S_L/2
\end{equation}

The quantum potential term on the right-hand side of equation~(\ref{eq:vir})
can be re-written as
\begin{align}
  -\int_V \rho \nabla Q \cdot \bm{r}~ d^3\bm{r} &= \frac{\hbar^2}{2m^2}\int_S \left(\sqrt{\rho}\nabla^2 \sqrt{\rho}\right) \bm{r}\cdot d\bm{S} \nonumber \\
  &+ \frac{\hbar^2}{2m^2}\int_V \left(-\frac{\nabla^2 \sqrt{\rho}}{\sqrt{\rho}}\right)(\bm{r}\cdot \nabla \rho + 3\rho) ~d^3\bm{r}
\end{align}
Let us denote the first term as $S_Q^{(1)}$, while the second
term can be further reduced to yield
\begin{align} \label{kin1}
   \frac{\hbar^2}{2m^2}&\int_V \left(-\frac{\nabla^2 \sqrt{\rho}}{\sqrt{\rho}}\right)(\bm{r}\cdot \nabla \rho + 3\rho) ~d^3\bm{r} \nonumber\\
   &= 3K_Q - \frac{\hbar^2}{2m^2}\int_V 2 \left(\nabla^2 \sqrt{\rho}\right) \bm{r} \cdot \nabla \sqrt{\rho}~ d^3\bm{r}
\end{align}
  with the quantum-kinetic energy
\begin{equation} \label{quantumkin}
  K_Q \equiv \frac{\hbar^2}{2m^2} \int_V (\nabla \sqrt{\rho})^2 ~d^3\bm{r}
\end{equation}
In order to evaluate the second term in equation~(\ref{kin1}), we use the short-hand notation
$(\nabla \sqrt{\rho})_i = a_i$, and write the integral as
\begin{align}
  \int_V \left(\nabla^2 \sqrt{\rho}\right) \bm{r} \cdot \nabla \sqrt{\rho}~ d^3\bm{r} = \int_V (x_i \partial_j(a_i a_j) - x_i a_j\partial_j a_i) d^3\bm{r}
\end{align}
Using $(\bm{r} \times \nabla \sqrt{\rho})(\nabla \times \nabla
\sqrt{\rho}) = 0$ implies $x_ia_j\partial_ja_i = x_ia_j\partial_ia_j$.
Thus, we have
\begin{align}
  \frac{\hbar^2}{2m^2} 2 &\int_V (x_i \partial_j(a_i a_j) - x_i a_j\partial_j a_i)~d^3\bm{r} \nonumber\\
  &= \frac{\hbar^2}{2m^2} 2 \int_V \left(\partial_j (x_i a_i a_j) - a_i a_j\partial_j x_i - \frac{1}{2} x_i \partial_i(a_j^2)\right)~d^3\bm{r} \nonumber\\
  &=  \frac{\hbar^2}{2m^2} 2 \int_V \partial_j(x_i a_i a_j)~d^3\bm{r} \nonumber \\
  &\quad + \frac{\hbar^2}{2m^2} 2 \int_V \left(-a_i a_j\delta_{ij} - \frac{1}{2}\partial_i(x_ia_j^2) + \frac{3}{2}a_j^2\right) ~d^3\bm{r} \nonumber\\
  &\equiv S_Q^{(2)} + \frac{\hbar^2}{2m^2}\int_V (\nabla \sqrt{\rho})^2  ~d^3\bm{r} \nonumber\\
  &= S_Q^{(2)} + K_Q
\end{align}
with another surface term $S_Q^{(2)}$. Collecting all
terms, the contribution from the quantum potential to the virial is given by
\begin{equation}\label{eq:quanta}
    -\int_V \rho \nabla Q \cdot \bm{r}~d^3\bm{r} = 2K_Q + S_Q^{(1)} - S_Q^{(2)}
\end{equation}
The gravitational term in equation~(\ref{eq:vir}) is calculated as usual, using
\begin{align}
   \mathcal{W} &\equiv -\int_V \rho \bm{r} \cdot \nabla \Phi d^3\bm{r} \nonumber \\
               &= \frac{1}{2}\int_V \rho \Phi ~d^3\bm{r} \nonumber \\
               &\quad - \frac{1}{4\pi G}\sum_i \int_S \left(x_jg_jg_i - \frac{1}{2}x_ig_jg_j - \frac{1}{2}\Phi g_i\right)~dS_i \nonumber \\
               &\equiv W + S_G
\end{align}
where we denoted $\bm{g} \equiv - \nabla \Phi$. $W$ and $S_G$ represent
the gravitational potential energy and its associated surface
term, respectively.

Finally, we get to the contribution from the repulsive SI in equation~(\ref{eq:vir}), which is of importance to our work in this paper. It can
be calculated as
\begin{align}
   -\int_V \bm{r} \cdot \nabla P_\text{SI} ~d^3\bm{r} &= -\int_S P_\text{SI}\bm{r}\cdot d\bm{S} + 3\int_V P_\text{SI} ~d^3\bm{r} \nonumber\\
   &\equiv S_\text{SI} + 3U_\text{SI} \label{eq:SIvir}
\end{align}
Again, $S_\text{SI}$ is a surface term and $U_\text{SI}$ corresponds to the energy due to SI.  

Collecting all terms from above, we shall arrive at the
following expression for the virial,
\begin{equation} \label{generalvir}
    \mathcal{V} = 2\mathcal{T} + \mathcal{W} + 2K_Q + 3U_\text{SI} + S_Q^{(1)} - S_Q^{(2)} + S_\text{SI}
\end{equation}

We immediately observe the following differences from the CDM
case (i.e. collisionless, self-gravitating gas, following the CBE). First, we have three additional surface terms, one from
self-interaction $S_\text{SI}$ and the other two from the quantum potential.
They all depend non-trivially on the
density at the surface.  Virial equilibrium
demands that $\mathcal{V} = 0$ in equation~(\ref{generalvir}). For haloes embedded in a cosmological background density field, all the relevant surface terms must be included. For isolated haloes (i.e. $\rho = 0$ outside of the surface $S$),    
we recover the virial theorem in the form of
\begin{equation} \label{isolatedvir}
  2T + W + 2K_Q + 3U_\text{SI} = 0
\end{equation}
a relation which has been applied extensively in the literature previously (e.g. in \cite{RDS12} we neglected the quantum potential, $K_Q = 0$, while $T$ described bulk rotation of halo cores). 
Differing from the usual case for which $T=|W|/2$, the appearance of $K_Q$ and $U_\text{SI}$ also leads to the following inequalities:
$T+K_Q < |W|/2$ for repulsive SI ($g > 0$, i.e. the case considered in this work),
while $T+K_Q > |W|/2$ for attractive SI ($g < 0$).  

In \S\ref{sec:fluidapprox}, we derived a fluid approximation that replaces the NLSE (as well as its QHD formulations) by the familiar hydrodynamical conservation equations of mass, momentum, and energy for an ideal gas whose pressure accounts for the large-scale dynamical effect of quantum pressure, but avoids the necessity of resolving its complex behavior on the small scales of the de~Broglie wavelength and below. These approximate fluid equations differ from the exact QHD equations of the Madelung formulation, in which the Euler equation, equation (\ref{eq:newmom}), represents the effect of quantum pressure by the ($\nabla Q$)-term. In particular, by smoothing the NLSE in phase space over scales larger than the de~Broglie wavelength, the large-scale behavior of quantum pressure was shown to be encoded in a velocity-dispersion pressure, $P_{\!\sigma}$, which replaces the ($\nabla Q$)-term of the Euler equation for the exact QHD formulation with a familiar ($\nabla P_{\!\sigma}$)-term. In spherical symmetry, this  $P_{\!\sigma}$ obeys the equation of state and internal energy equation of an ideal gas with a ratio of specific heats $\gamma = 5/3$, for which the ``temperature'' is given by the ``microscopic'' velocity dispersion of the field.
To derive the virial theorem in a form that is appropriate for this fluid approximation, we start with the approximate momentum equation obtained from the 1$^\text{st}$ moment of the CBE, equation~(\ref{eq:1momPPSI}):
\begin{equation}
    \rho\frac{\partial v}{\partial t} + \rho v \frac{\partial v}{\partial r} = - \frac{\partial P_{\!\sigma}}{\partial r} - \rho\frac{\partial \Phi}{\partial r} - \frac{\partial P_\text{SI}}{\partial r}
\end{equation}
This equation is identical to equation~(\ref{eq:newmom}) (in spherical symmetry) except that the quantum potential term has been replaced by the velocity-dispersion pressure term. 
Therefore, each term's contribution to the total virial can be calculated using the same procedure as above, with the velocity-dispersion pressure term following the same steps as for the SI pressure (equation~\ref{eq:SIvir}):
\begin{align}
   -\int 4\pi r^3 \frac{\partial P_{\!\sigma}}{\partial r} dr &= -4\pi r^3 P_{\!\sigma} + 3\int 4\pi r^2 P_{\!\sigma} dr \nonumber\\
   &\equiv S_{\!\sigma} + 2U_{\!\sigma} \label{eq:gasvir}
\end{align}
where $U_{\!\sigma}$ is the internal energy of a $\gamma = 5/3$ ideal gas and $S_{\!\sigma}$ is the corresponding surface pressure.
Thus, the total virial in our approximation scheme reads as
\begin{equation}
    \mathcal{V} = 2\mathcal{T} + \mathcal{W} + 2U_{\!\sigma} + 3U_\text{SI} + S_{\!\sigma} + S_\text{SI}
\end{equation}


\bsp
\label{lastpage}
\end{document}